\documentclass{emulateapj}

\def\mj{$M_{\rm J}\ $}
\def\rj{$R_{\rm J}\ $}
\def\etal{{et~al.\,}}

\def\mp{M$_{\rm p}$}
\def\rp{R$_{\rm p}\,$}

\def\sles{\lower2pt\hbox{$\buildrel {\scriptstyle <}
   \over {\scriptstyle\sim}$}}
\def\sgreat{\lower2pt\hbox{$\buildrel {\scriptstyle >}
   \over {\scriptstyle\sim}$}}

\slugcomment{Accepted to Ap.J.}

\begin{document}

\title{Theoretical Spectra and Light Curves of Close-in 
Extrasolar Giant Planets and Comparison with Data} 

\author{A. Burrows\altaffilmark{1}, J. Budaj\altaffilmark{1,2} \& I. Hubeny\altaffilmark{1}} 

\altaffiltext{1}{Department of Astronomy and Steward Observatory, 
                 The University of Arizona, Tucson, AZ \ 85721;
                 burrows@zenith.as.arizona.edu, budaj@as.arizona.edu, hubeny@aegis.as.arizona.edu}

\altaffiltext{2}{Astronomical Institute, Tatranska Lomnica, 05960 Slovak Republic}

\begin{abstract}

We present theoretical atmosphere, spectral, and light-curve models for
extrasolar giant planets (EGPs) undergoing strong irradiation for which
{\it Spitzer} planet/star contrast ratios or light curves have been published (circa June 2007).
These include HD 209458b, HD 189733b, TrES-1, HD 149026b, HD 179949b, and $\upsilon$ And b.
By comparing models with data, we find that a number of
EGP atmospheres experience thermal inversions and have stratospheres.
This is particularly true for HD 209458b, HD 149026b, and $\upsilon$ And b.
This finding translates into qualitative changes in the planet/star contrast
ratios at secondary eclipse and in close-in EGP orbital light curves.
Moreover, the presence of atmospheric water in abundance is fully consistent with
all the {\it Spitzer} data for the measured planets.  For planets
with stratospheres, water absorption features invert into emission features
and mid-infrared fluxes can be enhanced by a factor of two. In addition,
the character of near-infrared planetary spectra can be radically
altered. We derive a correlation between the importance of such stratospheres
and the stellar flux on the planet, suggesting that close-in EGPs bifurcate
into two groups: those with and without stratospheres. From the finding
that TrES-1 shows no signs of a stratosphere, while HD 209458b does,
we estimate the magnitude of this stellar flux breakpoint.
We find that the heat redistribution parameter, P$_n$, for the
family of close-in EGPs assumes values from $\sim$0.1 to $\sim$0.4.
This paper provides a broad theoretical context for the future direct
characterization of EGPs in tight orbits around their illuminating stars.

\end{abstract}


\section{Introduction}
\label{intro}

To date, more than 250 extrasolar planets have been discovered and
more than 29 of them are transiting their primary star\footnote{See J. Schneider's
Extrasolar Planet Encyclopaedia at http://exoplanet.eu, the Geneva Search Programme at
http://exoplanets.eu, and the Carnegie/California compilation at http://exoplanets.org}.  
One transiting planet is a ``Neptune" (GJ 436b), but the rest are giant planets 
with an impressively wide range of masses
and radii that speak to the heterogeneity of the family of close-in 
EGPs ({\bf E}xtrasolar {\bf G}iant {\bf P}lanets).  Table \ref{t1} lists
these transiting EGPs and some of their relevant properties, along with
many of the references to the observational and discovery papers from which 
these data were taken.  Table \ref{t2} lists useful data for the corresponding 
primary stars, including their masses, luminosities, radii, and approximate 
ages. Both tables are in order of increasing planetary semi-major axis and,
considering the pace of the field, both should be considered provisional. 
Not shown are the eccentricities, which are generally small, but which for HAT-P-2b, GJ 436b, and 
XO-3b are $\sim$0.507, $\sim$0.14, and $\sim$0.22, respectively. For these three 
close-in EGPs, significant tidal heating and, perhaps, forcing by an unseen 
companion are implied.  

Radial-velocity measurements for a non-transiting EGP provide a lower
limit to its mass, but little else.  However, the transiting EGPs yield
radii as well, and resolve the $\sin{i}$ ambiguity to reveal the planets' masses.
These data provide physical constraints with which detailed evolutionary and structural 
models that incorporate irradiation and migration can be tested (see, e.g., 
Burrows et al. 2007a; Guillot et al. 2006).  With superb photometric accuracy, the wavelength-dependence 
of the transit radii can in principle provide a measure of a planet's atmospheric 
composition (Fortney et al. 2003).  In this way, sodium has been 
detected in HD 209458b (Charbonneau et al. 2002) and water has been identified
in both HD 189733b (Tinetti et al. 2007, but see Ehrenreich et al. 2007) and HD 209458b (Barman 2007).  
Moreover, high-precision optical photometry has constrained
(perhaps, measured) the geometric albedo of the close-in EGP HD 209458b
(Rowe et al. 2006, 2007).  HD 209458b's optical albedo is very low ($\sim$3.8$\pm4.5$\%), in keeping with 
the predictions of Sudarsky et al. (2000) when the alkali metals, and not 
clouds, dominate absorption in the atmosphere and Rayleigh scattering
dominates scattering.

Nevertheless, using current technology, transit measurements have 
limited utility in characterizing the atmospheres and compositions of these 
planets.  Astronomers require more direct detections of the planet's spectrum to 
probe its chemistry and atmospheric properties. This is in the tradition of 
remote sensing in the solar system and of the Earth.  Until recently, it had 
been thought that the light from an extrasolar planet had to be separated from
under the glare of its parent star using high-angular resolution, extremely high-contrast 
imaging. This is still the case in the optical for the cool ``wide-separation" EGPs 
(Burrows et al. 2004; Sudarsky et al. 2005; Burrows 2005) and terrestrial planets,
for which the planet-star contrast ratio is $\sim$10$^{-9}$ to $\sim$10$^{-10}$,
but such performance has not yet been demonstrated. 

However, for the hot close-in 
EGPs, the planet-star contrast ratios in the mid-infrared are 
much more favorable (Burrows, Sudarsky, \& Hubeny 2003,2004),  
oftimes exceeding 10$^{-3}$.  This capability has led to a breakthrough
in the study of exoplanets. With the infrared space telescope 
{\it Spitzer} (Werner \& Fanson 1995), using its IRAC and MIPS cameras 
and the IRS spectrometer, one can now measure the summed light of the 
planet and the star in and out of secondary eclipse, and thereby,
from the difference, determine the planet's spectrum at superior conjunction.  
Moreover, for a subset of the closest EGPs it is possible to use {\it Spitzer} to measure 
their flux variations with planetary phase between transit and secondary eclipse.
Hence, for the close-in EGPs in the near- to mid-infrared, and without
the need to separately image planet and star, the direct detection of
planetary atmospheres via low-resolution spectroscopy and precision infrared (IR) photometry
is now a reality.  

The secondary eclipse fluxes have now been measured for five transiting EGPs (HD 189733b,
TrES-1, HD 209458b, HD 149026b, GJ 436b), but not yet in all {\it Spitzer} bands.
In addition, using the IRS spectrometer, spectra between $\sim$7.5
$\mu$m and $\sim$15 $\mu$m of two transiting EGPs at
secondary eclipse have been obtained (HD 189733b [Grillmair et al. 2007]
and HD 209458b [Richardson et al. 2007]). Though at very low resolution, these are the
first measured spectra of any extrasolar planet.
Furthermore, light curves have been measured for three EGPs ($\upsilon$ And b 
[Harrington et al. 2006: at 24 $\mu$m], HD 179949b [Cowan et al. 2007: at 
8 $\mu$m], and HD 189733b [Knutson et al. 2007b: at 8 $\mu$m]). Only one of these (HD 189733b) is transiting
and has an absolute calibration. For none of the latter three are there 
extant light-curve measurements for more than one {\it Spitzer} band;  
for some of these EGPs only upper limits in a few of the other bands have been 
determined.  Table \ref{t3} summarizes all the direct 
detection data for the EGP family obtained to date (circa June 1, 2007), along 
with associated references, comments, and table notes. Clearly, in the next 
year or two we can expect a great deal more secondary eclipse and light-curve 
data in the various {\it Spitzer} IRAC and MIPS bands.  However,  
there has already been significant progress in measuring
EGP atmospheres.

This paper is a continuation of our series of interpretative studies (e.g., Burrows et al. 2005, 2006, 2007b) 
of the direct measurements of close-in EGPs.  Here, we analyze the secondary eclipse and light-curve
data summarized in Table \ref{t3} for the EGPs HD 189733b, HD 209456b, TrES-1,
HD 149026b, HD 179949b, and $\upsilon$ And b and make theoretical predictions in support of
future {\it Spitzer} planet measurement campaigns. Importantly, by fitting the current data we  
extract physical information concerning the atmospheres, compositions, and thermal 
profiles of these first six objects listed in Table \ref{t3}.  We have explored the 
dependence of the spectra and light curves on the heat redistribution factor 
P$_n$ (Burrows et al. 2006), on atmospheric metallicity, and on the possible
presence of a stratospheric absorber.  The recent analysis by Burrows et al. (2007b)
of the IRAC data of Knutson et al. (2007c) indicates that HD 209458b boasts a thermal inversion
that radically alters the {\it Spitzer} fluxes and their interpretation.  In fact, 
in Burrows et al. (2007b), we speculate that thermal inversions and stratospheres
may play a role in the atmospheres of many close-in EGPs and are a new feature in the 
study of transiting planets. A similar conclusion was reached by Fortney et al. (2006),
specifically in the context of HD 149026b.

We find that we can fit all the secondary eclipse and light-curve data, 
except for the nightside flux of HD 189733b and its day/night contrast.  
While we can fit its dayside secondary eclipse flux, we suspect that 
HD 189733b will require a more sophisticated day/night 
redistribution model than we now employ (\S\ref{tech}).  
We note that Knutson et al. (2007b) conclude that the dimmest and 
brightest spots on HD 189733b reside on the same 
hemisphere and that the dimmest spot is shifted from the anti-stellar
point by as much as $\sim$30$^{\circ}$. Our current light curve models
are symmetric about the peak. 

We find that the degree of longitudinal 
heat redistribution (P$_n$) may vary from planet to planet, hinting at a variety
of meteorological conditions and day/night contrasts within the family of close-in 
EGPs.  Moreover, as also concluded in Burrows et al. (2005), we can not obtain
good fits at secondary eclipse without the presence of water in abundance in the 
atmospheres of these irradiated EGPs.  This is particularly true for 
TrES-1 and HD 209458b. Though we will not dwell in this paper 
on metallicity, we find that the metallicity dependence of the secondary-eclipse fluxes 
is not strong, but that it is in principle measurable, and that the metallicity 
dependence of the variation of the planetary flux with phase is only modest.
Importantly, we also conclude that upper-atmosphere absorption in the optical
by an as-yet unknown molecule, and the concommitent thermal inversions 
and stratospheres, provide better fits to some of the data.

In \S\ref{tech}, we describe our numerical techniques, the new heat redistribution model,
and how we generate stratospheres. This section is supported with 
Appendices \S\ref{redist}, \S\ref{app2}, \S\ref{app3}, and \S\ref{app5}, 
in which we provide details concerning the heat redistribution model and derive 
some analytic formulae concerning atmospheric physics with day-night 
coupling.  In particular, in \S\ref{app5}, we address the enhancement at 
secondary eclipse in the integrated planetary flux at Earth over and above 
what would be expected if the planet emitted isotropically. For a radar antenna,
this would be its ``gain" factor.  
In \S\ref{tp}, we discuss the derived temperature-pressure
profiles on the day and the night sides for all six EGPs highlighted in this investigation. We show
that the $\tau = 2/3$ decoupling layers for the {\it Spitzer} IRAC and MIPS (24-$\mu$m) 
band fluxes are all above the isothermal region of an irradiated EGP's atmosphere 
and, hence, that {\it Spitzer} does not probe these deeper regions.  For the same reasons, we
find that the presence of a thermal inversion at altitude and of a stratosphere in
some EGP atmospheres can significantly alter these {\it Spitzer} fluxes and their relative strengths.  
In \S\ref{tp}, we also provide representative planet spectra to demonstrate that most of the planet's 
flux emerges at shorter wavelengths than are accessible to {\it Spitzer}, and, hence,
that {\it Spitzer} probes only a small tail of the emergent flux distribution. 
This may be of relevance when JWST is available to follow up on the 
{\it Spitzer} EGP data and, even earlier, as the JWST exoplanet campaign is being designed.  

In \S\ref{planetstar}, we present the best-fit planet-star contrast ratios at secondary 
eclipse for four of the transiting EGPs for which these have been measured (all 
but GJ 436b, for which see Deming et al. 2007 and Demory et al. 2007), as well 
as various comparison models to gauge a few parameter dependences.  
Then, in \S\ref{curves} we match our theoretical phase light curves with 
the three measured light curves and derive approximate planetary parameters.  
The paper is brought to a close in \S\ref{conclusions} with a synopsis 
of our results and a general discussion of the issues raised.

\section{Numerical Techniques}
\label{tech}

Our model atmospheres are computed using the updated code {\sc CoolTLUSTY}, described
in Sudarsky, Burrows, \& Hubeny (2003), Hubeny, Burrows, \& Sudarsky (2003),
and Burrows, Sudarsky, \& Hubeny (2006), which is a variant of the
universal spectrum/atmosphere code {\sc TLUSTY} (Hubeny 1988; Hubeny \& Lanz 1995).
The molecular and atomic opacities are taken from Sharp \& Burrows (2007)
and the chemical abundances, which include condensate rainout, are derived 
using the thermochemical model of Burrows \& Sharp (1999), updated as described 
in Sharp \& Burrows (2007) and in Burrows et al. (2001).

To handle convection, we use standard mixing-length theory,
with a mixing length equal to the pressure scale height.  The stellar
irradiation boundary condition is numerically challenging, and has
not been done properly by some workers in the past.  To ensure
an accurate numerical solution with a non-zero incoming specific intensity, 
we use the formalism discussed in Hubeny, Burrows, \& Sudarsky (2003).
The stellar spectral models are taken from Kurucz (1994) for the six stars
listed in Table \ref{t2} that are the primaries of the EGPs upon which we 
focus in this paper.  The day and night sides are approached differently,
with the nightside, quite naturally, experiencing no incident flux, but
receiving heat from the irradiated dayside using the new algorithm described in 
Appendix \S\ref{redist}.  An important additional feature of our new heat redistribution
formalism is the capacity to match both the entropy and the gravity at the base
of both the day- and the night-side atmospheres.  Since the inner convective zone,
which constitutes most of the planet, is isentropic, this is the physically
correct procedure.  For a dayside calculation, we can assume a given interior
flux effective temperature, $T_{\rm int}$ (a standard number could be 75 K).
For a given gravity and irradiation regime, this leads to an atmosphere
solution on the dayside.  This solution incorporates an entropy in
the convective zone.  For the nightside atmosphere, we can adjust the nightside
$T_{\rm int}$ until the entropy in its convective interior matches that
found in the dayside convective zone.  One product of this procedure is a connection
between dayside and nightside $T_{\rm int}$ that has a bearing on overall
planet cooling and shrinkage (Burrows et al. 2007a).  However, since for a given
measured planet radius, this mapping does not have a significant effect on the 
close-in planet's spectrum, we do this here only approximately and leave 
to a later paper a general discussion of this topic.  

The simple parametrization we use to simulate the effects of
an extra stratospheric absorber entails placing an absorber
with constant opacity, $\kappa_{\rm e}$, in the optical frequency
range $(\nu_0, \nu_1)$ = ($3\times 10^{14}$ Hz, $7\times 10^{14}$ Hz)
and high up at altitude, where the pressure is below a prescribed value, 
generally take to be 0.03 bars.  Hence, $\kappa_{\rm e}$ 
is the most important parameter to be adjusted.  We could 
easily introduce a specified frequency and/or depth
dependence, but this would add free parameters
which we feel are not justified at this stage.  We have also
generated models in which TiO and VO are allowed to assume their
chemical-equilibrium upper-atmosphere abundances (Sharp \& Burrows 2007),
uncorrected for the cold-trap effect (\S\ref{conclusions}; Burrows et al. 
2007b), and to generate a stratosphere. These TiO/VO models produce 
qualitatively the same effects as do our ad hoc models.  However, in this paper
we prefer the flexibility of the $\kappa_{\rm e}$ prescription.

In Burrows et al. (2006), once the day- and night-side atmospheres were 
calculated, we used a 2D radiative transfer code to determine the 
integrated emissions ``at infinity" at a given viewing angle from 
the planet-star axis for the day- and night-side hemispheres.  These numbers were then 
transformed into a light curve as a function of wavelength and planetary phase angle using the
methods described in Sudarsky et al. (2000, 2005).  However, we have found that
since the detailed shape of the light curve connecting the day- and the night-side fluxes 
is likely to be only poorly constrained for the foreseeable future and since our model is 
symmetric about secondary eclipse, it is inappropriate to invest a disproportionate 
amount of effort in performing expensive 2D transfer calculations.  Rather, we
invest our efforts in obtaining state-of-the-art day- and the night-side fluxes
at secondary and primary eclipse and then connect them with a simple, though well-motivated, curve.
Therefore, our light curve model is:
\begin{equation}
{\cal{C}} = \frac{D + N}{2} + \frac{D - N}{2}\cos{\alpha}\sin{i}\, ,
\label{phasel}
\end{equation}
where $\cal{C}$ is the planet/star flux ratio, $D$ is the dayside flux (see Fig. \ref{fig4}), $N$ is the 
nightside flux, $\alpha$ is the phase angle, and $i$ is the inclination angle ($\sim{90^{\circ}}$
if in transit).  This is the form adopted in Cowan et al. (2007).  Using this simpler
approach, one does not imply more precision than is warranted at this preliminary
stage of inquiry. 

We have revisited the question of what type of averaging of
the incoming radiation from the star over the surface of the
planet is best suited to describe the planetary spectrum close
to the secondary eclipse. We had demonstrated earlier (Sudarsky
et al. 2005) that detailed 2D phase-dependent spectra averaged
over the phase are equal, within a few percent, to the spectrum
computed for a representative model atmosphere that is
constructed assuming that the incoming flux is distributed
evenly over the surface of the dayside.
In the usual terminology, this corresponds to the flux distribution
factor $f=1/2$ (Burrows et al. 2000)\footnote{We allow the planet to be irradiated
by the full flux received from the star; it is only deeper in the
atmosphere where the energy is transported to the nightside. See Appendix \ref{redist}.}.

However, the spectrum of a planet observed close to the secondary
eclipse should be biased toward a higher flux than that obtained
using the $f=1/2$ model. This is because the hottest part of the
planet, the substellar point, is seen as emerging from the planet
perpendicularly to the surface, and, thus, with the lowest
amount of limb darkening. Therefore, the contribution of the
hottest part of the planet is maximized. To study this effect,
we have computed a series of model atmospheres corresponding to
a number of distances from the substellar point, and have integrated
the individual contributions to get the flux received by an
external observer at a phase close to superior conjunction. It
turns out that the flux is very close to that computed for
$f=2/3$, which is the value we subsequently use in all simulations
presented here. There is a simple analytic argument
why $f$ should be approximately equal to $2/3$ which we present 
in Appendix \S\ref{app5}.

The formalism for $D$ and $N$ is the best we have fielded to date.  Nevertheless, 
a 3D general circulation model (GCM) that incorporates state-of-the-art opacities, 
compositions, and radiative transfer will be needed to properly address
day-night heat redistribution, the vortical and zonal mass motions, and the positions
of the hot and cold spots.  Such a model is not yet within reach, but there have been
preliminary attempts to treat this physics (Showman \& Guillot 2002; Cho et al. 2003; 
Menou \etal 2003; Williams et al. 2006; Cooper \& Showman 2005; Lunine \& Lorenz 2002).  The proper GCM 
physics remains the major uncertainty in current planetary secondary eclipse and 
light-curve modeling.  

\section{Temperature$-$Pressure Profiles}
\label{tp}

We have used the techniques outlined in \S\ref{tech} and in 
Appendix \S\ref{redist} and the data in Tables \ref{t1} and \ref{t2} to create models
of six of the close-in EGPs in Table \ref{t3} for which there are {\it Spitzer}
secondary eclipse or phase light curve data.  These planets are HD 209458b, HD 189733b, TrES-1,
HD 179949b, HD 149026b, and $\upsilon$ And b.  The product of our
investigation is an extensive collection of atmosphere models, with associated
spectra, for many combinations of planet, P$_n$, metallicity, and values of $P_0$/$P_1$ (\S\ref{redist}).
We have, however, settled on presenting in this paper only the central and
essential results for each planet, in the knowledge that the data are not yet 
exquisitely constraining.

We focus on models with solar-metallicity (Asplund,
Grevesse, \& Sauval 2006) opacities, ($P_0, P_1$) = (0.05, 0.5) bars,
and an interior flux $T_{\rm int}$ of 75 K.  These are our baseline model parameters.
For a given measured planet radius, the dependence of the models on
$T_{\rm int}$ is extremely weak. We find that the specific pressure range ($P_0, P_1$)
in which most of the heat carried from the dayside to the nightside is conveyed 
plays a role in the planet-star flux ratios, but a subtle one.  Therefore, in 
lieu of a comprehensive, and credible, 3D climate model, we prefer not to claim too much
concerning the details of atmospheric circulation and heat redistribution.
We also explore the effects of a stratospheric absorber with an optical 
opacity of $\kappa_{\rm e}$ (see also Burrows et al. 2007b).
We find that such models will be most important for close-in EGPs with the greatest
stellar insolation and guided by this principle, particularly relevant for
HD 209458b, HD 149026b, and $\upsilon$ And b, we explore the consequences. 

\begin{figure}
\centerline{
\includegraphics[width=6.cm,angle=-90,clip=]{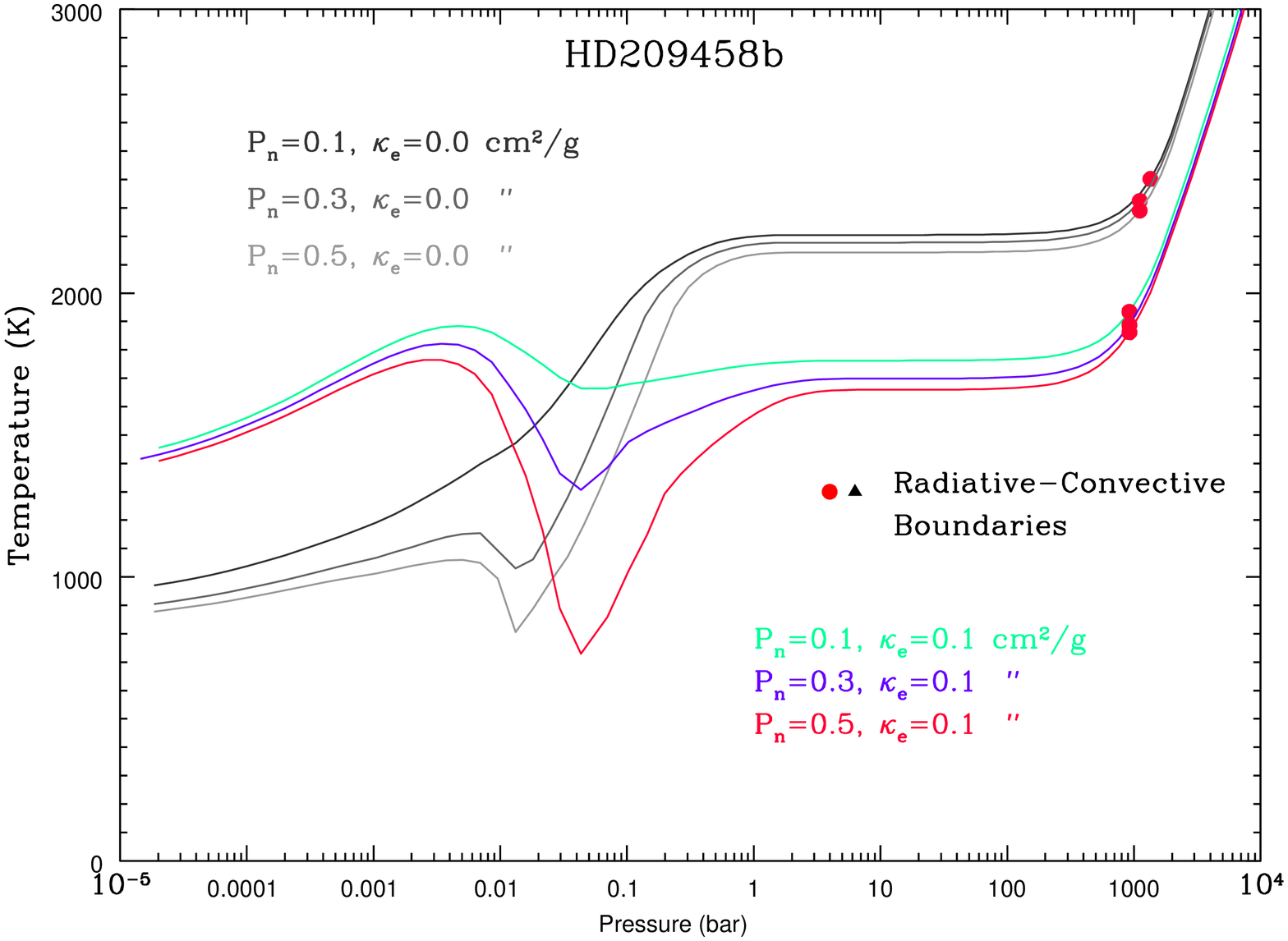}
\includegraphics[width=6.cm,angle=-90,clip=]{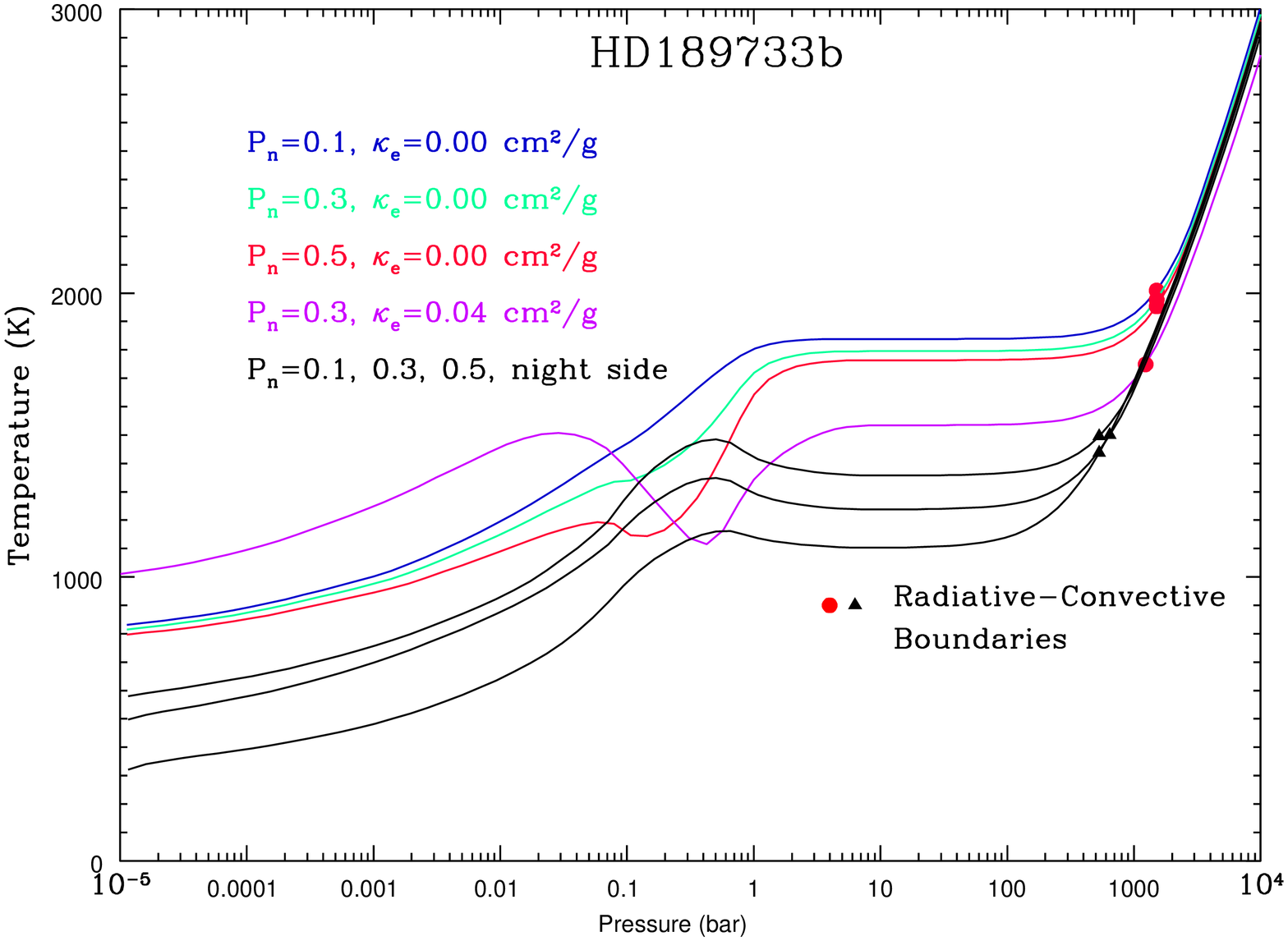}}
\centerline{
\includegraphics[width=6.cm,angle=-90,clip=]{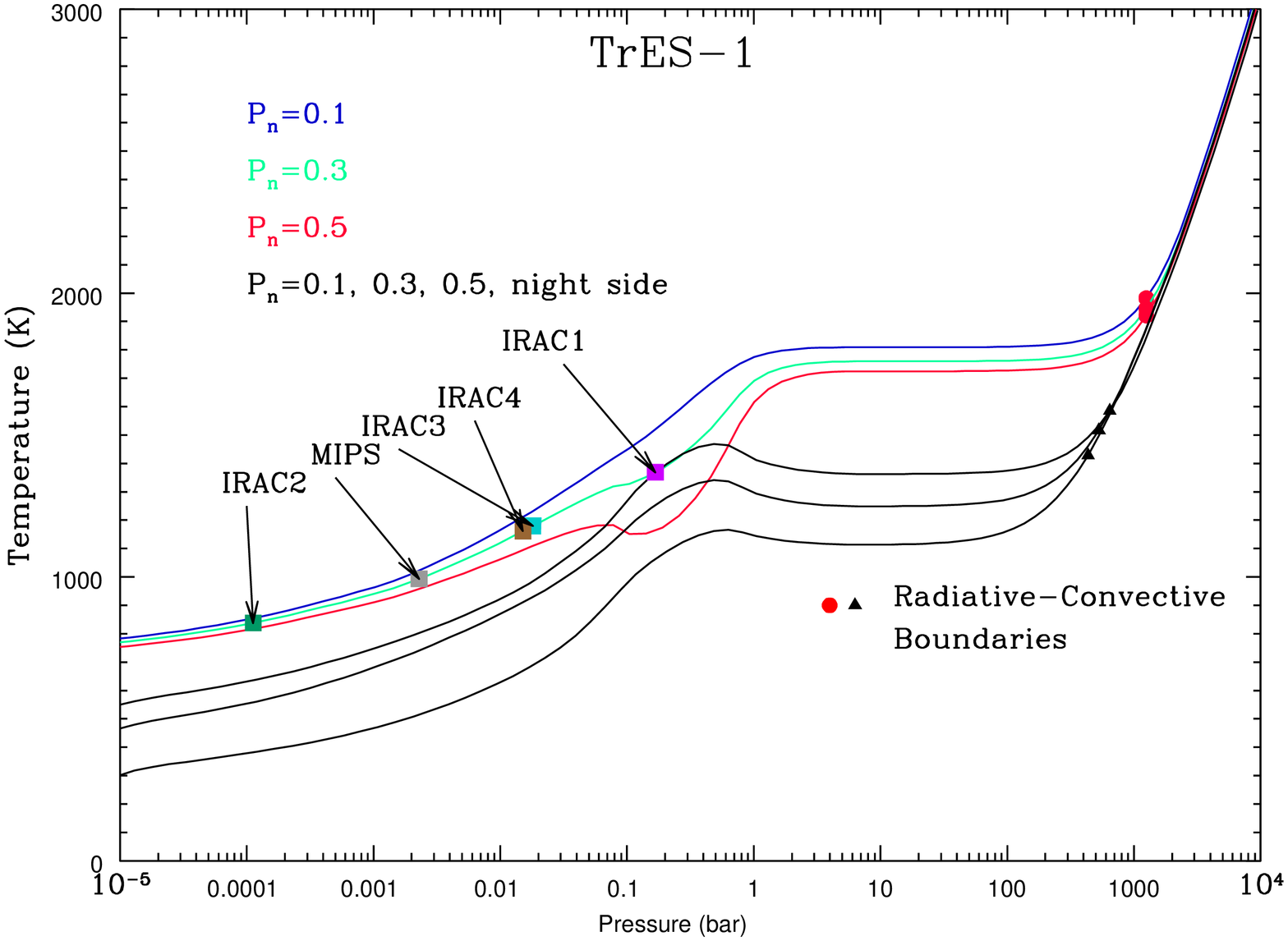}
\includegraphics[width=6.cm,angle=-90,clip=]{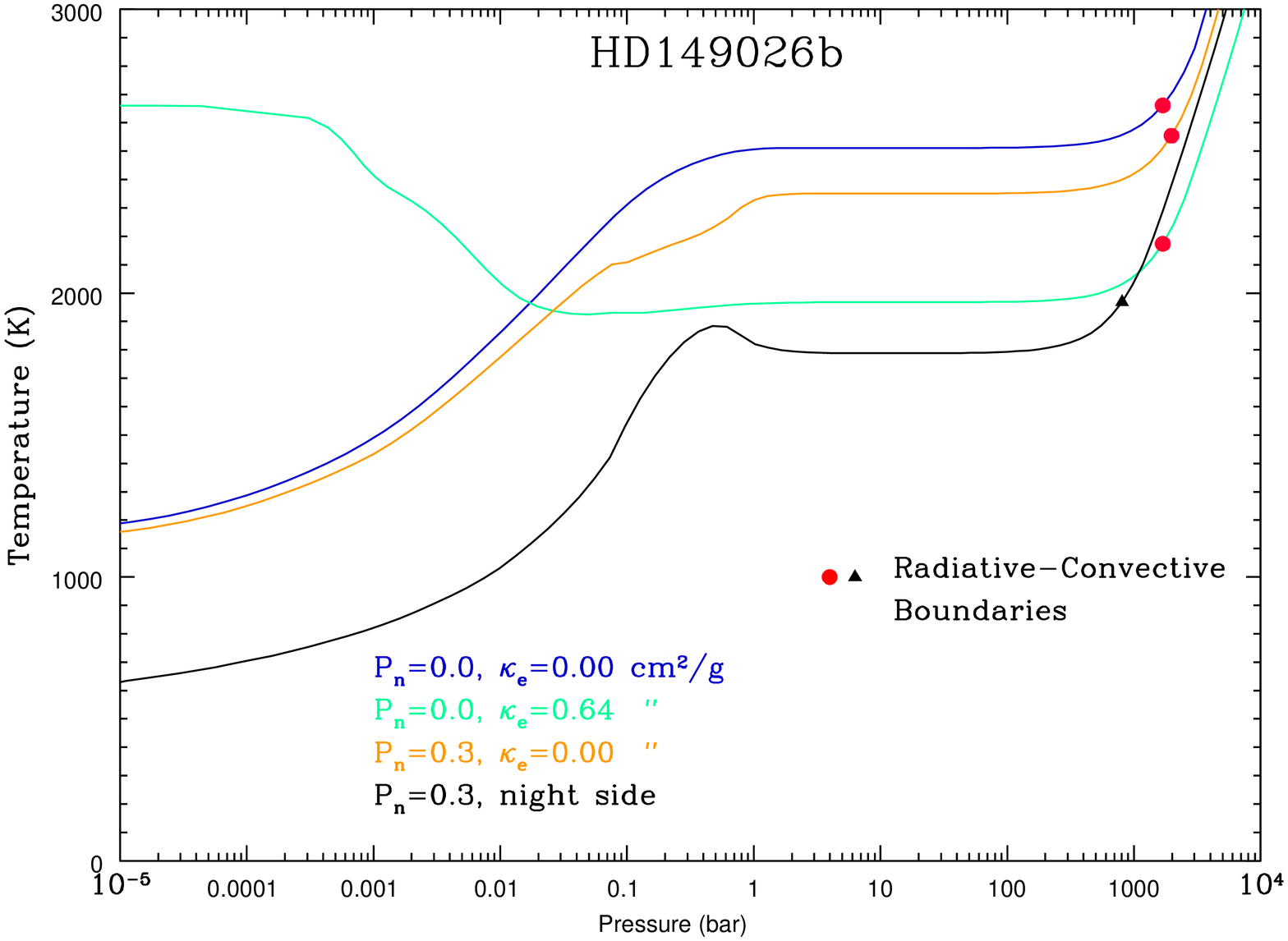}}
\centerline{
\includegraphics[width=6.cm,angle=-90,clip=]{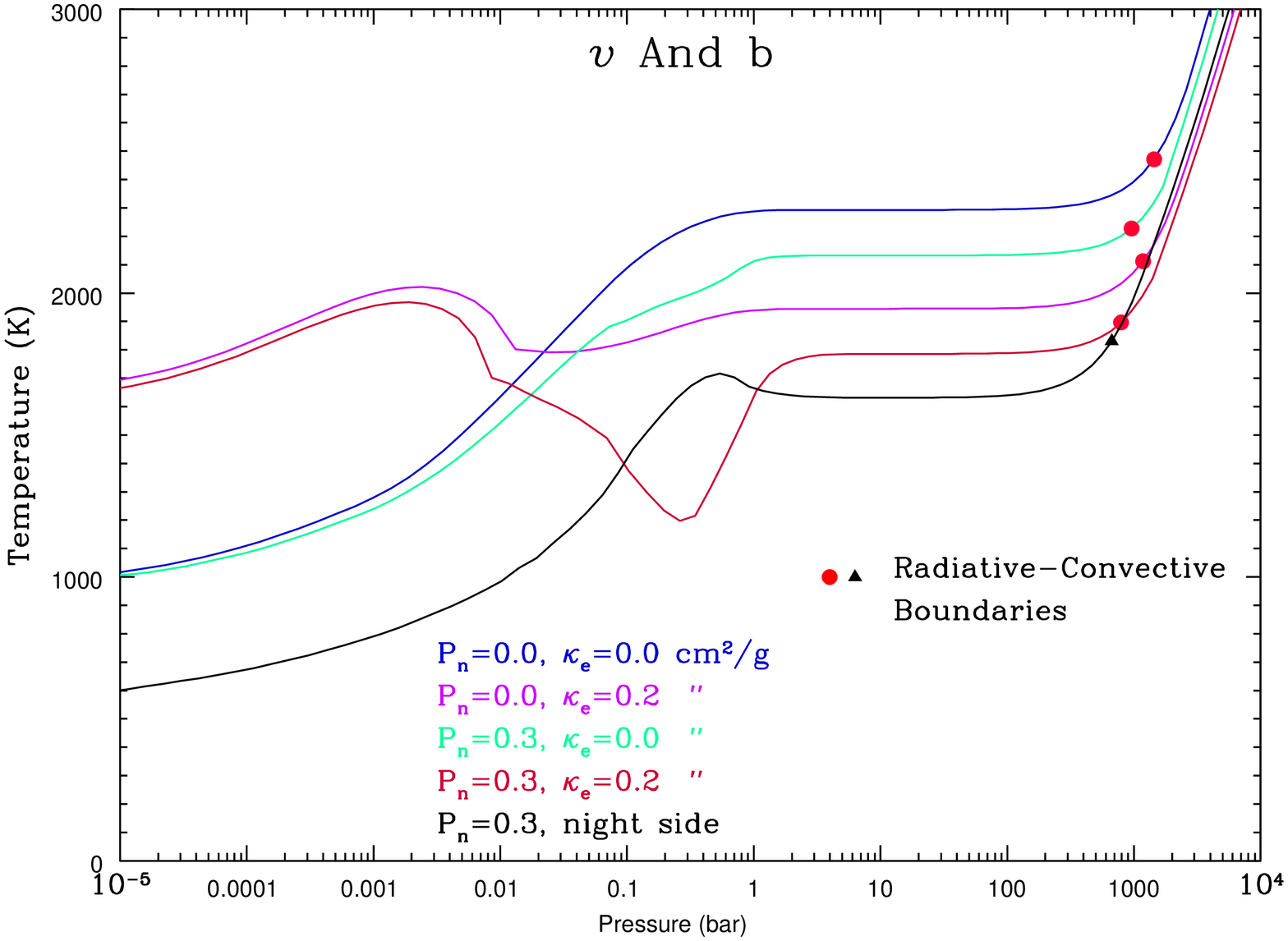}
\includegraphics[width=6.cm,angle=-90,clip=]{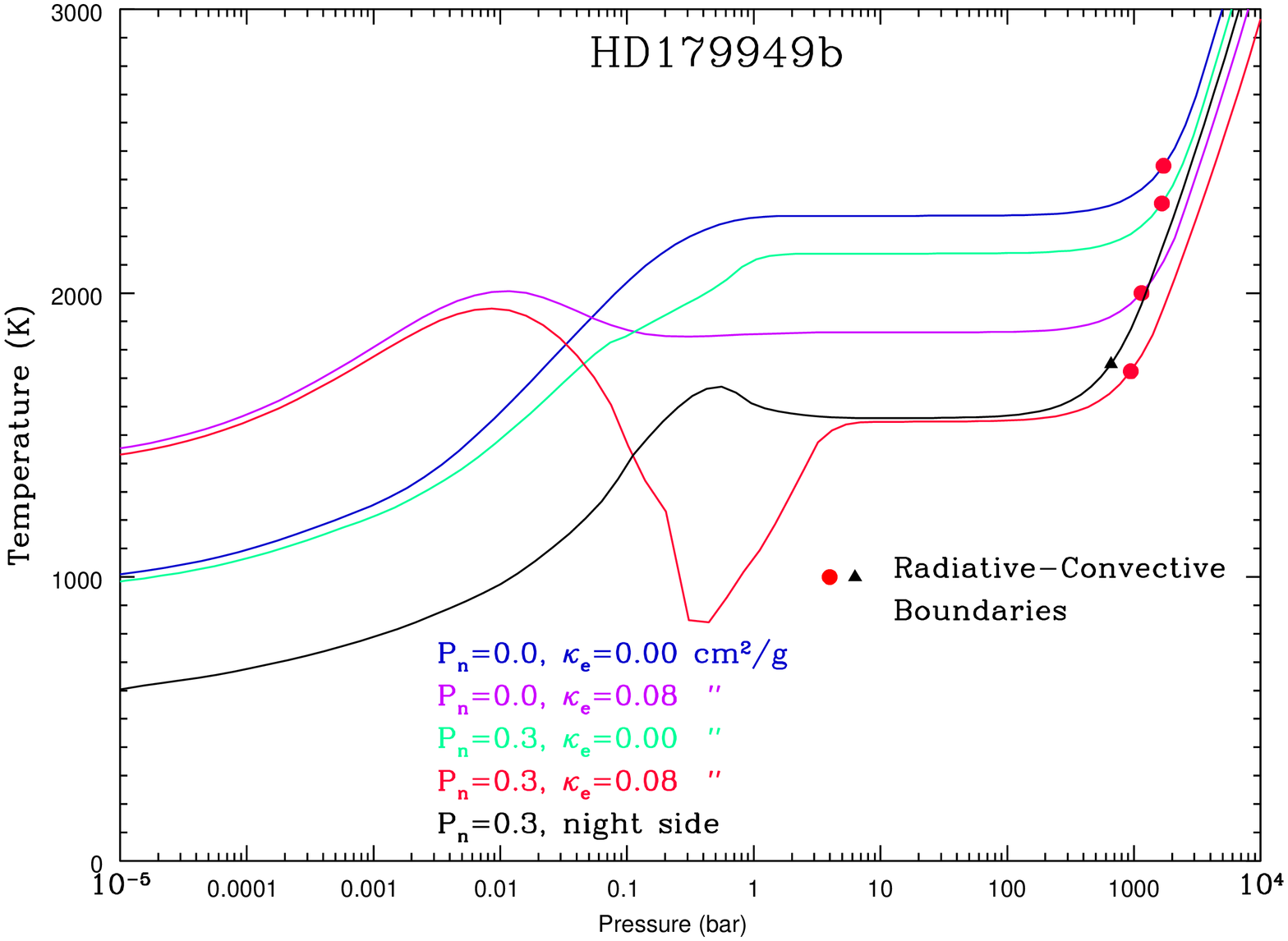}}
\caption{
Temperature-pressure profiles for the six close-in planets
studied in this paper. Dayside profiles incorporate the external substellar 
irradiation/flux given in Table \ref{t1} and an internal 
flux for the planet corresponding to the temperature of 75 K at 
the lower boundary. A sink of energy corresponding to the particular 
value of P$_{n}$ was introduced between pressures of 0.05 and 0.5 bars.
The nightside is calculated without irradiation, assuming
an energy source corresponding to the same value of P$_{n}$
at the same pressures employed for the dayside sink.
Entropies at the bases of the convection zones on the day-  
and night-sides were approximately matched by adjusting the internal 
planetary flux at the bottom of the nightside model at the same gravity.   
Models with an extra upper-atmosphere absorber in the optical 
are included for HD 209458b, HD 189733b, HD 149026b, $\upsilon$ And b, and HD 179949b.
See text in \S\ref{tp} for a discussion of these panels.
}
\label{fig1}
\end{figure}

Figure \ref{fig1} portrays in six panels the temperature-pressure ($T/P$) profiles
of a representative collection of dayside and nightside models of the six close-in EGPs of this study.
For the dayside, the different curves correspond to different values of P$_n$ from
0.0 (no redistribution) to 0.5 (full redistribution) and to 
models with and without stratospheric absorbers. For the nightside, P$_n$ ranges
from 0.1 to 0.5.  For all models, the radiative-convective boundaries are identified
and are quite deep (on the far right of each panel).  When $\kappa_{\rm e} \ne 0$, the $T/P$
profiles show distinct thermal inversions.

There are quite a few generic features in evidence on these panels.  The first is that
the atmospheres are never isothermal.  Since the opacities in the optical, where most of
the stellar irradiation occurs, and in the infrared, where most of the reradiation occurs,
are very different, a quasi-isothermal inner region interior to $\sim$1 bar is always bounded by
an outer region in which the temperature decreases (Hubeny, Burrows, \& Sudarsky 2003).
As Fig. \ref{fig1} indicates, the magnitude of the temperature decrease  
from the plateau to the $\sim$10$^{-5}$ bar level for dayside models 
without stratospheres is $\sim$1000 K.  With a stratosphere, the outward increase 
from a pressure of $\sim$0.1 bars can be correspondingly large.  For our nightside models,
the monotonic decrease is $\sim$500$-$1000 K.  Models with temperature inversions due
to a strong absorber at altitude clearly stand out in the panels of Fig. \ref{fig1} and may
result from the presence of a trace species, TiO/VO, or a non-equilibrium species 
(Burrows et al. 2007b).  The possible effect of such upper-atmosphere absorbers 
on the $T/P$ profiles and the resultant dayside spectra are exciting new features
of the emerging theory of irradiated EGPs.

\begin{figure}
\centerline{
\includegraphics[width=14.cm,angle=-90,clip=]{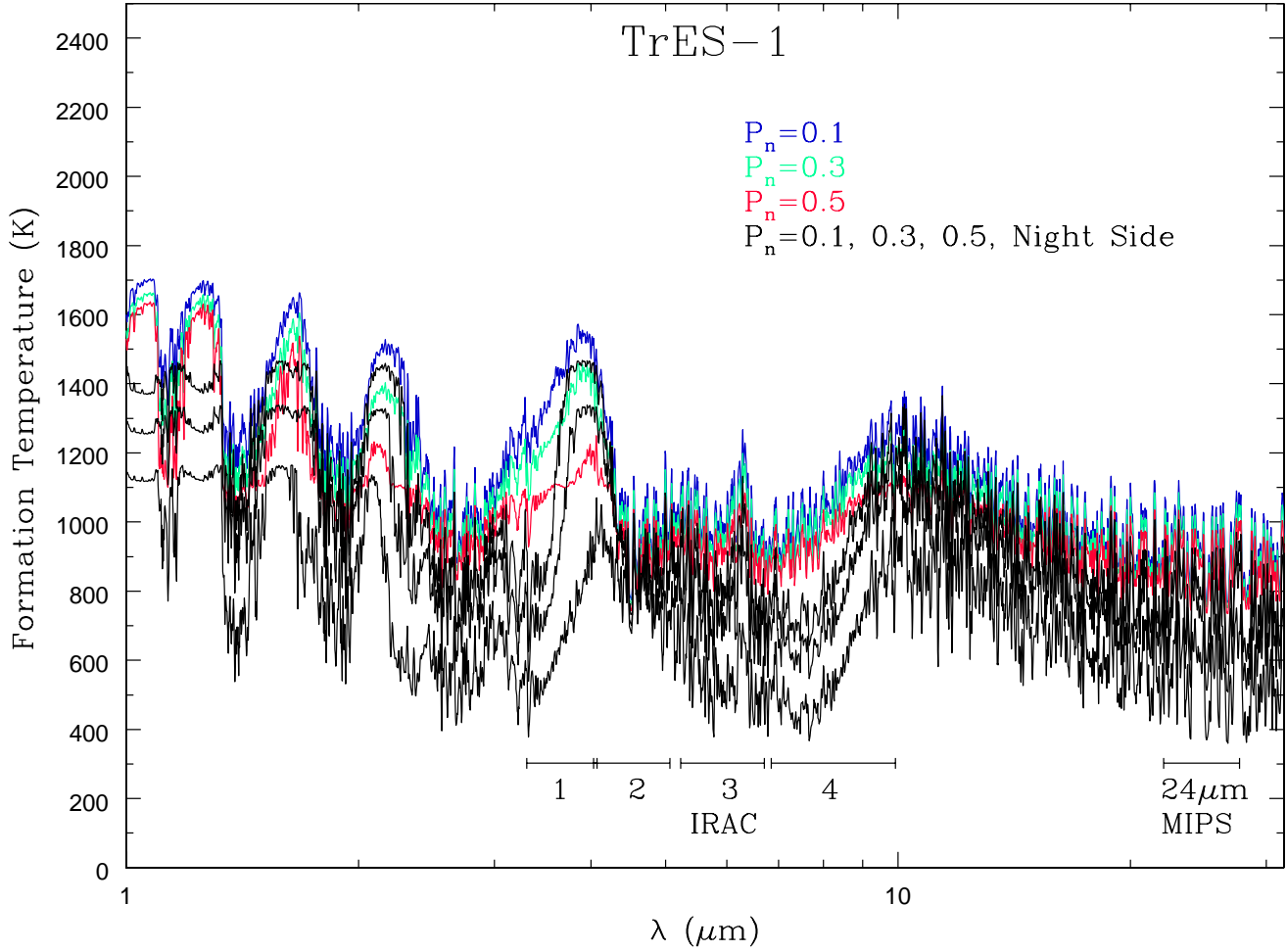}
}
\caption{
``Brightness" or ``formation" temperature spectra on the dayside for three models (P$_n$ = 0.1, 0.3, and 0.5)
of TrES-1.  The formation temperature at the particular wavelength is the temperature
where the optical depth at this wavelength reaches 2/3. See text for a discussion.
}
\label{fig2}
\end{figure}

The discussion above is made more germane when we note that the decoupling surfaces for the 
IRAC and MIPS (24-$\mu$m) channels, the effective photospheres where $\tau_{\lambda} \sim 2/3$, 
are all in the outer zone.  Figure \ref{fig1} indicates their positions for the dayside P$_n = 0.3$ 
model of TrES-1.  They are at similar pressures for all other models.  The photospheres  
for shorter wavelengths not accessible to {\it Spitzer} are deeper in.
Figure \ref{fig2} portrays these ``formation," ``brightness," or photospheric temperatures as a 
function of wavelength for three models of TrES-1 with different values of P$_n$,
and for both the dayside and nightside, and illustrates this fact. The 
approximate wavelength intervals of the {\it Spitzer} bands are superposed.  
Though the IRAC 1 flux can decouple at interesting depths, the photospheres for the $Y$, $J$, $H$, 
and $K$ bands are generally deeper. The photospheres in the far-IR beyond $\sim$10 $\mu$m
are high up at altitude and we repeat that {\it Spitzer} photometry does not probe the isothermal 
region so characteristic of theoretical close-in EGP atmospheres.  Moreover, since the {\it Spitzer}
observations are probing the outer regions of the atmosphere most affected by stellar 
irradiation and which can have inversions, the treatment of the outer boundary 
condition due to incoming stellar flux must be accurate. Slight errors or uncertainties 
in the outer boundary condition of the transfer solution, or in the upper-atmosphere opacities, can 
translate into significant errors in the predicted {\it Spitzer} fluxes. This is particularly true 
longward of $\sim$10 $\mu$m. As a result, measured fluxes in both the near-IR and 
mid-IR are useful diagnostics of upper-atmosphere absorbers and thermal 
inversions (Hubeny, Burrows, \& Sudarsky 2003; Burrows et al. 2006; Burrows 
et al. 2007b; Knutson et al. 2007c).  All these caveats and points must be 
borne in mind when interpreting the {\it Spitzer} EGP data.

\begin{figure}
\centerline{
\includegraphics[width=14.cm,angle=-90,clip=]{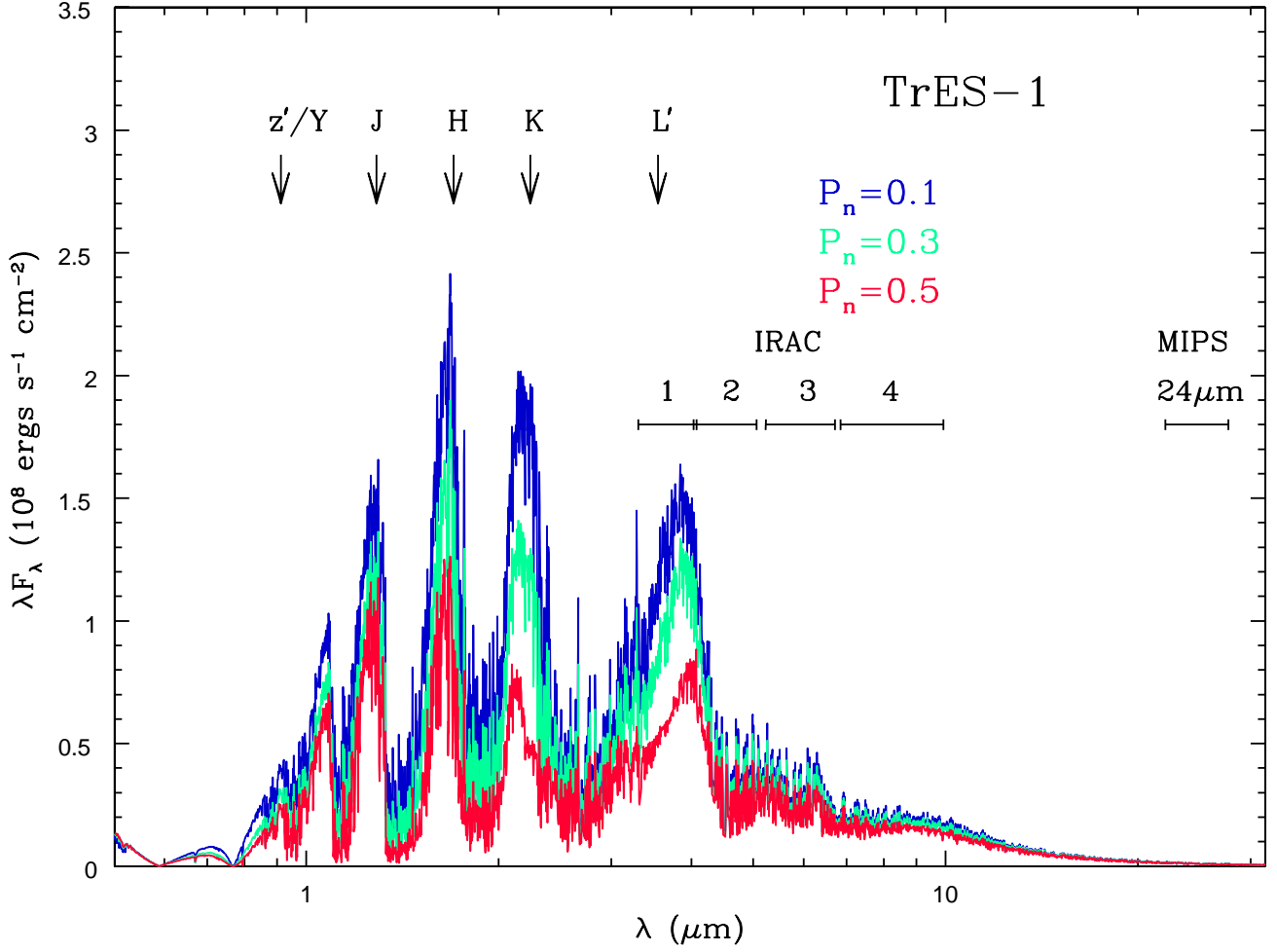}
}
\caption{
Spectral energy distribution (SED) ($\lambda$F$_{\lambda}$ versus log$_{10}$($\lambda$)) 
for the dayside of TrES-1 for three values of the day-night heat redistribution
parameter P$_{n}$. Notice that observations in IRAC and MIPS bands cover 
only a small fraction of the SED of the planet.
}
\label{fig3}
\end{figure}

As a prelude to our discussions in \S\ref{planetstar} and \S\ref{curves} of the {\it Spitzer} 
planet-star flux ratios at secondary eclipse and during an orbital traverse, 
and to emphasize the fact that {\it Spitzer} does not comprehensively probe the 
irradiated planet's atmosphere, we plot in Fig. \ref{fig3} theoretical 
dayside spectra ($\lambda$F$_{\lambda}$ versus log$_{10}$($\lambda$)) for three models of TrES-1  
at zero phase angle (superior conjunction). Superposed on the plot are the positions
of the near-IR, IRAC, and MIPS bands. Such a figure allows one to determine at a glance 
the wavelengths at which most of the flux is radiated (at least, theoretically).  
As Fig. \ref{fig3} suggests, most of the planet's flux 
emerges in the near-IR, not in the IRAC or MIPS channels. In fact, depending 
on the planet, no more than one fifth to one third of the planet's flux comes 
out longward of $\sim$3.6 $\mu$m (IRAC 1), while no more than one twentieth to one 
tenth emerges longward of $\sim$6.5 $\mu$m, the ``left" edge of the IRAC 4 channel.
Since much of the best EGP data have been derived in IRAC channel 4, one must acknowledge
that they may represent very little of the total planetary emissions.    

Finally, we call the reader's attention to the slight bumps (on the nightside) 
and depressions (on the dayside) between 0.05 and 1.0 bars in the $T/P$ profiles depicted 
in Fig. \ref{fig1}.  This region is near where we imposed heat redistribution using 
the formalism described in \S\ref{redist}.  The actual shapes of these profiles 
are determined by this mathematical procedure and other algorithms will produce 
different local thermal profiles.  Note that with this formalism it is possible
at the higher P$_n$s ($\ge 0.35$) for the nightside to be hotter than the dayside 
at the same pressure levels in the redistribution region.  While this 
may seem at odds with thermodynamics, what is essential is that energy 
is conserved and is redistributed at optical depths that are not either
too low or too high.  If the former, the absorbed stellar heat would be reradiated
before it can be carried to the nightside.  If the latter, then the stellar radiation
can not penetrate to the conveyor belt. For our default choice of $P_0$ and $P_1$, $\tau_{\rm Rossland}$
is generally between $\sim$0.3 and $\sim$6.  These depths are not unreasonable, but 
our redistribution algorithm is clearly only a stopgap until a better GCM 
can be developed and justified.

\section{Planet-Star Flux Ratios$-$Comparison with Data}
\label{planetstar}

We discuss below and in turn model fits for each transiting EGP at secondary eclipse.
However, first we present our results collectively and in summary fashion.
Figure \ref{fig4} in four panels portrays for the four transiting EGPs the correspondence between  
the secondary eclipse data and representative models of the planet-star 
flux ratio as a function of wavelength from 1.5 $\mu$m to 30 $\mu$m. This figure 
summarizes our major results. The models are for values of P$_n$ of 
0.1, 0.3, and 0.5 and various values of $\kappa_{\rm e}$. The data 
include 1-$\sigma$ error bars and can be found in Table \ref{t3}. 
As Fig. \ref{fig4} indicates, we can fit all the published
data. The P$_n$ dependence for both stratospheric models and models 
without inversions is strongest in the $K$ band and in IRAC 1. In fact, 
in the near-IR, models with inversions depend very strongly on P$_n$.  
Fig. \ref{fig5} for HD 209458b in the near-IR indicates this most clearly.  This finding  
implies that measurements at these shorter IR wavelengths are good diagnostics of  
P$_n$, particularly if inversions are present. 

Importantly, including a non-zero $\kappa_{\rm e}$ 
and generating a stratosphere results in a pronounced enhancement 
longward of IRAC 1, particularly in IRAC 2 and 3, but also at MIPS/24 $\mu$m and at
the 16-$\mu$m peak-up point of {\it Spitzer}/IRS.  Hence, fluxes at the longer IR wavelengths
might be good diagnostics of thermal inversions.  The models in Fig. \ref{fig4} for
HD 209458b and HD 149026b demonstrate this feature best. 

No attempt has been made to achieve refined fits, but the correspondence between
theory and measurement, while not perfect, is rather good for all the planets.  Moreover, 
different EGPs seem to call for different values of P$_n$ and $\kappa_{\rm e}$, and, hence, perhaps,
different climates, degrees of heat redistribution, compositions, and upper-atmospheric physics. 
The light-curve analyses in \S\ref{curves} also suggest this. A goal is to relate 
these measured differences with the physical properties of the star and planet and these 
infrared secondary eclipse data allow us to begin this program in earnest. 

We note that comparisons between model and data must actually be 
made after the band-averaged flux-density ratios of the detected
electrons are calculated.  Performing this calculation slightly mutes
the predicted variation from channel to channel in the IRAC regime. 
This is particularly true when comparing IRAC 1 and IRAC 2, 
even if a pronounced spectral bump at and near $\sim$3.6 $\mu$m 
obtains, as it does for models with modest or no thermal 
inversion.  However, to avoid the resultant clutter and 
confusion, we do not plot these bandpass-averaged predictions 
on Figs. \ref{fig4} and \ref{fig5}.  We now turn to case-by-case 
discussions of the secondary eclipse measurements and models.

\subsection{HD 209458b}
\label{hd209sub}

The first transiting EGP discovered was HD 209458b (Henry et al. 2000; Charbonneau et al. 2000)
and it has since been intensively studied. The direct-detection data of relevance 
to this paper are summarized in Table \ref{t3}.   The most relevant data are the geometric albedo 
constraints in the optical from MOST (Rowe et al. 2006,2007), a $K$-band upper limit
using IRTF/SpeX from Richardson, Deming, \& Seager (2003), a MIPS/24-$\mu$m photometric point 
from Deming et al. (2005) (and its possible update), a low-resolution {\it Spitzer}/IRS spectrum from 
Richardson et al. (2007), and, importantly, photometric points in IRAC channels 1 through 4 
from Knutson et al. (2007c).  These data collectively provide useful information
on the atmosphere of HD 209458b.  

Motivated by the recent data of Knutson et al. (2007c),  
Burrows et al. (2007b) provide partial theoretical explanations for HD 209458b's 
atmosphere.  Much of the discussion in the current paper concerning HD 209458b is taken 
from Burrows et al. (2007b), so we refer the reader to both the Burrows et al. (2007b)
and Knutson et al. (2007c) papers for details.  However, here we expand upon 
the discussion in those works where it is necessary to put the HD 209458b findings in the broader 
context of the EGPs listed in Table \ref{t3}.  The major conclusion of Burrows et al. (2007b) is
that the atmosphere of HD 209458b has a thermal inversion and a stratosphere, created 
by the absorption of optical stellar flux by a strong absorber at altitude, 
whose origin is currently unknown.  This converts absorption features into emission
features, while still being consistent with the presence of water in abundance. 

\begin{figure}

\centerline{
\includegraphics[width=6.cm,angle=-90,clip=]{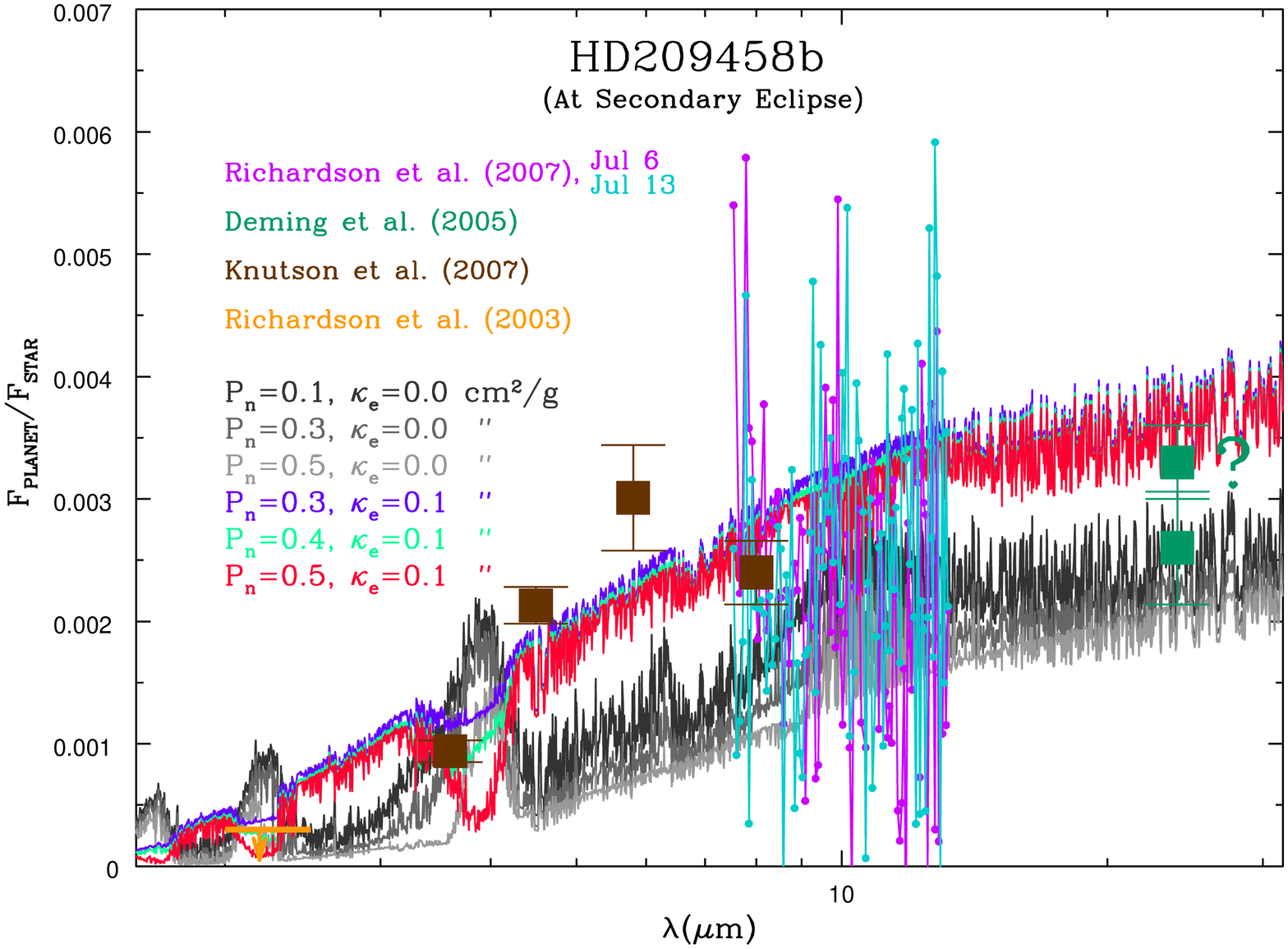}
\includegraphics[width=6.cm,angle=-90,clip=]{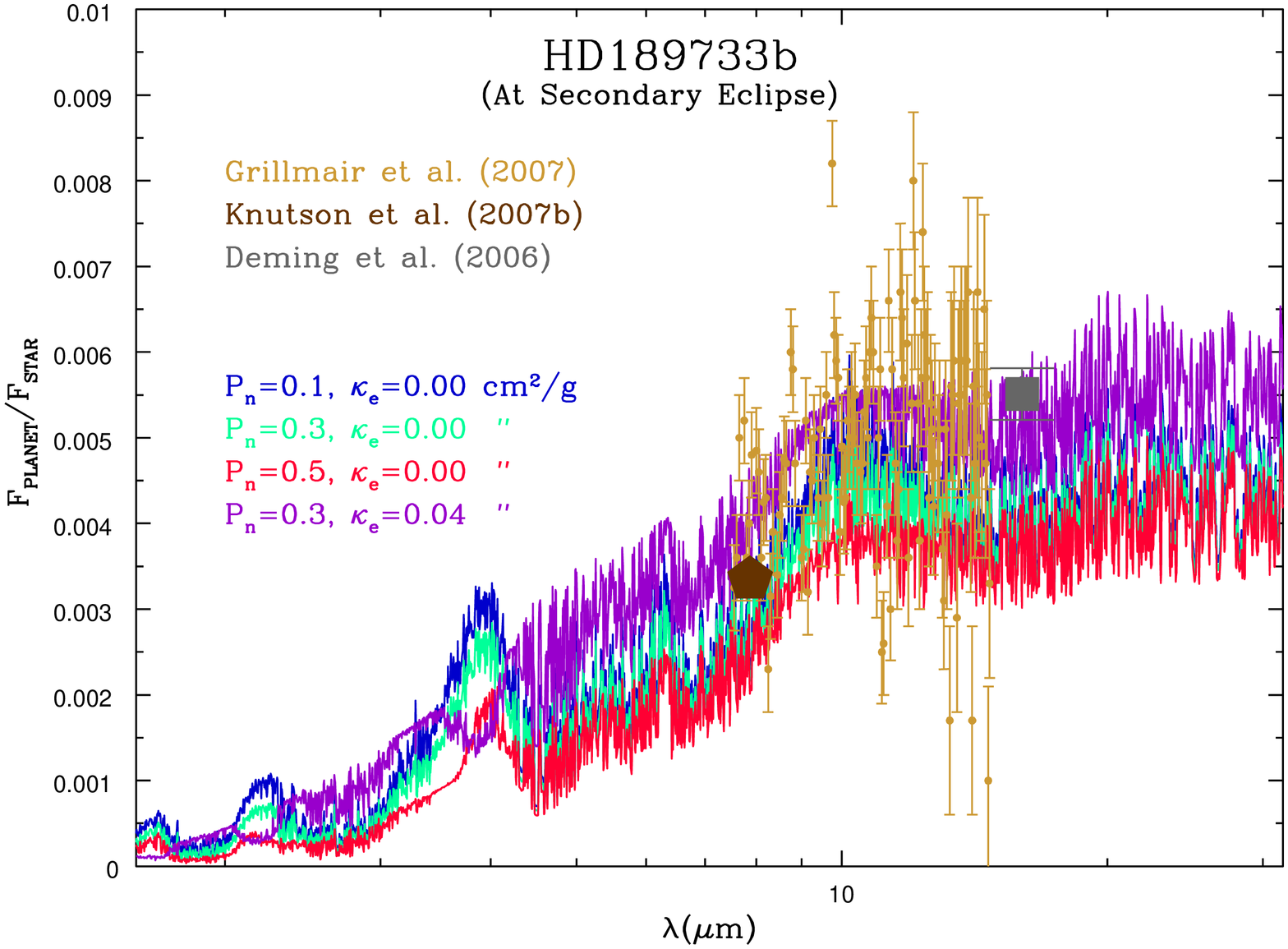}}
\centerline{
\includegraphics[width=6.cm,angle=-90,clip=]{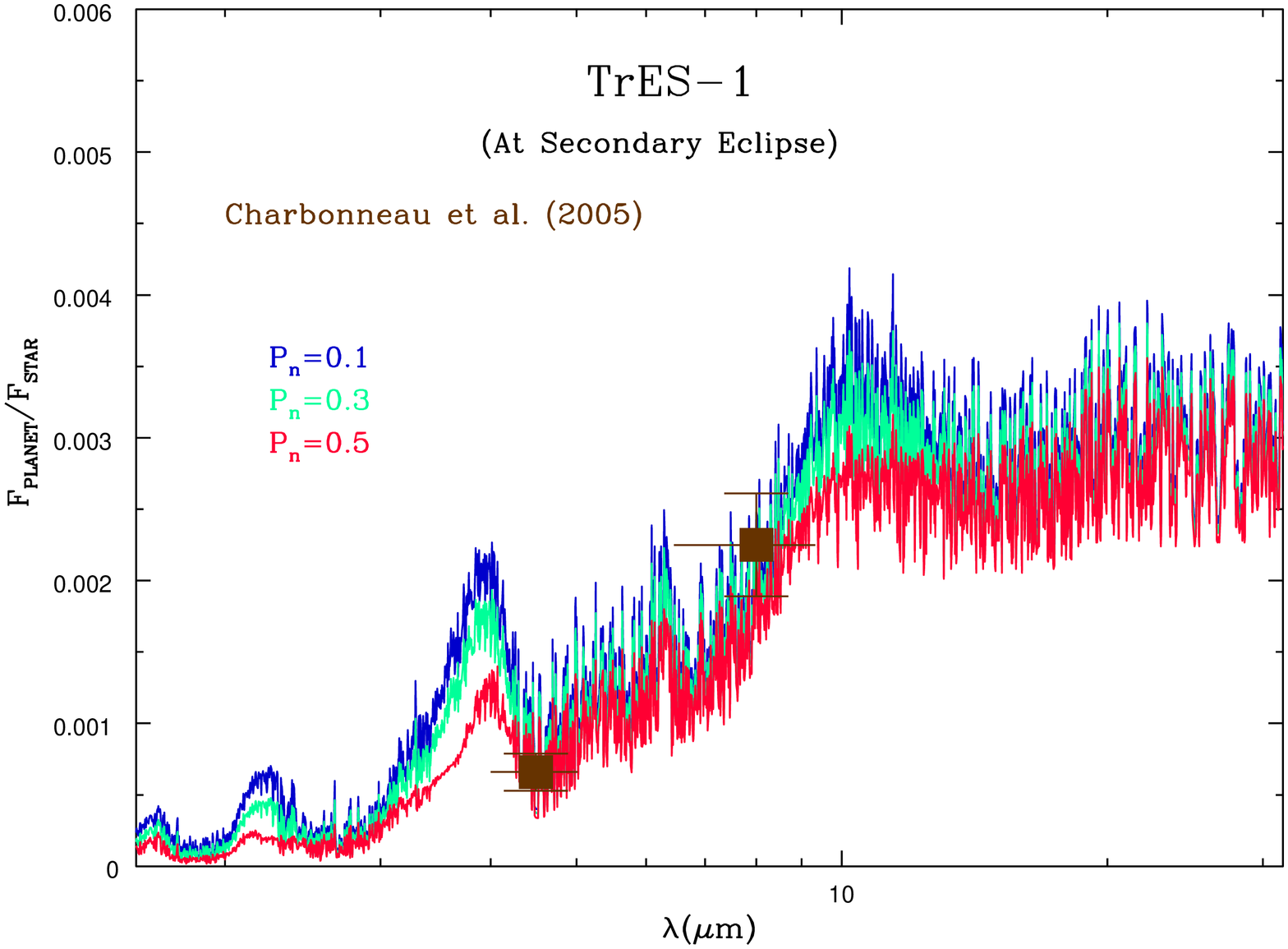}
\includegraphics[width=6.cm,angle=-90,clip=]{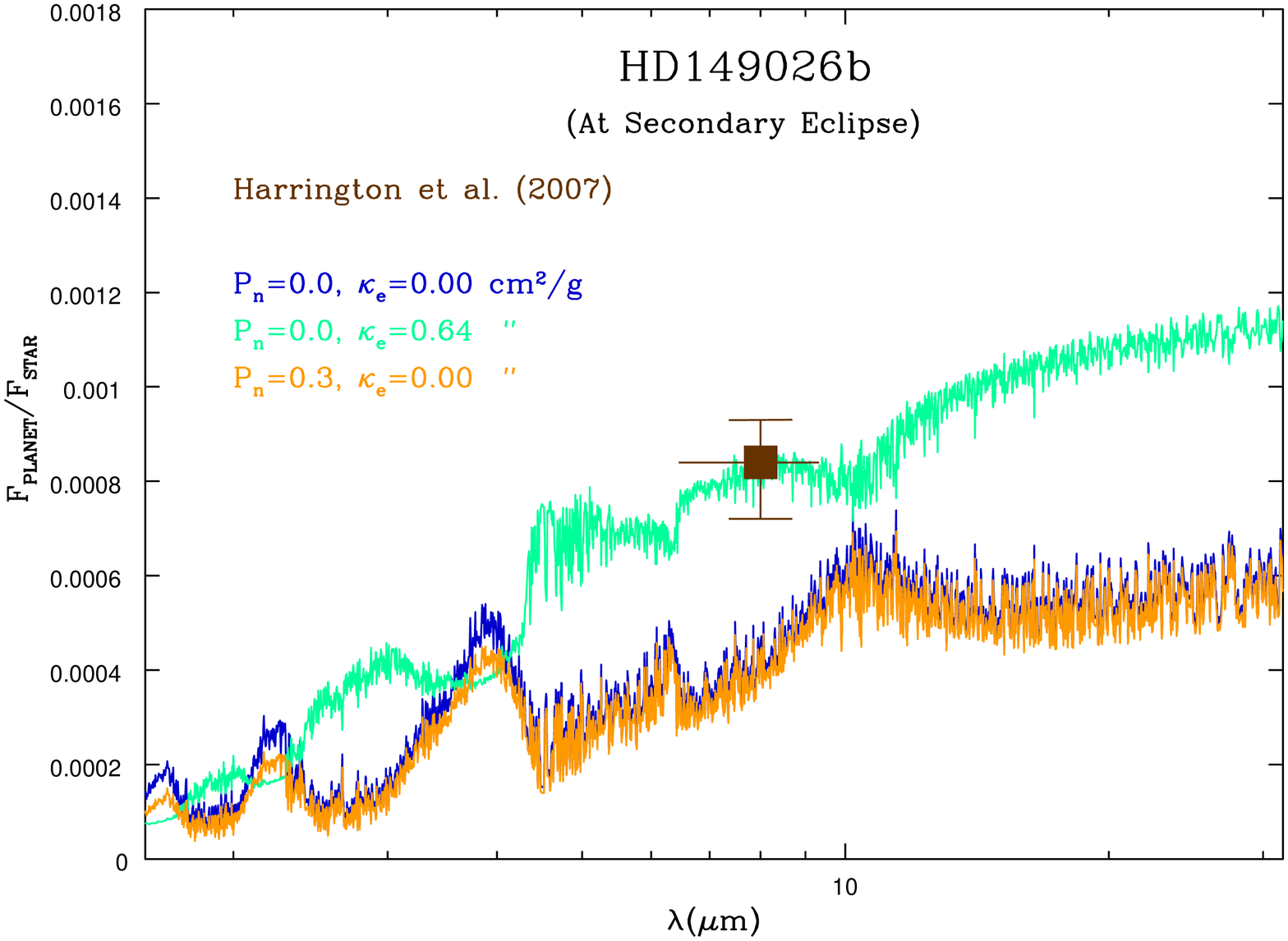}}
\caption{
The planet/star flux ratios versus wavelength from $\sim$1.5 $\mu$m
to 30 $\mu$m for various models of four transiting
EGPs measured by {\it Spitzer} at secondary eclipse.  
Notice the different scales employed in each panel.  Models
for different values of P$_n$ (\S\ref{redist}) and $\kappa_{\rm e}$ 
%
%
are provided where appropriate and the data 
from Table \ref{t3} for each planet are superposed.  The plot legends
indicate the color schemes used for the different EGPs.  On the upper-left
panel (HD 209458b), models with the lighter gray shade(s) are for 
the higher value(s) of P$_n$. Notice also that two different values 
for the flux at 24 $\mu$m (green) are shown on this same panel.
The one with the question mark is a tentative update to 
the Deming et al. (2005) 24-$\mu$m measurement, kindly provided 
by Drake Deming (private communication). If the flux at
24 $\mu$m is indeed $\sim$0.0033$\pm{0.0003}$, then our model(s)
with inversions provide the best fit at that wavelength as well.  
Note that the comparison between model and data should 
be made after the band-averaged flux-density ratios of the detected 
electrons are calculated.  This not only incorporates the significant 
widths of the {\it Spitzer} bandpasses, but the fact that one should 
compare photon counts (or detected electrons), and not monochromatic
fluxes.  The result is that the theoretical IRAC predictions do not
actually vary on the figures as much as do the plotted spectra and that the predicted 
contrasts, for instance between IRAC 1 and IRAC 2, are more muted, even
when the pronounced bump at and near $\sim$3.6 $\mu$m obtains.  
However, to avoid the resultant clutter, we do not put these 
bandpass predictions on these spectral plots. See text for a discussion 
of each irradiated planet and the inferences drawn.
}
\label{fig4}
\end{figure}

\begin{figure}

\centerline{
\includegraphics[width=14.cm,angle=-90,clip=]{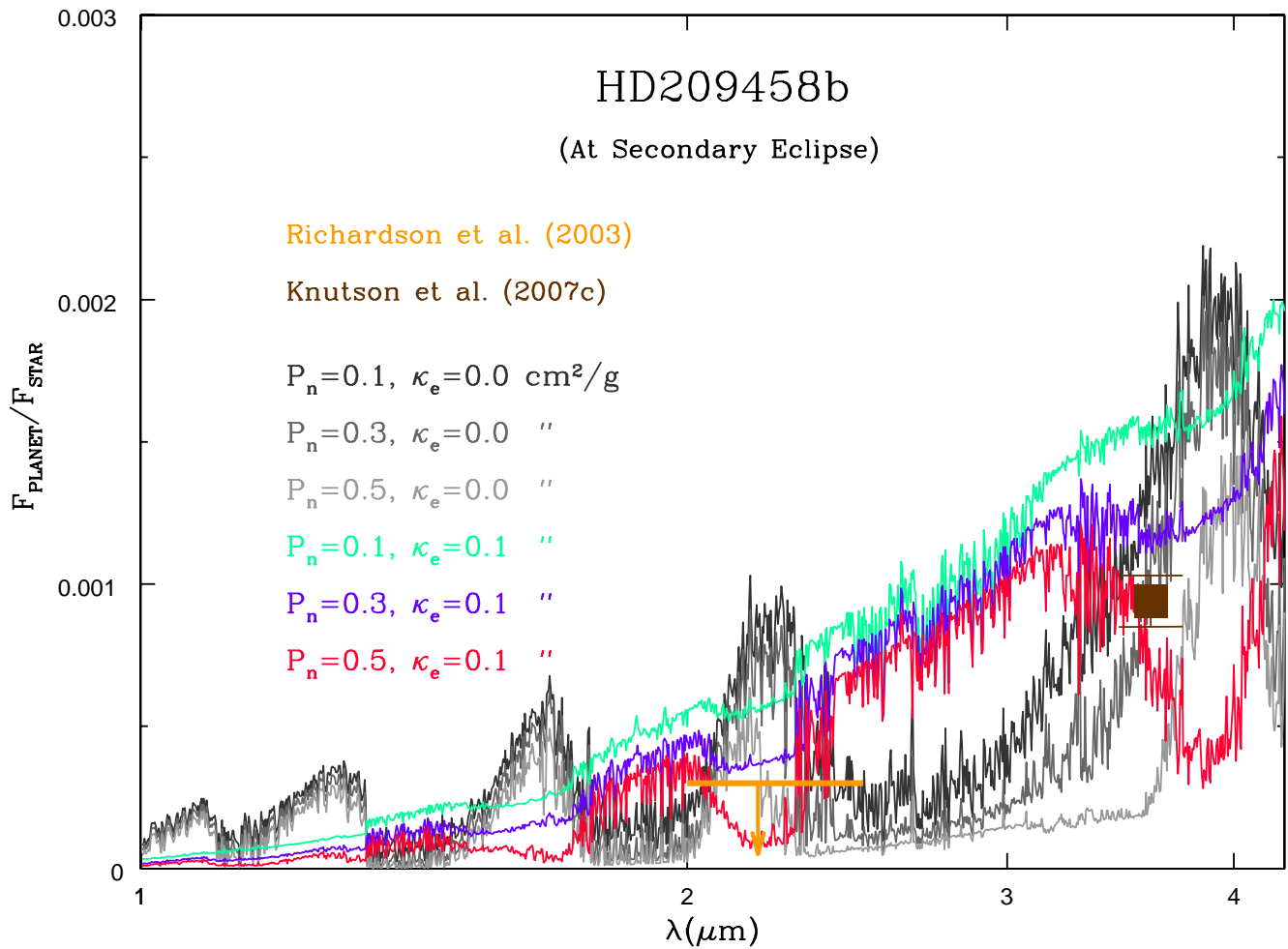}}
\caption{
The same as Fig. \ref{fig4}, but for various models of HD 209458b between 1 $\mu$m and 4 $\mu$m.
The dependence on both P$_n$ and the presence and strength of a thermal inversion
is greatest in this wavelength region.  Note that a thermal inversion
flips what would be water absorption features into emission features, altering
the interpretation of any data in this spectral region significantly.
Superposed is the Knutson et al. (2007c) data point for IRAC 1 and the
Richardson et al. (2003) upper limit near 2.2 $\mu$m.  See text 
for a discussion.
}
\label{fig5}
\end{figure}

All relevant data, save the albedo constraint in the optical, are displayed in the upper-left panel
of Fig. \ref{fig4}.  Figure \ref{fig5} includes the Knutson et al. (2007c) IRAC 1
point and the Richardson, Deming, \& Seager (2003) upper limit in $K$ and focuses on the
near-IR.  Also provided on both figures are models for 
P$_n$ = 0.1, 0.3, and 0.5, without and with an extra stratospheric absorber. 
The latter is implemented using the formalism in outlined in \S\ref{tech}   
and a $\kappa_{\rm e}$ of 0.1 cm$^2$/g.  

Figure \ref{fig4} shows that the low upper limit of Richardson, Deming, \& Seager (2003) in the $K$ band
that was problematic in the old default theory (Burrows, Hubeny, \& Sudarsky 2005; Fortney et al. 2005;
Barman, Hauschildt, \& Allard 2005; Seager et al. 2005; Burrows et al. 2006) is consistent with the models with an extra
upper-atmosphere absorber in the optical, particularly for higher values of P$_n$.
This is more clearly seen in Fig. \ref{fig5}.
Moreover, the theoretical peak near the IRAC 1 channel ($\sim$3.6 $\mu$m) in the old 
model without an inversion is reversed with the extra absorber into a deficit
that fits the Knutson et al. (2007c) point. The theory without an extra absorber
at altitude predicts that the planet-star flux ratio in the IRAC 2 channel should be lower than
the corresponding ratio at IRAC 1.  However, with the extra absorber the relative strengths in
these bands are reversed, as are the Knutson et al. (2007c) points.  This reversal
is a direct signature of a thermal inversion in the low-pressure regions of 
the atmosphere, and an indirect signature of the placement of the heat redistribution 
band (see Appendix \S\ref{redist}).  The top-left panel of Fig. \ref{fig1} depicts the corresponding 
temperature-pressure profiles and the thermal inversion at low pressures 
introduced by the presence of an extra absorber in the optical 
that is indicated by the data.

As Fig. \ref{fig4} also demonstrates,
there is a significant difference in the IRAC planet-star
flux ratios between the old default model without an inversion
and the new models with a stratospheric absorber, and 
that the models with a stratosphere fit the IRAC
channel 1, 2 and 4 flux points much better.  However, the height of the
IRAC 3 point near 5.8 $\mu$m is not easily fit, while maintaining the good
fits at the other IRAC wavelengths and consistency with the $K$-band limit.
Theoretically, the positions of the IRAC 3 and IRAC 4 photospheres should be close to one
another, so this discrepancy is surprising. Nevertheless, the IRAC 2, 3, and 
4 data together constitute a peak, whereas in the default theory 
an absorption trough was expected.  

The 24-$\mu$m MIPS point obtained by Deming et al. (2005) is lower 
than the prediction of our best-fit model.  However, the flux at this point 
is being reevaluated and may be closer to $\sim$0.0033$\pm{0.0003}$ 
(D. Deming, private communication).  If the new number supercedes the old
published value, then our best-fit model(s)
with inversions fit at this mid-IR point as well (Fig. \ref{fig4}). Higher
planetary fluxes longward of $\sim$10 $\mu$m are generic features of stratospheric
inversions.

The 1-$\sigma$ optical albedo limit from Rowe et al. (2007) is a very low 8.0\%.  For comparison, the
geometric albedos of Jupiter and Saturn are $\sim$40\%.  However, such a low number was predicted
due to the prominence in the optical of broadband absorptions by the alkali metals sodium
and potassium in the hot atmospheres of irradiated EGPs (Sudarsky et al. 2000;
their ``Class IV"). The associated planet-star flux ratios are $\sim$10$^{-5}$$-$10$^{-6}$.
This low albedo is consistent with the identification of sodium
in the atmosphere of HD 209458b using HST/STIS transit spectroscopy (Charbonneau
et al. 2002). Both these datasets suggest that any clouds that might reside in the
atmosphere of HD 209458b are thin.  A thick cloud layer would reflect light 
efficiently, leading to a high albedo. If the extra stratospheric absorber 
is in the gas phase, and there is no cloud, then our new thermal inversion models 
are easily consistent with the low albedo derived by Rowe et al. (2006,2007).
If the extra absorber is a cloud, the cloud particles must have a low scattering albedo
and can not be very reflecting.  This rules out pure forsterite, enstatite, and iron
clouds.

The IRS data are noisy, but their flattish shape is consistent with 
our model(s) with thermal inversions and a stratosphere.  Richardson et al. (2007) suggest
that there is evidence in the IRS data for two spectral features: one near
7.78 $\mu$m and one near 9.67 $\mu$m.  However, we think the data are too noisy to 
draw this conclusion.  Richardson et al. (2007) also suggest that 
the flatness and extension of their data to shorter wavelengths implies 
the near absence of water, since previous theoretical models predicted
a spectral trough between $\sim$4 $\mu$m and $\sim$8 $\mu$m.  However, 
if there is an outer thermal inversion, as we here and 
in Burrows et al (2007b) argue is the case for HD 209458b,
a trough is flipped into a peak for the same water abundance.
This renders moot the use of the spectral slope at the edge 
of the IRS spectrum to determine the presence or absence of water.
One of our major conclusions, implicit in Figs. \ref{fig4} and \ref{fig5}, 
is that water is not depleted at all in the atmosphere of HD 209458b.

The recent controversies surrounding such an interpretation occasion 
the following remarks. One thing to bear in mind concerning the use of these IRS spectra
to infer compositions is that they are very low resolution.  The use of classical
astronomical spectroscopy to identify constituents stems from the ability
at much higher resolution to see characteristic features at precise wavelengths 
and patterns of absorption or emission lines to high accuracy. This allows one to make
element and molecule identifications in a narrow wavelength range without
a global view across the whole spectrum.  However, at the low resolution of the IRS, 
no individual water features are accessible.  There are water features near $\sim$10 $\mu$m,
but they would require a $\lambda/\Delta\lambda$ is excess of $\sim$2000 to identify.  Otherwise,
all one sees is the collective effect of millions of lines and the resulting pseudo-continuum
(the band structure).  Clearly, when the data are low-resolution, a {\it global} photometric and spectral fit
is necessary to address the issue of composition. The signature of water's presence comes from the goodness 
of the global fit across the entire spectrum from the optical to the mid-IR.  
The good fit we obtain in Fig. \ref{fig4},
lead us to conclude that the IRS, IRAC 4, and MIPS data for HD 209458b are
consistent with the presence of water in abundance.

Note that if the $T/P$ profile were entirely flat (but see Fig. \ref{fig1}), whatever 
the opacity and molecular abundances the emergent spectrum would be a perfect black body
and would give no hint concerning composition.  Transit spectra would then be our
only reliable means of determining atmospheric composition (Fortney et al. 2003; 
Barman 2007; Tinetti et al. 2007; Ehrenreich et al. 2007).  However, as we have argued, there is every indication that
the $T/P$ profiles of strongly irradiated EGPs are not flat (Fig. \ref{fig1}). As a result,
spectral measurements of irradiated EGPs can be usefully diagnostic of 
both composition and non-trivial thermal profiles.

\subsection{HD 189733b}
\label{hd189}

Models for HD 189733b at secondary eclipse, with and without an 
extra upper-atmosphere absorber, are portrayed in the upper right-hand panel of 
Fig. \ref{fig4}.  They include $\kappa_{\rm e}$ = 0.0 cm$^2$/g models with P$_n$ = 0.1, 0.3, and 0.5,
and one $\kappa_{\rm e}$ = 0.04 cm$^2$/g model with P$_n$ = 0.3. The IRAC 4 data at 
8 $\mu$m from Knutson et al. (2007b) [brown], the IRS peak-up 
point at 16 $\mu$m obtained by Deming et al. (2006) [gray],
and the IRS spectrum between $\sim$7.5 $\mu$m and $\sim$13.5 $\mu$m
from Grillmair et al. (2007) [gold] are superposed on the figure.  
Though data in the other IRAC channels and at 24 $\mu$m have been
taken and reduced, they have yet to be published. 

As this panel indicates, the IRAC 4 point can be fit by models that include 
most values of P$_n$, with a very slight preference for lower values from 0.1
to 0.3.  The IRS data are not well-calibrated, but evince the slight turndown 
at the shorter wavelengths characteristic of atmospheres with weak or no stratospheric
absorber.  This turndown is in contrast with the behavior of the Richardson et al. (2007)
IRS data for HD 209458b, and reinforces the conclusion that a thermal inversion,
if present in HD 189733b, is very slight (see the 
upper-right panel of Fig. \ref{fig1}).  However, the 16-$\mu$m point of
Deming et al. (2006) is a bit higher than models with $\kappa_{\rm e}$ = 0,
whatever the value of P$_n$.  This suggests that there may be some extra heating in the
upper atmosphere of HD 189733b, but that it is weaker than in the atmosphere of HD 209458b.
The $\kappa_{\rm e}$ = 0.04 cm$^2$/g model shown in the HD 189733b panels of 
Figs. \ref{fig1} and \ref{fig4} indicates the possible magnitude of such stratospheric
heating, if present.   Note that to fit the 16-$\mu$m point we require
a smaller value of $\kappa_{\rm e}$ than used to fit the IRAC channel data for 
HD 209458b ($\kappa_{\rm e}$ = 0.1 cm$^2$/g).  This may not be surprising, since,
as Table \ref{t1} indicates, the stellar flux at the substellar point of
HD 189733b is lower by more than a factor of two than the corresponding
number for HD209458b. Perhaps, this indicates a systematic trend for the family
of strongly irradiated transiting EGPs (see Tables \ref{t1} and \ref{t3}), with
planets with the higher values of F$_{p}$ possessing stratospheres and atmospheres with
pronounced inversions. 

Be that as it may, we can predict, using the logic employed in \S\ref{hd209sub},
that if the IRAC 1 to IRAC 2 ratio turns out to be greater than or close to one,
any thermal inversion in the atmosphere of HD 189733b is either not pronounced,
or is absent. Under these circumstances, we certainly would then expect the 
``brightness" temperature at IRAC 1 to be demonstrably higher than 
that at IRAC 2 (see, e.g., Fig. \ref{fig2}).  Conversely, if the IRAC 1 to IRAC 2 
ratio is much less than one (as for HD 209458b), then a thermal inversion in the 
atmosphere of HD 189733b would be strongly suggested. The same can be said of 
the IRAC 4 to IRAC 3 ratio.  If the IRAC channel 3 planet-to-star flux 
ratio is higher than the Knutson et al. (2007b) point
at 8 $\mu$m, then a stratosphere would be indicated for HD 189733b.
Since the irradiation regime of HD 189733b is a bit more benign than that
of HD 209458b, and given the contrast in the short-wavelength behavior 
of the IRS data for each EGP, we hypothesize that HD 189733b does 
not boast much of a stratosphere.  Note that our HD 189733b models all 
have water in abundance and that if the unpublished IRAC 1, 2, and 3 data for HD 189733b 
prove to decrease monotonically with wavelength shortward of IRAC 4, this
would be fully consistent with the presence of the water band between
$\sim$4 $\mu$m and $\sim$8 $\mu$m in absorption.

\subsection{TrES-1}
\label{tres1}

Charbonneau et al. (2005) obtained IRAC 2 ($\sim$4.5 $\mu$m) 
and IRAC 4 ($\sim$8.0 $\mu$m) data for TrES-1 and these data were 
analyzed by Burrows et al. (2005).  A major conclusion of that paper
was that water is indeed seen in absorption.  Our new models, depicted in 
the bottom-left panel of Fig. \ref{fig4} with the two IRAC data points 
superposed, reinforce this finding. As can be seen in the figure, the 
models with different values of P$_n$ (here, all with $\kappa_{\rm e}$ = 0) can not 
easily be distinguished using these two IRAC points.  This fact emphasizes 
the need to obtain more {\it Spitzer} photometric data to help better 
constrain the properties of the atmosphere of TrES-1. However, it is 
clear from the significant drop in planet/star flux ratio from IRAC 4 
to IRAC 2 that the atmosphere of TrES-1 is qualitatively different from  
that of HD 209458b.  In particular, this behavior is a signature of a strong 
water absorption trough.  There are no signatures of a thermal inversion in
the atmosphere of TrES-1, or of water in emission, and the old, default models 
with a monotonic temperature profile (see left-middle panel of Fig. \ref{fig1}) are 
perfectly suitable.  Given this, we predict that the IRAC 1 point 
(when obtained) will be slightly higher than the IRAC 2 point.  


\subsection{HD 149026b}

As is suggested by the relative values of F$_p$ found in Table \ref{t1} for HD
209458b, HD 189733b, and TrES-1 and the different thermal profiles
inferred for their atmospheres, there appears to be a correlation between
the character of a planet's atmosphere and its 
value of F$_p$, or a related quantity (UV insolation?).
This possibility is intriguing, but not yet explained.  In particular,
we have yet to identify the extra stratospheric absorber in that subset
of close-in EGPs ``clearly" manifesting thermal inversions.  However, 1) HD 149026b's F$_p$ 
is almost twice that of HD 209458b, 2) it has one of the hottest atmospheres 
among those listed in Table \ref{t1} (see middle-left panel 
of Fig. \ref{fig1}), and 3) we can not fit the IRAC 4 data point obtained  
by Harrington et al. (2007) without a strong temperature inversion. 
The latter conclusion agrees with that of Fortney et al. (2006),
who predicted a mid-infrared flux for HD 149026b near the value actually 
measured by Harrington et al. (2007) by allowing TiO/VO to reside at low 
pressures for the hottest atmospheres (Hubeny et al. 2003). The lower-right 
panel of Fig. \ref{fig4} depicts three models for HD 149026b, one of 
which has $\kappa_{\rm e} = 0.64$ cm$^2$/g.  This is much larger than the $\kappa_{\rm e}$ 
employed to fit HD 209458b.  The stratospheric model shown has P$_n = 0.0$,
which minimizes the value of $\kappa_{\rm e}$ necessary to fit the lone 
Harrington et al. (2007) data point, and it is the only model 
among the three depicted in Fig. \ref{fig4} that does fit.  Therefore, the trend
in ``inversion" strength" with F$_p$, seen in the sequence TrES-1, HD 189733b, 
and HD 209458b, continues with HD 149026b.  Not only do the EGP atmospheres
grow hotter with F$_p$ (a not unexpected result), but the importance
of a thermal inversion and a stratosphere in explaining the extant data 
increases with it as well.  Given the current paucity
of data for HD 149026b, we urge that HD 149026b be a priority target
so as to help discriminate the various models only partially represented on
the HD 149026b panel of Fig. \ref{fig4}.  We predict that the pattern
of the four IRAC flux ratios for HD 149026b will mimic that found for 
HD 209458b, and that its flux ratios from $\sim$10 $\mu$m to $\sim$30 $\mu$m 
will comfortably exceed those of models without obvious thermal inversions,
perhaps by large margins.

\section{Light Curves$-$Comparison with Data}
\label{curves}

Measuring the infrared planet-star contrast ratio as a function of orbital 
phase, i.e., the planet's light curve, provides the best constraints
on the longitudinal distribution of planetary emissions.  In principle,
phase-dependent light curves at different wavelengths can be inverted to determine the 
``brightness" temperature and composition distributions over the surface of the planet, 
including its night side.  Contrast ratios obtained not just at secondary eclipse 
($\alpha = 0^{\circ}$), but also at other phase angles, help reveal and quantify 
zonal winds and establish their role in redistributing stellar energy 
(i.e., P$_n$) and matter around the planet.  They can help identify transitions
at the terminator (Guillot \& Showman 2002; Showman \& Guillot 2002), 
shifts in the substellar hot spot (Cooper \& Showman 2005; Williams et al. 2006), asymmetries 
in the thermal distributions (Knutson et al. 2007b), and persistent atmospheric structures.  
In sum, light curve measurements probe both atmospheric dynamics and the 
planet's climate and are the key to the bona fide remote sensing of exoplanets.

Having said this, since full light curves require many more pointings and much more 
telescope time to obtain, and mostly address the dimmer phases of a planet's orbital traverse, 
obtaining them is much more difficult than measuring the contrast ratio at secondary eclipse.
As a result, to date there are only three published light curves for irradiated EGPs
(for $\upsilon$ And b, HD 179949b, and HD 189733b), despite numerous observational forays.   
All of these are for only one {\it Spitzer} waveband each and none  
covers a complete orbit.  There do exist recent upper limits (e.g., Cowan et al. 2007),
but these are not usefully constraining and we will not address them here\footnote{However, a few of these
limits are listed in Table \ref{t3}.}.

Below, we discuss the three systems for which light curves, however sparse, 
have been obtained and try to extract physical information by comparison with our 
light-curve models (eq. \ref{phasel}).  Before we do so, we note the following.  Classic light curve studies
are in the optical and measure geometric albedos, phase functions (Sudarsky et al. 2005),
and polarizations, i.e. they measure reflected stellar light. Albedos and polarizations 
are significantly affected by the presence of clouds, and so these traditional optical 
campaigns focus on reflection by condensates or surfaces.  The planet-star flux ratios in the optical
range from $\sim$10$^{-10}$ for EGPs at AU distances to $\sim$10$^{-5}$$-$10$^{-6}$ 
for the close-in EGPs near $\sim$0.05 AU (Table \ref{t1}).  However,   
in the near- and mid-IR, the planet-star contrasts are around $\sim$10$^{-3}$ (see Figs. 
\ref{fig4} and \ref{fig5}).  These larger numbers are why {\it Spitzer}
IR measurements, rather than optical measurements, have assumed center stage
in the direct study of EGPs.  The light seen is not reflected stellar light, but reprocessed 
stellar flux, emitted predominantly in the near- and mid-IR (Fig. \ref{fig3}) at the lower
temperatures ($\sim$1000$-$2000 K) of the resulting planetary atmospheres.  The ratio
between the optical and IR components, and thus the relative advantage of IR measurements, 
is roughly the square of the ratio between the orbital distance and the 
stellar radius, a number near $\sim$10$^2$ for most of the planets listed in Table \ref{t1}.

\subsection{$\upsilon$ And b}

Harrington et al. (2006) have measured the phase variation at 24 $\mu$m of the planet-star
contrast for the close-in EGP $\upsilon$ And b (Butler et al. 1997).  Since this planet is not transiting,
we know neither \mp\ nor $\sin{(i)}$, but only the combination M$_{\rm p}$$\sin{(i)}$ (= 0.69 \mj). 
Moreover, without a transit we don't have a measurement of \rp.  In fact, the models 
used to fit the five (!) data points obtained by Harrington et al. (2006), which are
not anchored by absolute calibration, depend on P$_n$, $\kappa_{\rm e}$, \rp, and $\sin{(i)}$ (see eq. \ref{phasel}). 
In addition, all interpretations hinge upon only the two extreme points in the Harrington et al. (2006) dataset.
Therefore, we have too many degrees of freedom to allow us to draw strong conclusions
concerning planetary and atmospheric parameters and must make do with limits and general
correlations.

\begin{figure}
\centerline{
\includegraphics[width=4.7cm,angle=-90,clip=]{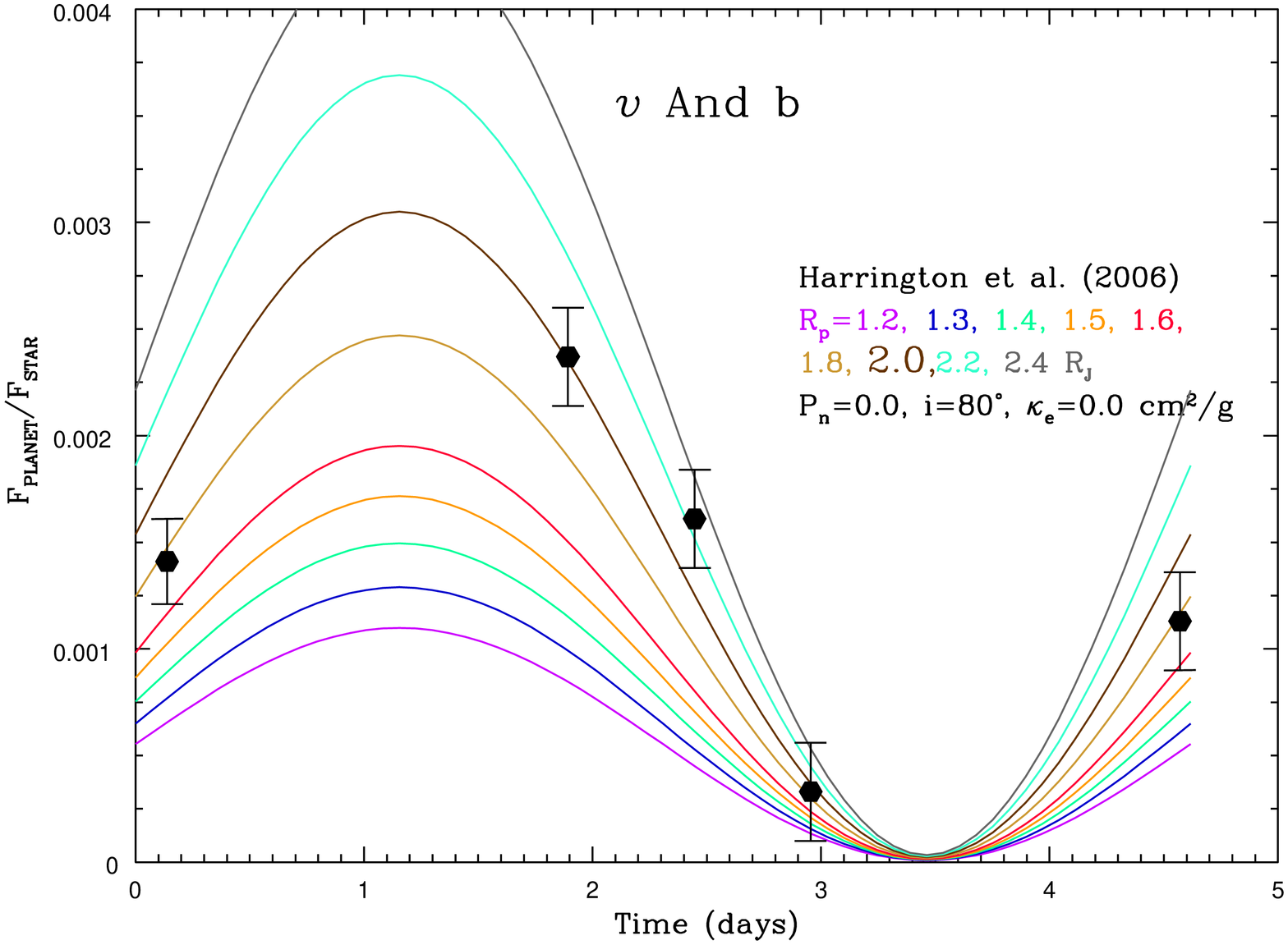}
\includegraphics[width=4.7cm,angle=-90,clip=]{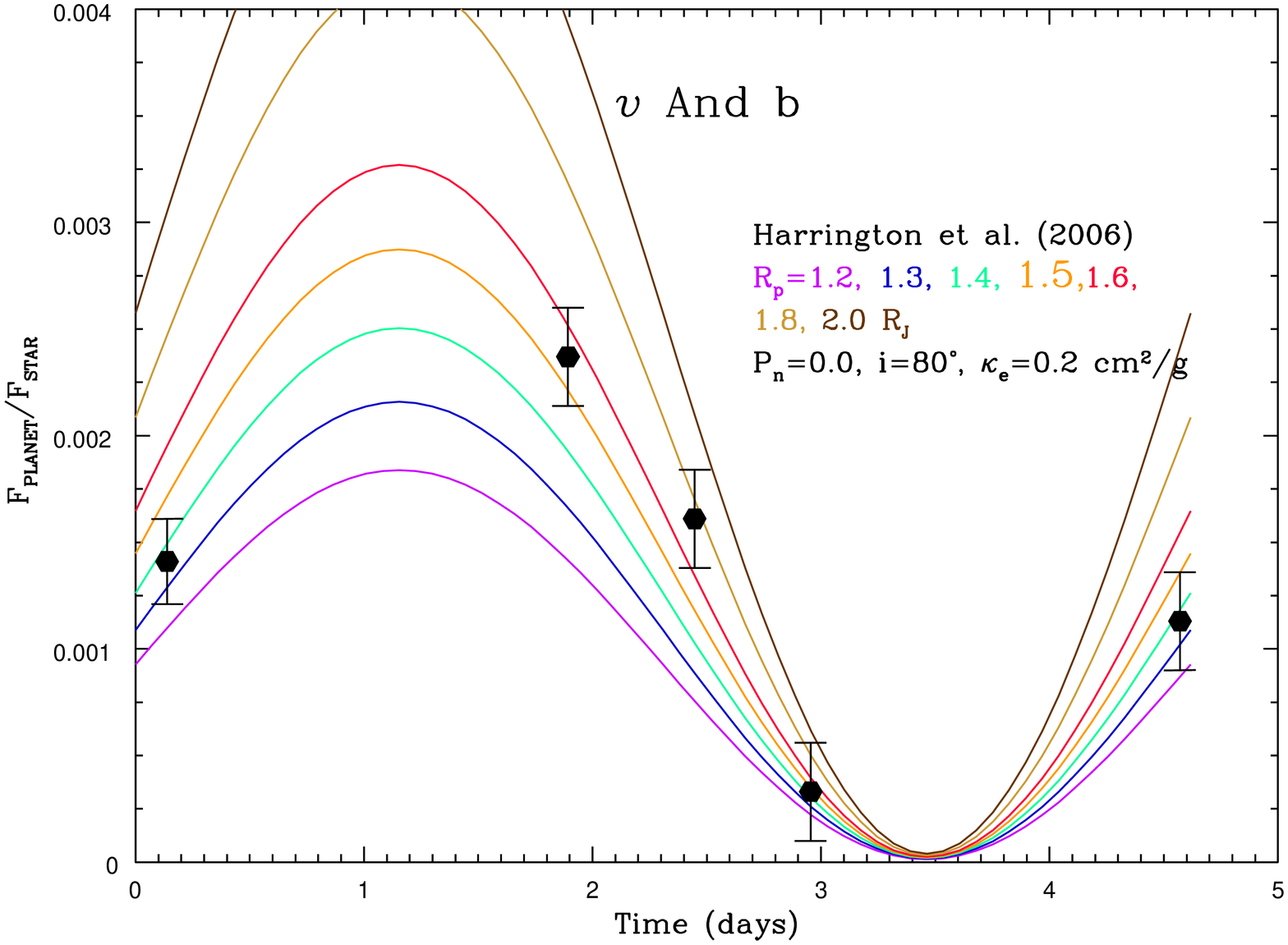}}
\centerline{
\includegraphics[width=4.7cm,angle=-90,clip=]{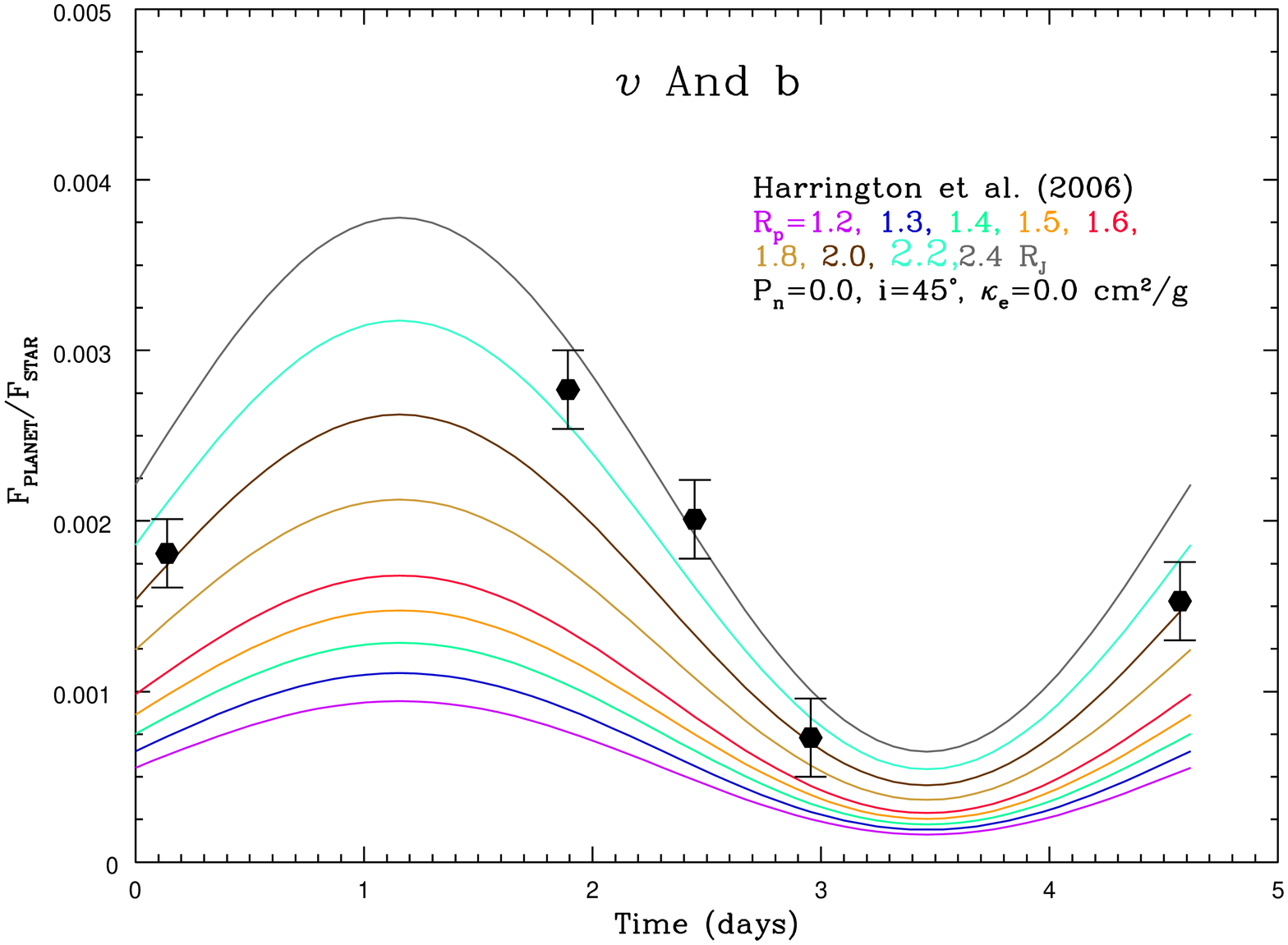}
\includegraphics[width=4.7cm,angle=-90,clip=]{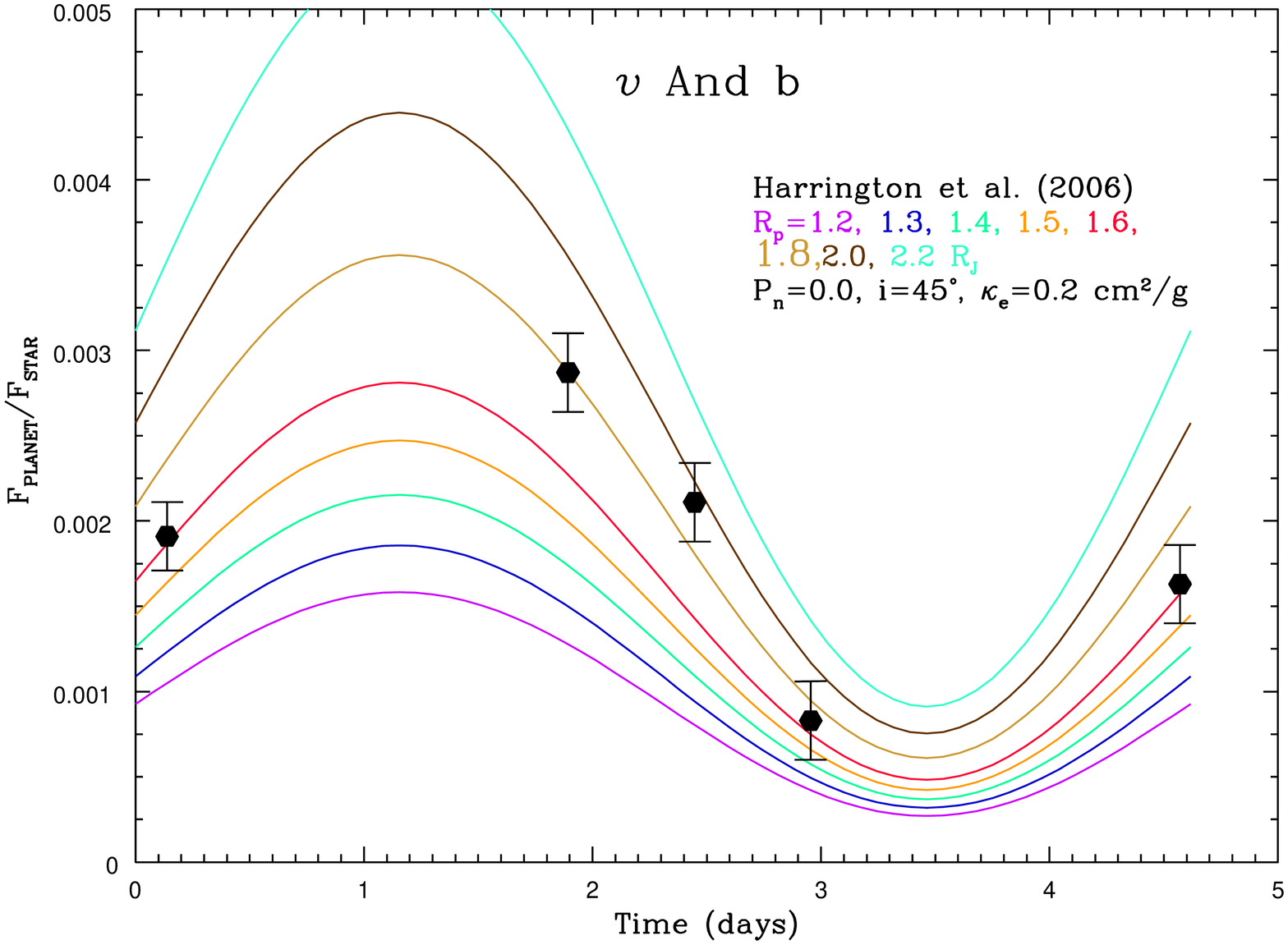}}
\centerline{
\includegraphics[width=4.7cm,angle=-90,clip=]{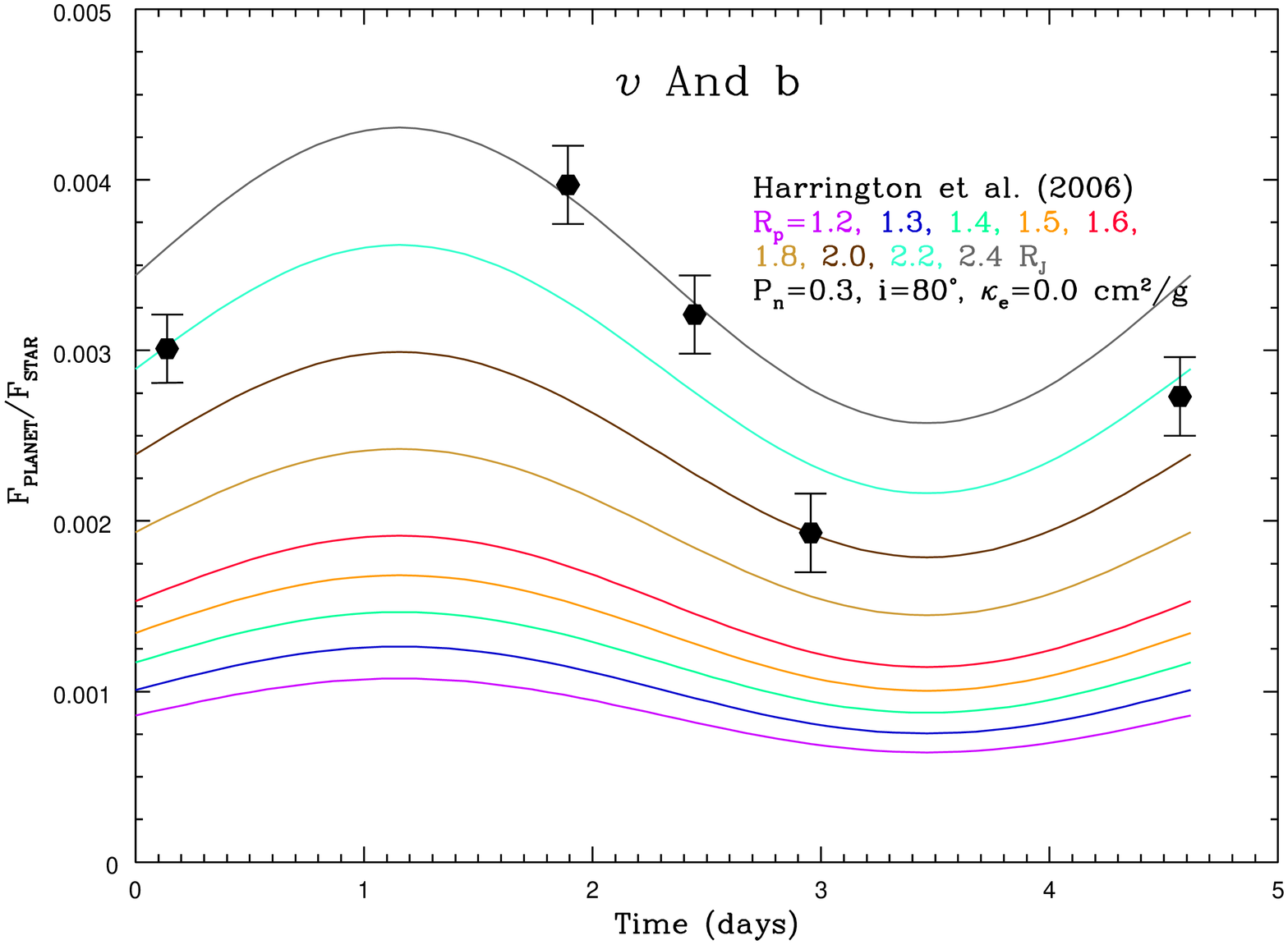}
\includegraphics[width=4.7cm,angle=-90,clip=]{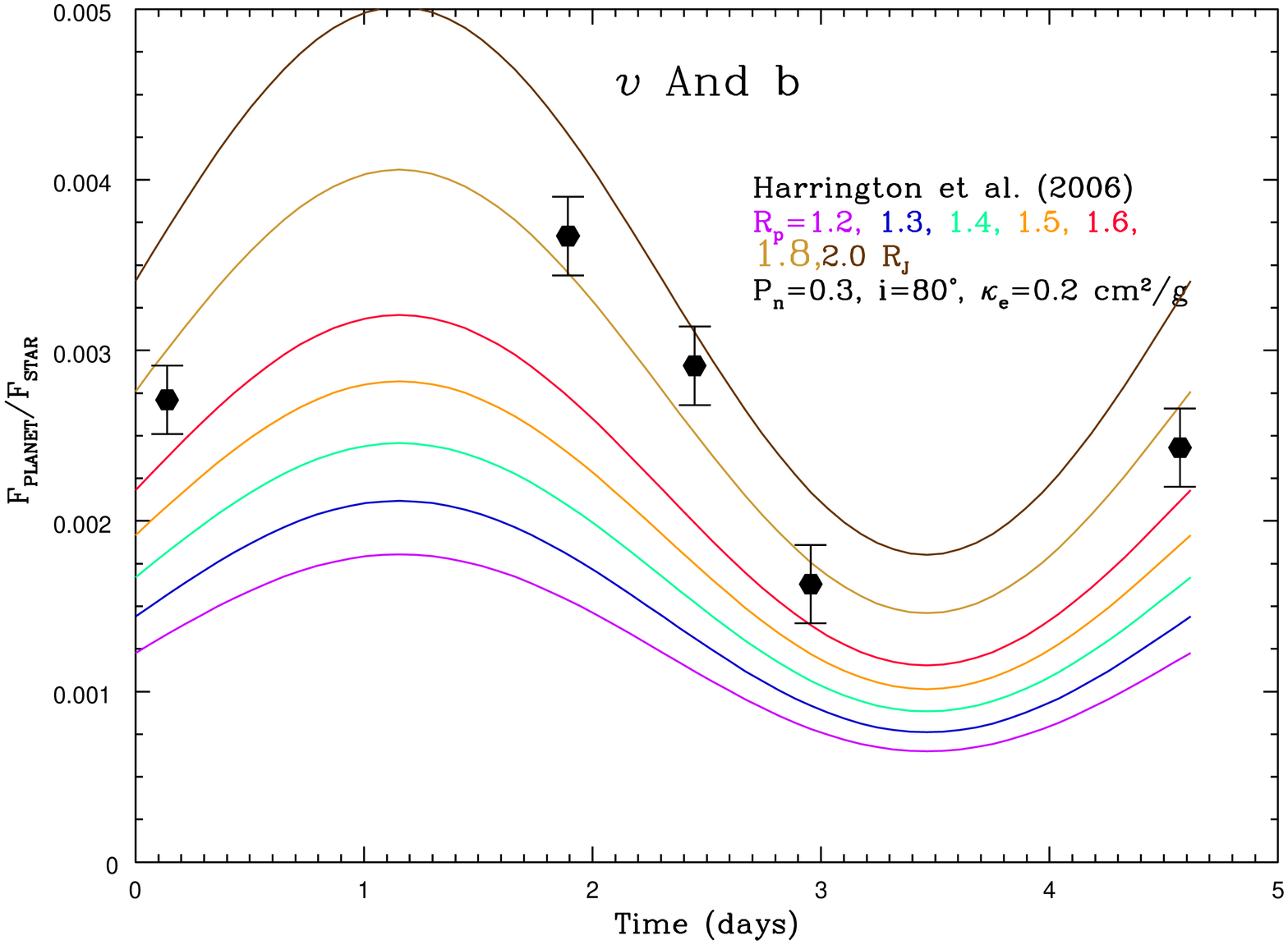}}
\centerline{
\includegraphics[width=4.7cm,angle=-90,clip=]{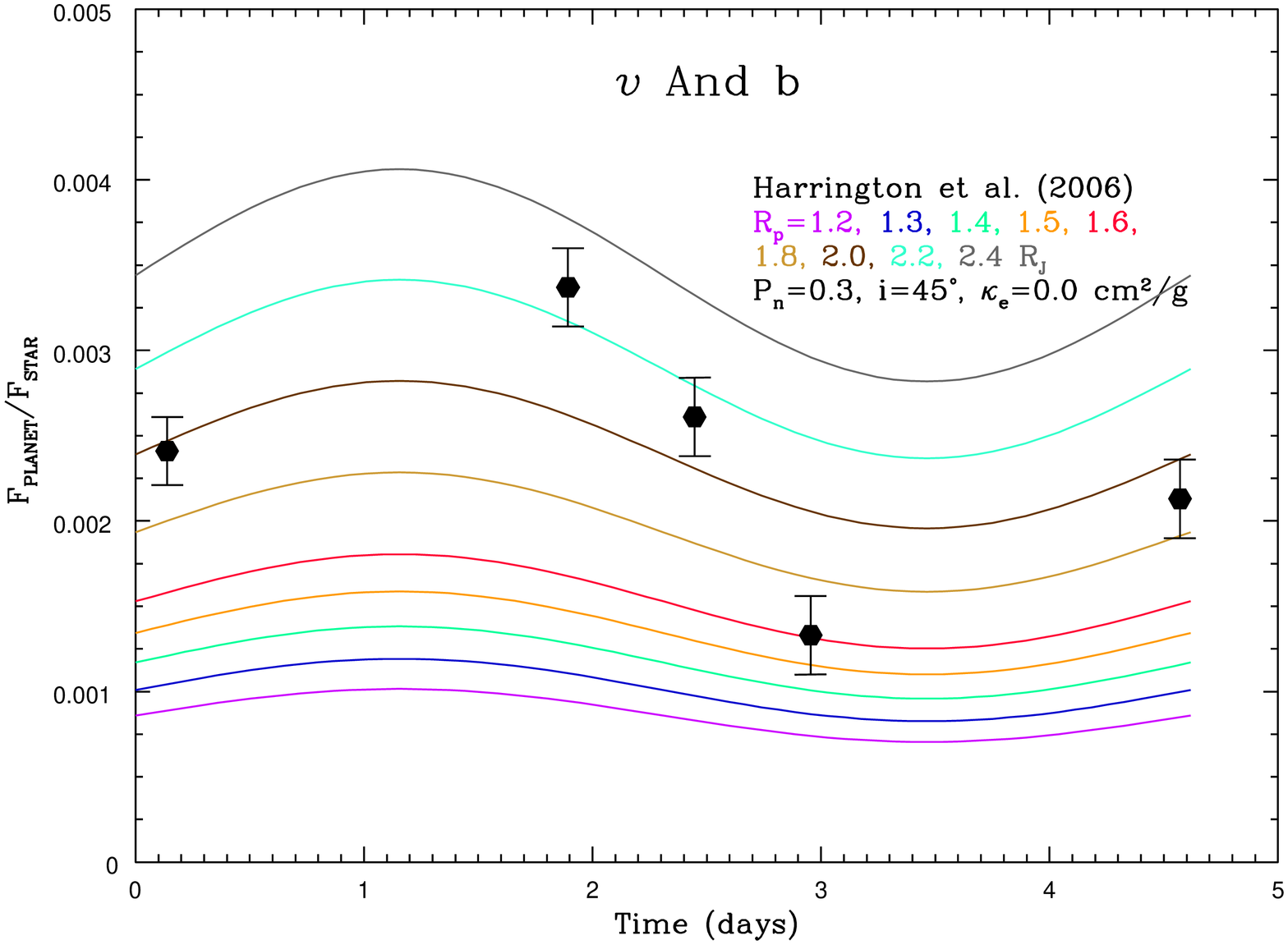}
\includegraphics[width=4.7cm,angle=-90,clip=]{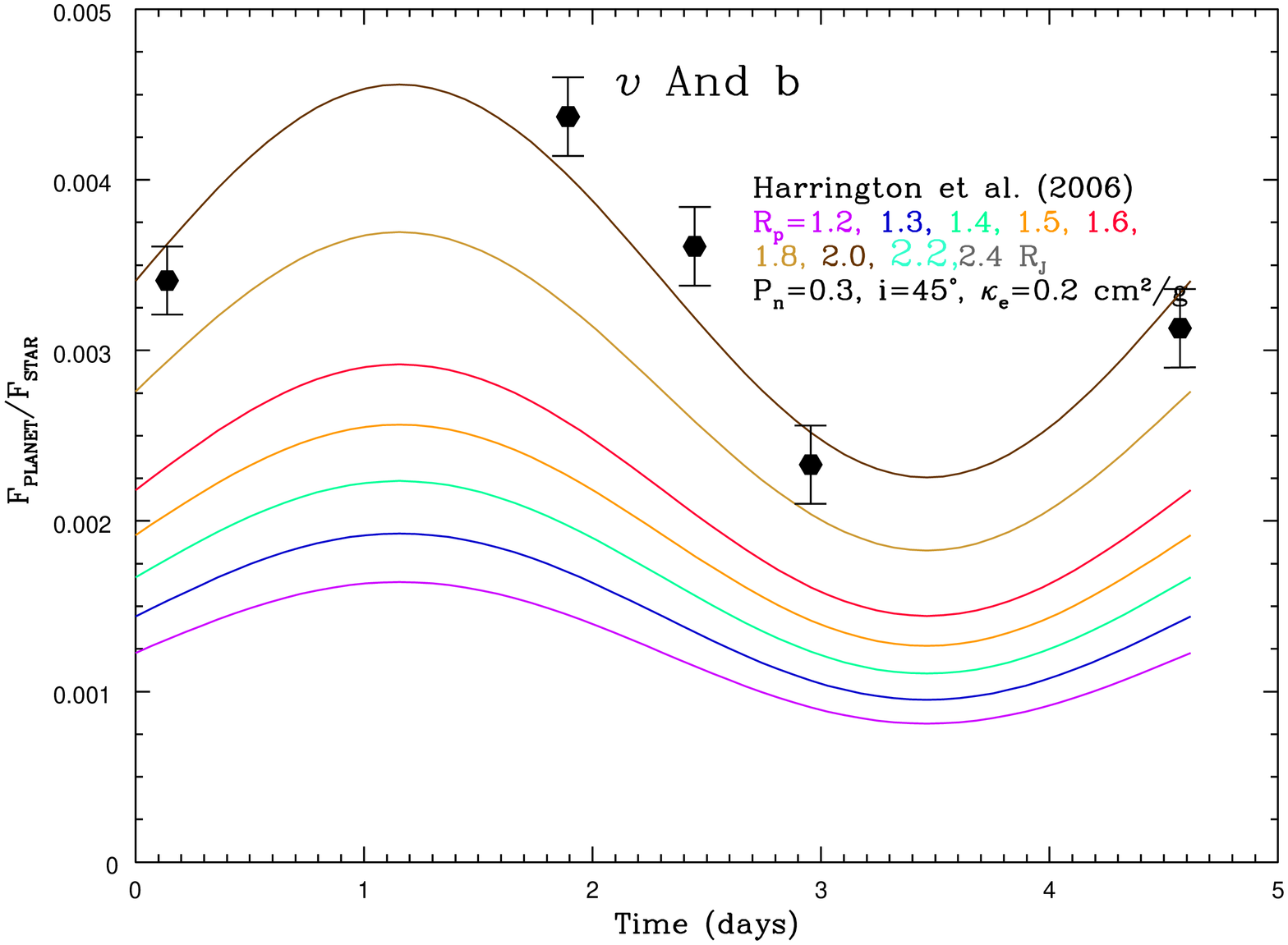}}
\caption{
Theoretical light curves for the non-transiting EGP $\upsilon$ And b in 
the MIPS 24-$\mu m$ band for different inclinations ($i$ = 45$^{\circ}$ 
and 80$^{\circ}$), values of P$_{n}$ (0.0 and 0.3), values of $\kappa_{\rm e}$ 
(0.0 and 0.2 cm$^2$/g), and a range of planetary 
radii.  Superposed are the light-curve measurements of Harrington et al. (2006).
Note that Harrington et al. (2006) obtain only relative contrast values, not
absolute values. Therefore, the data on each panel are shifted in absolute 
contrast space, while maintaining the measured relative values.  The day/night
contrast difference is preserved.  In this way, we find the corresponding best fits
consistent with the observations, but for different inclinations, etc. Note that the scales 
for the different panels can be different, though the right panel that faces each left panel
has the same scale as that left panel.  The left panels depict models without
a thermal inversion, while the right panels depict models that have thermal inversions
created using $\kappa_{\rm e}$ = 0.2 cm$^2$/g. 
%
%
Together, this set of 
panels illustrates the dependence on the three most important
free parameters: P$_{n}$, $i$, R$_{p}$, and $\kappa_{\rm e}$. See text for a discussion.
}
\label{fig6}
\end{figure}

Figure \ref{fig6} portrays in eight panels comparisons of theoretical 24-$\mu$m phase curves
with the $\upsilon$ And b data. The left panels contain models with $\kappa_{\rm e}$ = 0,
and the right panels contain models with $\kappa_{\rm e}$ = 0.2 cm$^2$/g.  The models in the top four panels 
have P$_n$ = 0.0, and those in the bottom four panels have P$_n$ = 0.3. Inclinations of both 45$^{\circ}$
and 80$^{\circ}$ (near eclipse) are employed.  On each panel, we provide models
with a wide range of planetary radii. Since the data have no absolute calibration,
we are free to move the data points up and down, as long as their relative values 
are maintained, and we have done so in an attempt to provide on each panel the best fit
to the overall shape and the day/night difference.  The corresponding $T/P$ profiles 
at $\alpha = 0^{\circ}$ (day) and $\alpha = 180^{\circ}$ (night) are displayed in the lower-left 
panel of Fig. \ref{fig1}.

From figures like Fig. \ref{fig6}, we can extract general trends and limits.  In their discovery
paper, the authors noted that the shift of the hot spot away from the substellar point was
small.  They also remarked on the large difference from peak to trough ($\sim$0.002).  Both
observations suggested that there is not much heat redistribution from the dayside 
to the nightside and that P$_n$ is small, perhaps near zero.  While at this stage this 
conclusion can not be refuted, our models suggest that there is a broader range of 
possible interpretations.  Importantly, as we have noted in \S\ref{planetstar}, the presence
of a stratosphere can enhance dayside planetary fluxes in the mid-IR, and certainly near 24 $\mu$m,
even for modest values of P$_n$ which would otherwise decrease $D/N$ (eq. \ref{phasel}).
$\upsilon$ And b experiences a substellar flux, F$_p$, of $\sim$1.3$\times$10$^9$ erg cm$^{-2}$ s$^{-1}$,
and this is larger than that impinging upon HD 209458b.  Therefore, we expect that
the atmosphere of $\upsilon$ And b will have a thermal inversion as well, thereby enhancing
the mid-IR day/night contrasts even for values of P$_n$ near 0.3\footnote{It is possible that
$\sin{(i)}$ is small, and, therefore, that the gravity and \mp\ are large. It is also possible that 
a large gravity can shift the breakpoint between EGP atmospheres with and without 
inversions. However, we suspect that F$_p$, more than gravity, is the crucial
parameter in determining this bifurcation.}. What is more, large values of \rp are 
becoming commonplace, and we can not eliminate this possibility
for $\upsilon$ And b.  A large value of \rp increases the amplitude swing 
from day to night.  Therefore, many parameter combinations can fit these data.

We summarize the lessons of Fig. \ref{fig6} as follows.  All else remaining the same, 
a change of P$_n$ from 0.0 to 0.3 results in an increase in the \rp required of
$\sim$0.3$-$0.5 \rj.  Substituting $\kappa_{\rm e} = 0.2$ cm$^2$/g for $\kappa_{\rm e} = 0.0$
decreases the \rp necessary by $\sim$0.4$-$0.5 \rj.  Replacing models with
$i$ = 45$^{\circ}$ by those with $i$ = 80$^{\circ}$ decreases the \rp required
by $\sim$0.2$-$0.3 \rj.  Specifically, a (P$_n$, $\kappa_{\rm e}$, $i$) 
= (0.0, 0, 80$^{\circ}$) model has a required radius of $\sim$2.0 \rj (top-left panel
of Fig. \ref{fig6}), while a (P$_n$, $\kappa_{\rm e}$, $i$) = (0.0, 0.2, 80$^{\circ}$)  model
has a required radius of $\sim$1.5 \rj (top-right panel of Fig. \ref{fig6}).  
And while a (P$_n$, $\kappa_{\rm e}$, $i$) = (0.3, 0, 80$^{\circ}$) model 
requires an \rp of $\sim$2.5 \rj, one with $\kappa_{\rm e} = 0.2$ cm$^2$/g requires
a radius of ``only" $\sim$1.8 \rj.  Clearly, introducing stratospheres into the mix allows
P$_n$ to assume a range of non-zero values for which heat redistribution would not be considered 
small.  However, the implied planetary radius would not be small either, though 
it would still be within the currently measured range.

Some of this degeneracy might be broken with light-curve measurements at many wavelengths
and with a more rapid cadence.  Furthermore, astrometric measurements of the stellar wobble
can provide $\sin{(i)}$ and the planet's mass, eliminating one important ambiguity. 
{\it Spitzer} can still be used to provide the former, while the latter is well within reach
of ground-based astronomy.  Finally, we would be remiss if we did not mention that JWST 
will inaugurate an era of stunning photometric improvement (by a factor of 
at least $\sim$10$^2$) over current IR platforms for the study of the light 
curves of both transiting and non-transiting EGPs.

\subsection{HD 179949b}

Cowan et al. (2007) obtained a light curve in IRAC channel 4 ($\sim$8 $\mu$m)
of another non-transiting giant planet, HD 179949b (Santos, Israelian \& Mayor 
2004; Wittenmyer, Endl \& Cochran 2007). It has an M$_{\rm p}$$\sin{(i)}$
of $\sim$0.95 \mj and a value of F$_p$ of $\sim$1.32$\times$10$^9$ erg cm$^{-2}$ s$^{-1}$).
This makes it similar to the $\upsilon$ And system, both in its general properties
and in the limitations on what can be uncovered.

\begin{figure}
\centerline{
\includegraphics[width=5.cm,angle=-90,clip=]{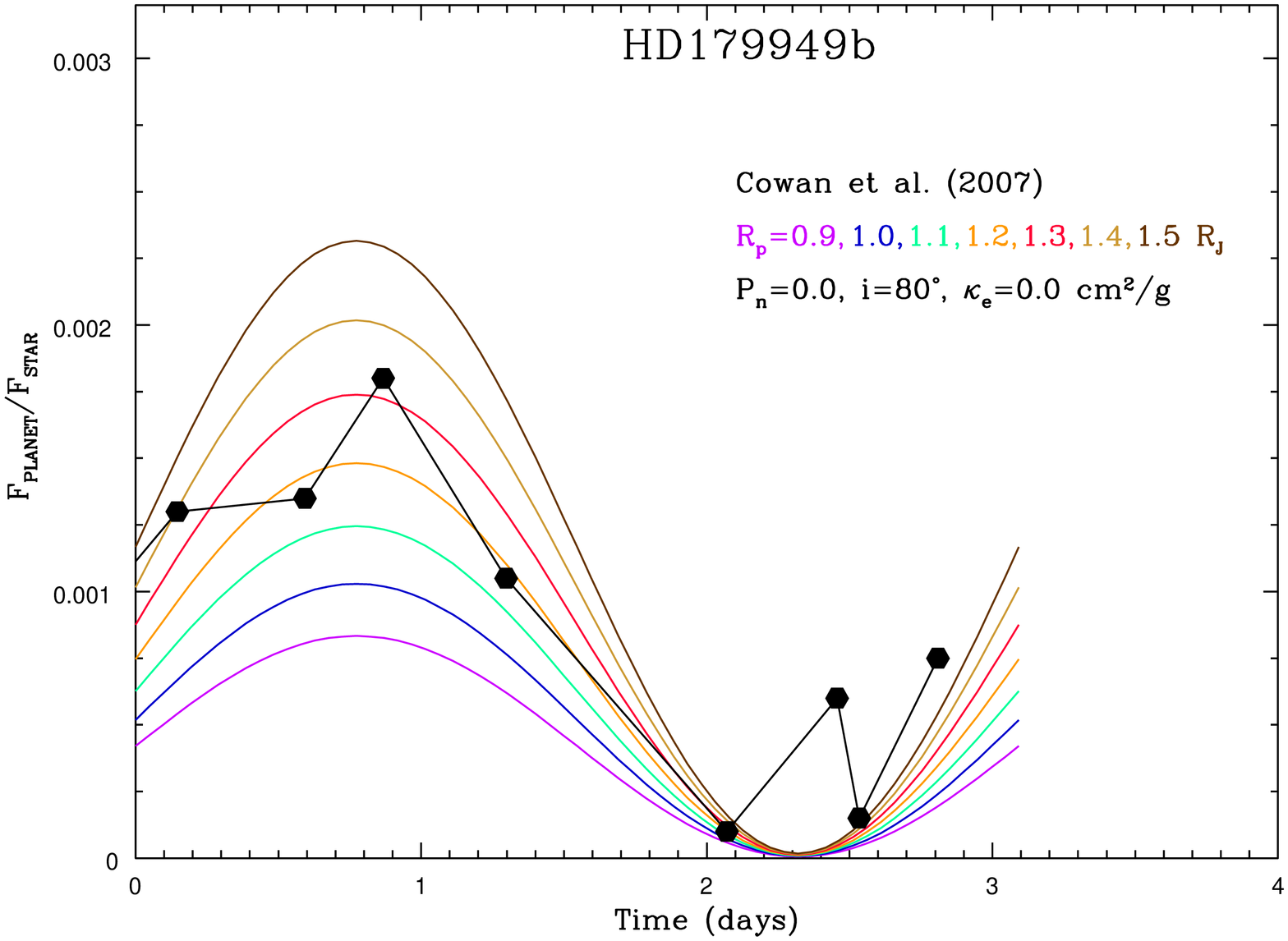}
\includegraphics[width=5.cm,angle=-90,clip=]{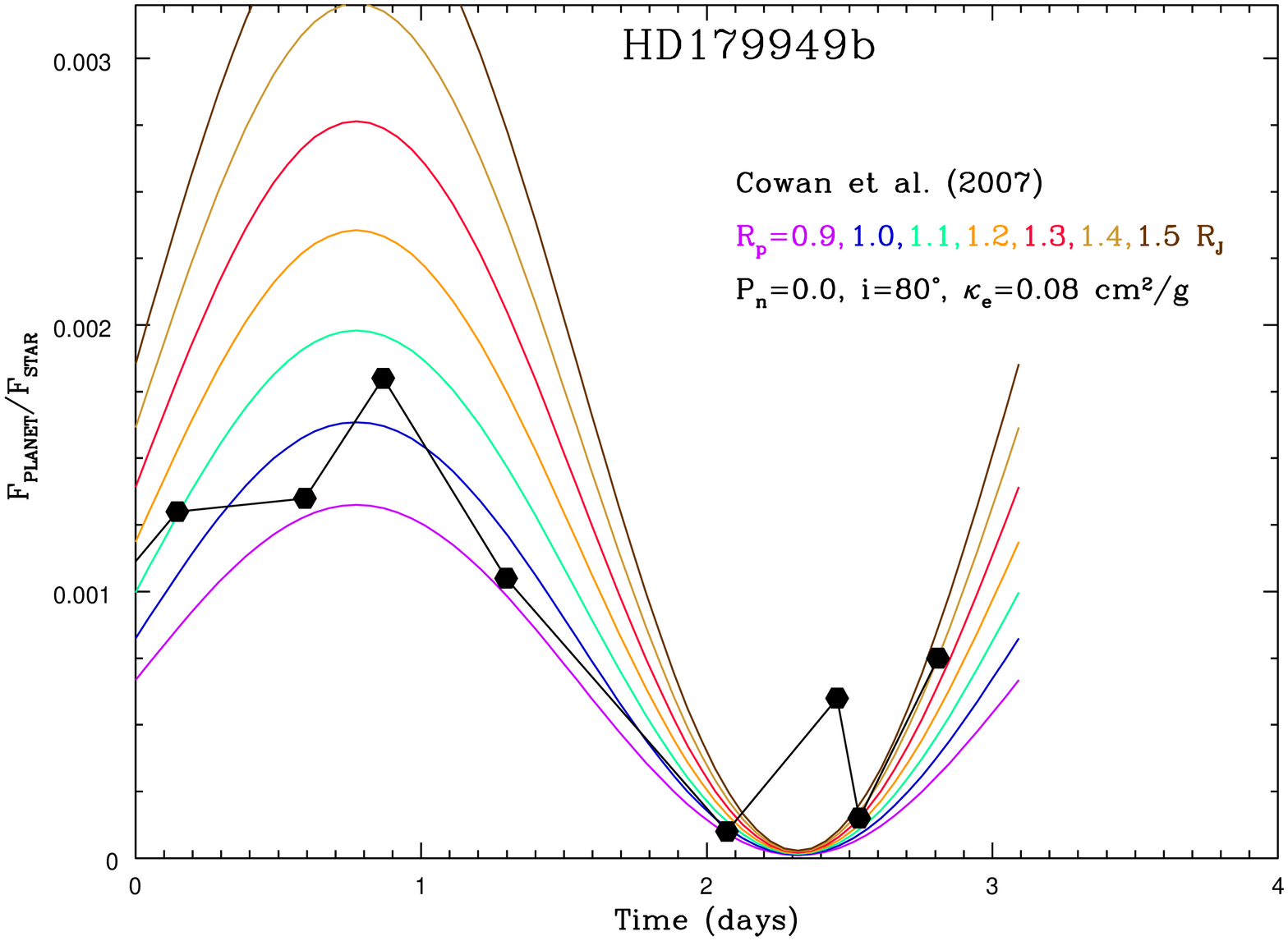}}
\centerline{
\includegraphics[width=5.cm,angle=-90,clip=]{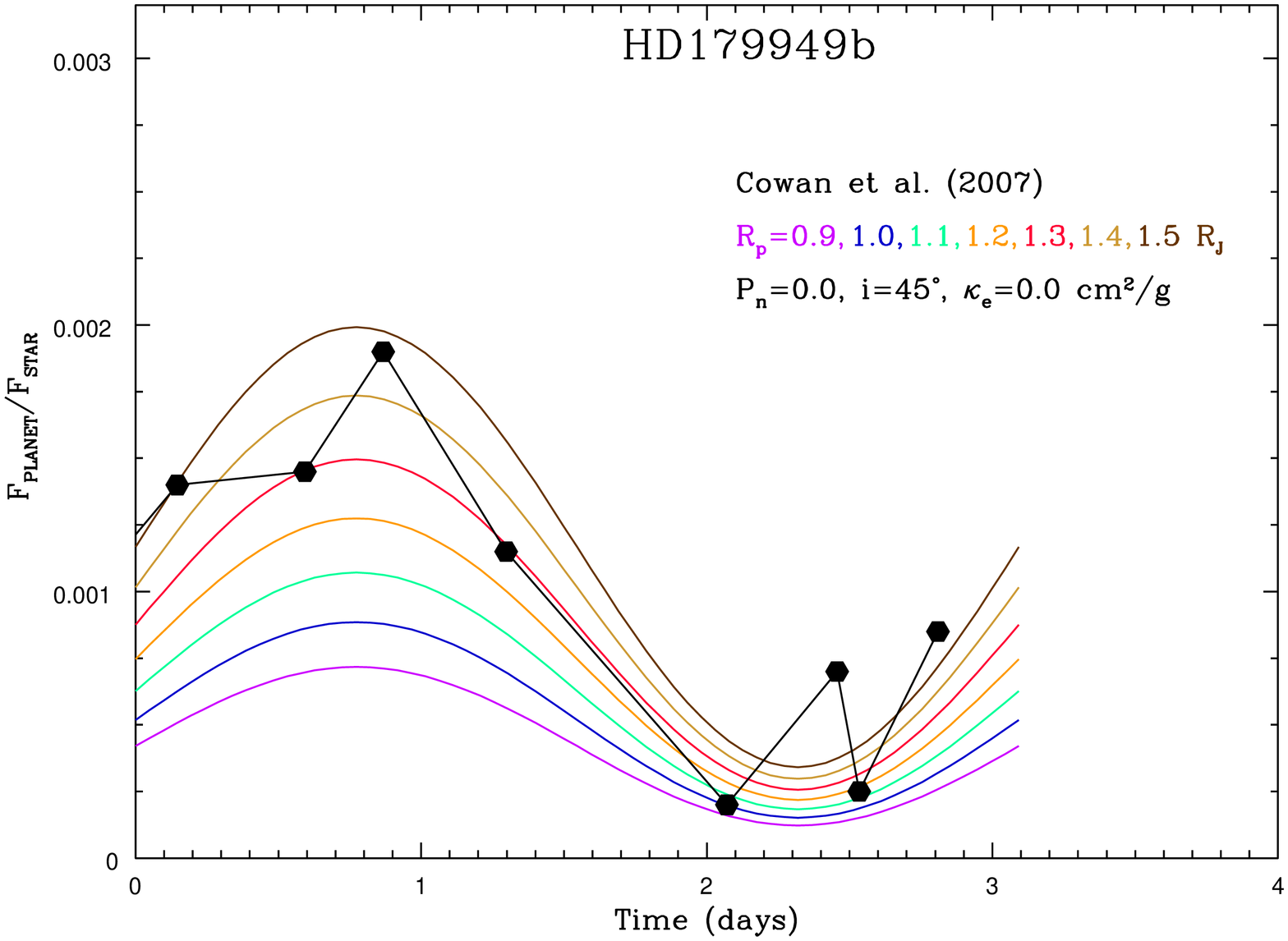}
\includegraphics[width=5.cm,angle=-90,clip=]{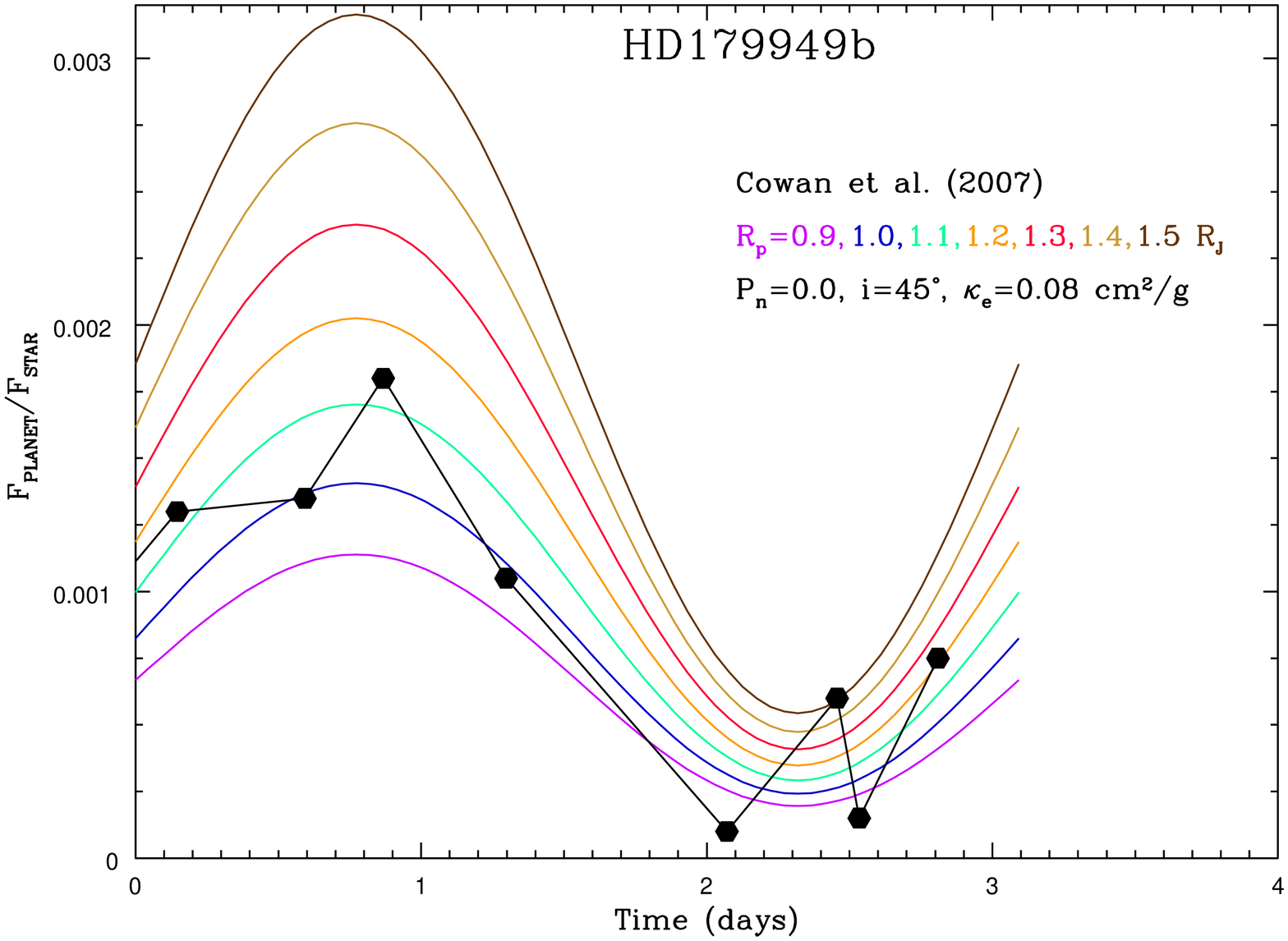}}
\centerline{
\includegraphics[width=5.cm,angle=-90,clip=]{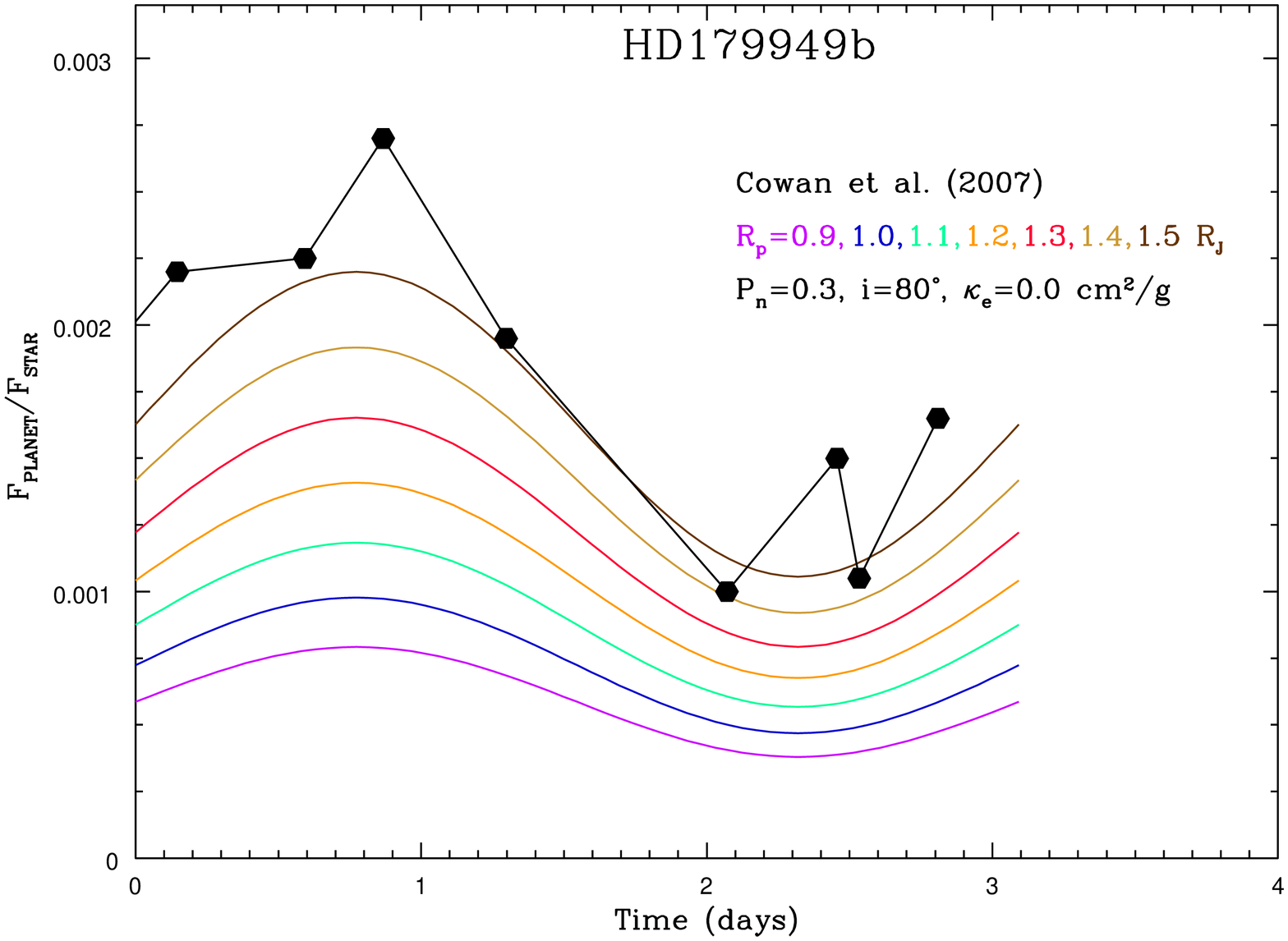}
\includegraphics[width=5.cm,angle=-90,clip=]{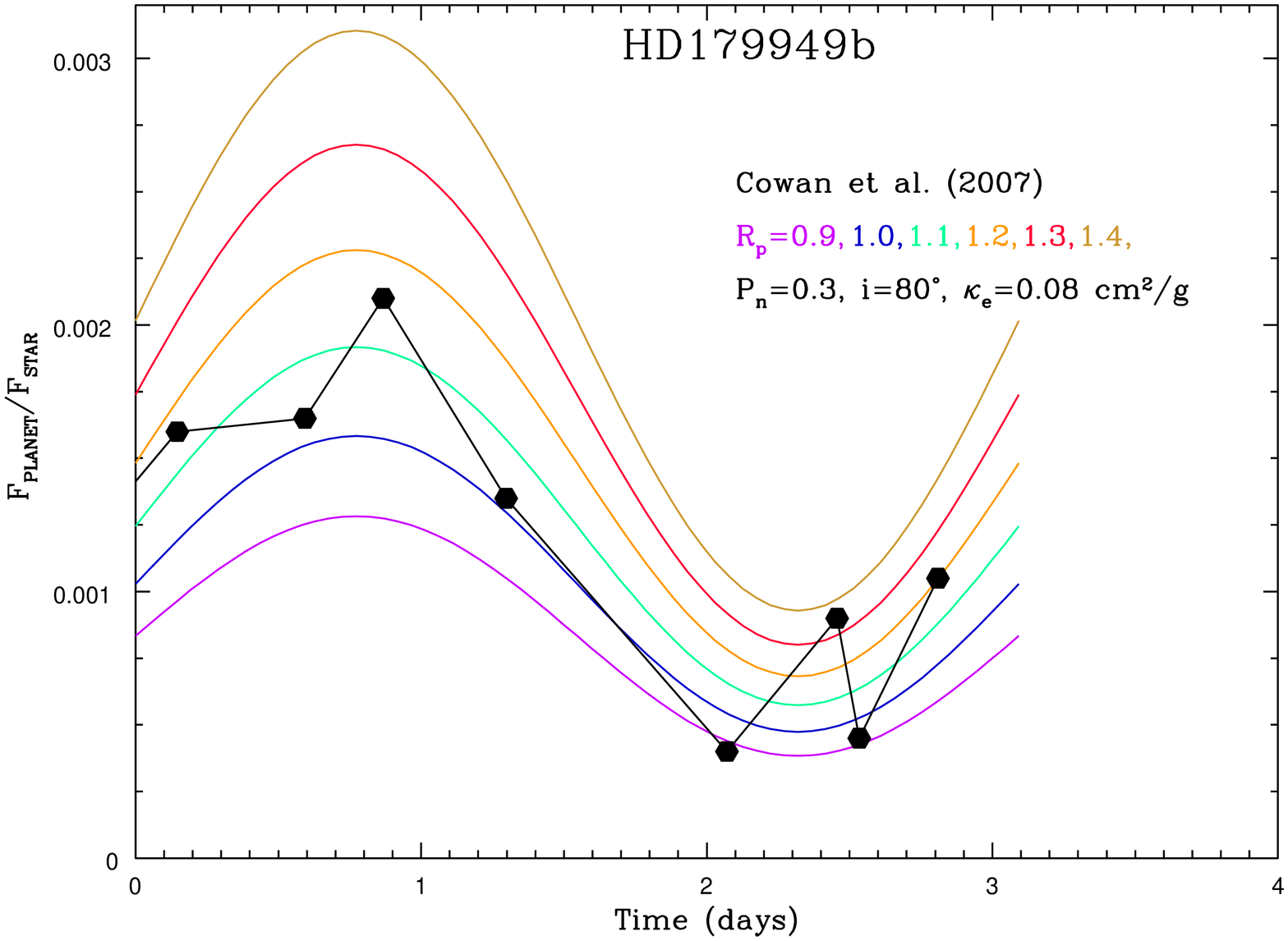}}
\centerline{
\includegraphics[width=5.cm,angle=-90,clip=]{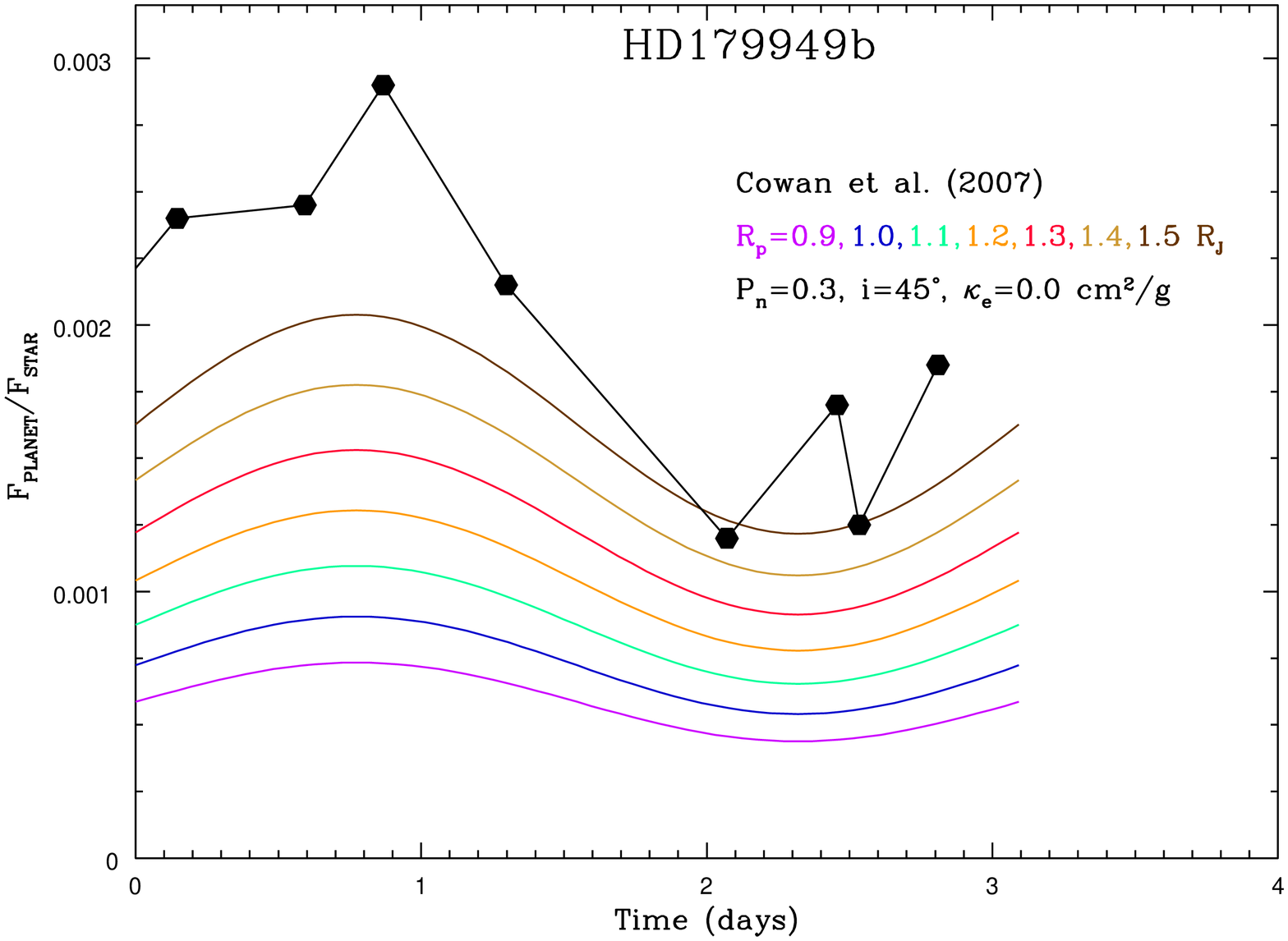}
\includegraphics[width=5.cm,angle=-90,clip=]{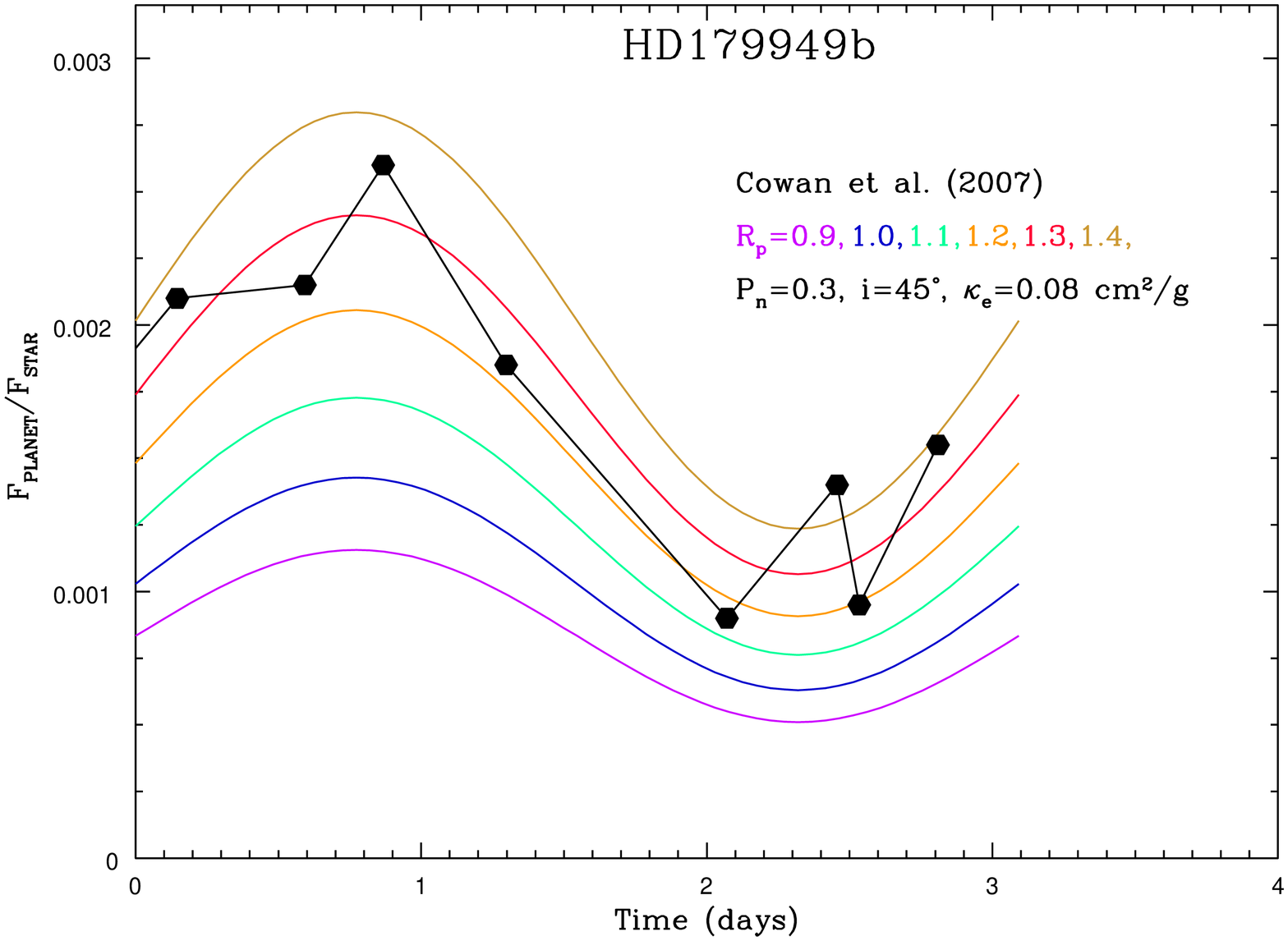}}
\caption{
The same as Fig. \ref{fig6}, but for HD 179949b in the IRAC 4 band
for two different inclination angles (45$^{\circ}$ and 80$^{\circ}$), two values of 
P$_{n}$ (0.0 and 0.3), a range of planetary radii, and two values of
$\kappa_{\rm e}$ (0.0 and 0.08 cm$^2$/g).  The light curve data from 
Cowan et al. (2006) are superposed on each panel. 
See text for a discussion.
}
\label{fig7}
\end{figure}

The lower right-hand panel of Fig. \ref{fig1} displays the dayside and 
nightside $T/P$ profiles of the light-curve models used to model these data. 
Figure \ref{fig7} portrays the corresponding eight-panel figure comparing
our theoretical models for various combinations of P$_n$, $\kappa_{\rm e}$, $\sin{(i)}$, and \rp
with the eight data points of Cowan et al. (2007).  Figure \ref{fig7}
is similar to Fig. \ref{fig6}, but the thermal inversion models are for 
$\kappa_{\rm e}$ = 0.08 cm$^2$/g and the range of model radii are different.
The data for HD 179949b are important, but no less ratty than those for $\upsilon$ And b.
Nevertheless, the values of \rp required to fit HD 179949b are systematically lower.

We summarize our conclusions from Fig. \ref{fig7} as follows. A 
change of P$_n$ from 0.0 to 0.3 requires \rp to increase by 
$\sim$0.1$-$0.2 \rj.  Replacing $\kappa_{\rm e} = 0.0$ cm$^2$/g by $\kappa_{\rm e} = 0.08$
decreases \rp by $\sim$0.2$-$0.4 \rj.  Substituting models with
$i$ = 45$^{\circ}$ for those with $i$ = 80$^{\circ}$ decreases \rp 
by $\sim$0.1$-$0.2 \rj.  A (P$_n$, $\kappa_{\rm e}$, $i$)
= (0.0, 0, 80$^{\circ}$) model that fits the data requires a radius of $\sim$1.2 \rj (top-left panel
of Fig. \ref{fig7}), while a (P$_n$, $\kappa_{\rm e}$, $i$) = (0.0, 0.08, 80$^{\circ}$)  model
requires a radius of $\sim$1.0 \rj (top-right panel of Fig. \ref{fig7}).
Whereas a (P$_n$, $\kappa_{\rm e}$, $i$) = (0.3, 0, 80$^{\circ}$) model
requires an \rp of $\sim$1.6 \rj, one with $\kappa_{\rm e} = 0.08$ cm$^2$/g requires
a much smaller radius, $\sim$1.1 \rj.  These radii are unexceptional, and in general 
the presence of a stratosphere substantially decreases the values required to fit 
these data.  Therefore, for HD 179949b we can fit the light-curve data with a quite
reasonable combination of parameters, though the various degeneracies 
still need to be broken.  

Finally, Fig. \ref{fig7} suggests that
the shift between the transit ephemeris and the light curve phases 
is not large, so we don't see any obvious advection downstream of the 
hot spot.  This is consistent with the interpretation by Harrington 
et al. (2007) of the $\upsilon$ And b phase curve (Fig. \ref{fig6}) and 
may be a feature of EGPs with stratospheres and/or hot upper atmospheres.  
While very tentative, this suggestion is reinforced by the observation that 
there is a definite displacement of the hot spot of HD 189733b, which 
seems to have a cooler upper atmosphere (Fig. \ref{fig1}). However,
the perception of meaningful differences in the displacements of hot spots 
could just as easily be false and be a consequence of having better 
data for HD 189733b. Moreover, we don't yet have a good model 
for the origin of such differences and possible correlations with P$_n$. 
Clearly, better sampled phase curve data would be very useful.

\subsection{HD 189733b}

Currently, the only light curve we have for a transiting EGP was obtained in IRAC 4 at 
$\sim$8 $\mu$m by Knutson et al. (2007b)\footnote{However, this is only the first of many anticipated.}. 
Not only do we have HD 189733b's radius (Table \ref{t1}), but these light-curve data have absolute calibrations.
In addition, there is dense coverage over a bit more than half the orbit, from just
before secondary eclipse to just after primary transit. Knutson et al. (2007b) 
derive the longitudinal dependence of the surface brightness
and find a hot spot shifted by $16\pm 6^{\circ}$ east of the substellar point, while
the coolest region is shifted about 30$^{\circ}$ west of the anti-stellar point.
Curiously, both the hot spot and the coolest spot are in the same hemisphere.
Nevertheless, this is the first ``map" of the surface of an exoplanet (Burrows 2007).
The authors also found an indication of a nonzero, but small, eccentricity with 
$e\, {\cos {\omega}}=0.0010\pm 0.0002$, where $\omega$ is the longitude of 
periastron, a transit radius at 8 $\mu$m of $1.137\pm 0.006\, R_{J}$ (slightly 
smaller than the optical radius), a stellar radius of $0.757\pm 0.003\, R_{\odot}$, 
and an inclination of $85.61\pm 0.04$ degrees.  

\begin{figure}
\centerline{
\includegraphics[width=13.cm,height=18.cm,angle=-90,clip=]{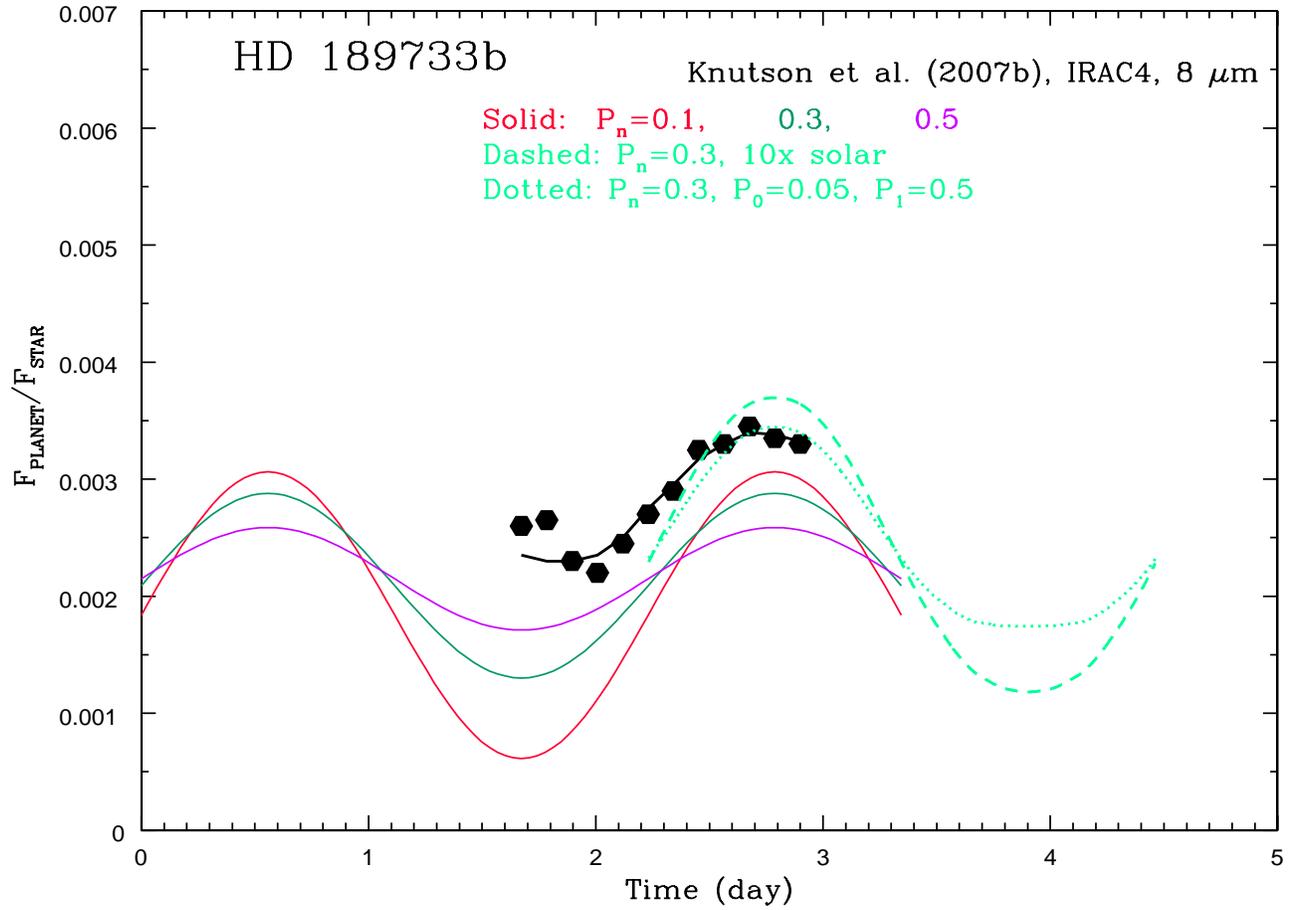}}
\caption{
A comparison between the light-curve measurements of HD 189733b in the 
IRAC 4 band (8 $\mu$m) performed by Knutson et al. (2007b, hexagons)
and our theoretical light curves for various values of P$_n$ (0.1, 0.3, 0.5).
Most of these models employ values for the redistribution pressure range (P$_0$ and P$_1$)
of 0.1 and 1.0 bars, but one model (dotted, and P$_n$ = 0.3) uses (P$_0$, P$_1$) = (0.05, 0.5) bars.
See figure legend for model parameters. Also included is a model with 10$\times$solar metallicity
(dashed).  All the P$_n = 0.3$ models are in green.  
See text for a discussion.
}
\label{fig8}
\end{figure}

These data are clearly the best of their kind and we have attempted to fit
them with our techniques and eq. \ref{phasel}.  The results are displayed
in Fig. \ref{fig8}.  The data are plotted as black hexagons, while the models
are for various values of P$_n$.  One model (dashed green) assumes 10$\times$solar
metallicity.  All these models, save one, assume (P$_0$: P$_1$) = (0.1, 1.0) bars,
not our default pair, but this makes little difference.  As indicated in \S\ref{hd189},
we can fit the contrast ratio in IRAC 4 at secondary eclipse rather easily, with a slight preference
for a small non-zero $\kappa_{\rm e}$.  As Fig. \ref{fig8} suggests, a super-solar
metallicity might also do the trick, but the metallicity dependence is
rather weak. Data in other bandpasses should break the degeneracy.

However, we can not fit the small day/night difference with any of our models.  The
data seem to imply a severe degree of heat redistribution, one that is still not
captured even with our P$_n = 0.5$ model.  We note in passing that models with P$_n = 0.5$
do not imply that the dayside and nightside should look the same, only that the integral
fluxes over the entire spectrum should be comparable.  Since the dayside is irradiated,
while the nightside emits into the blackness of space, as the upper right-hand panel 
of Fig. \ref{fig1} indicates, the $T/P$ profiles at $\alpha = 0^{\circ}$ and $\alpha = 180^{\circ}$ 
are different. This translates quite naturally into different day-night contrast ratio 
differences for different wavelengths, even for P$_n = 0.5$. 

What we seem to be seeing in the Knutson et al. (2007b) data are atmospheric inhomogeneities,
thermal structures (vortices?), on the surface of HD 189733b.  That the hot spot and the coolest spot
are in the same hemisphere, separated by only $\sim$45$^{\circ}$, suggests our symmetric
models are inadequate to fit this phase curve.  It is of paramount importance that 
a {\it full} light curve over all phase angles be taken in a variety of wavebands.  
Well-sampled data at longer wavelengths would be particularly welcome.
Moreover, model phase curves need to be sophisticated enough to incorporate temporal and 
3D spatial variations.  The light curves and night-side heating and 
thermal profiles depend centrally on jet streams, winds, and general thermal redistribution.
This puts a premium on developing GCMs with reasonable global dynamics coupled to 
realistic radiative transfer models.  Such models do not yet exist for the study of EGPs.

\section{Discussion and Conclusions}
\label{conclusions}

In this paper, we have constructed atmosphere and spectral models for all the close-in 
extrasolar giant planets for which direct-detection data from {\it Spitzer} have been 
published (except for the ``Neptune" GJ 436b).  These models incorporate the effects
of external stellar irradiation, detailed atmospheres, heat redistribution, and, 
for some, a model for stratospheric heating. Comparing the resulting suite of models with 
the data for these six EGPs, we have derived constraints on their atmospheric properties. 
We find, as did Burrows et al. (2007b), that many severely irradiated EGPs can have thermal
inversions at altitude which translate into qualitative changes in 1) the planet/star 
contrast ratios at secondary eclipse,  2) their wavelength dependences, and   
3) day-night flux contrasts during a planetary orbit.  Absorption features can flip
into emission features, planetary fluxes at long wavelengths can be enhanced, and 
the secondary-eclipse spectra in the near-IR can be altered significantly.  What is 
more, we find a correlation between the importance of such stratospheres and 
the flux at the substellar point on the planet.

Hubeny, Burrows, \& Sudarsky (2003) and Burrows, Sudarsky, \& Hubeny 
(2006) showed that strongly irradiated atmospheres can experience 
a solution bifurcation to an atmosphere with an inversion for 
which water spectral features are reversed from troughs (absorption) 
to peaks (emission).  This possibility is supported by the good fits 
obtained by Burrows et al. (2007b) to the HD 209458b IRAC data (Knutson 
et al. 2007c) and by our models in this paper for the subset of irradiated 
EGPs for which the presence of stratospheres is suggested (in particular 
HD 149026b and, perhaps, $\upsilon$ And b). In Hubeny, Burrows, \& Sudarsky (2003) 
and Burrows, Sudarsky, \& Hubeny (2006),  as well as in the prescient paper
by Fortney et al. (2006), the absorber was gas-phase TiO/VO, 
which for hot atmospheres in chemical equilibrium can exist at low pressures
at altitude and not just at high temperatures at depth.  The upper-atmosphere
absorber that is producing stratospheres for the higher values of F$_p$ might indeed 
be TiO/VO, but a ``cold-trap" effect can operate to deplete the upper atmosphere 
of TiO/VO.  However, when mass loss is ongoing, as we know to be 
the case for HD 209458b (Vidal-Madjar et al. 2003,2004), the atmosphere 
is constantly being replenished and TiO/VO at some non-zero abundance 
remains a viable option.  Such vigorous mass loss is expected for 
those planets with the highest values of F$_p$, and in this paper we have
discovered a possible correlation between F$_p$ and the existence of thermal
inversions and stratospheres.  Hence, the possible mass-loss/stratosphere 
connection may make for a compelling scientific narrative.

The tholins, polyacetylenes, or various non-equilibrium compounds discussed in the 
context of solar-system bodies could also be the necessary upper-atmosphere optical absorber.
Given the stellar UV and integral-flux regimes experienced by strongly-irradiated EGPs, 
such species might be photolytically produced with sufficient abundance (Burrows et al. 2007b; 
Marley et al. 2007).  However, to study these molecules requires a full 
non-equilibrium chemical network and we won't attempt this here.  Clearly, what 
the high-altitude absorber actually is, TiO/VO or some other compounds,
awaits investigation and is the primary reason we parametrized its opacity 
with $\kappa_{\rm e}$\footnote{However, for some of the models in Burrows et al. (2007b)  
we used equilibrium TiO/VO abundances and the corresponding molecular opacities 
(Sharp \& Burrows 2007). These models reproduced the HD 209458b IRAC data reasonably well.}.

The trend with F$_p$ we have uncovered suggests, however crudely, that those
EGPs with values of F$_p$ higher than HD 209458b's ($\sim$$10^{9}$ erg cm$^{-2}$ s$^{-1}$) 
may well have stratospheres.  Table \ref{t1} provides the needed numbers.  What this table suggests
is that TrES-2, TrES-3, TrES-4, HAT-P-2b, HAT-P-4b, HAT-P-5b, HAT-P-6b, 
OGLE-TR-10b, WASP-1b, WASP-3b, XO-3b, OGLE-TR-56b, OGLE-TR-211b, and OGLE-TR-132b,
in addition to HD 149026b, are strong candidates for having stratospheres, with all the consequences 
implied for their spectra and light curves (\S\ref{planetstar}; \S\ref{curves}).  Close-in,
though non-transiting, EGPs with high values of F$_p$ (such as $\tau$ Boo b, to name 
only one of many) are also likely to have thermal inversions. They too should manifest the 
spectral discriminants identified in Figs. \ref{fig4} and \ref{fig5} and the 
changes in the phase curves discussed in \S\ref{curves} and suggested by Figs. \ref{fig6} and
\ref{fig7}.  Note that by including in the above list HAT-P-2b, which has $\sim$10$\times$ 
the mass of the average close-in EGP, we have not addressed the possible role of 
gravity in these systematics. Though we suspect gravity is sub-dominant when compared 
with F$_p$, the dependence of upper-atmosphere physics and chemistry upon gravity should 
prove worth exploring.

It is unlikely that we will soon obtain spectral data for the OGLE planets.  
However, it is distinctly possible that EGPs listed in Table \ref{t1} with 
F$_p$s slightly lower than HD 208459b's will have ``weak" stratospheres, as we 
speculated may be the case for HD 189733b. This could include XO-2b, HAT-P-1b, 
WASP-2b, and, perhaps, XO-1b.  We note that F$_p$ for TrES-1 is lower still 
($\sim$0.43$\times10^{9}$ erg cm$^{-2}$ s$^{-1}$) and that this planet shows good evidence
for water in absorption and no appreciable stratosphere (\S\ref{tres1} and 
Burrows, Hubeny, \& Sudarsky 2005).  Hence, we have a hint at a breakpoint
between EGPs with and without significant stratospheres and thermal inversions. 

Despite speculation to the contrary, our models with abundant atmospheric water
are fully consistent with all the {\it Spitzer} data for all the measured
EGPs, though at times the water is in emission, not absorption.  This conclusion
is consistent with the possible identification of water in HD 189733b by 
Tinetti et al. (2007)  (though see Ehrenreich et al. 2007) and in HD 209458b by Barman (2007).

We find that the family of close-in EGPs probably boasts a range of values of P$_n$
from $\sim$0.1 to $\sim$0.4.  However, our constraints on this parameter are rather weak, 
particularly given the possible complicating effects in some EGP atmospheres of a stratospheric absorber. 
Currently, the presence of such an absorber makes it easier for values of P$_n$ that are not 
small to explain the data, even for $\upsilon$ And b.  However, the magnitude of these 
effects is hard to pin down with rigor, other than to say they are in evidence $-$ there remains 
a slight degeneracy between P$_n$ and ``$\kappa_{\rm e}$."  Without a more first-principles theory 
concerning the chemistry, spectroscopy, and abundance of this extra absorber at low 
pressures, and concerning the stellar environment in which it arises, usefully constraining 
P$_n$ may continue to be difficult.  This is the case even if TiO and VO fit the bill, 
since their steady-state abundances would still be an issue. Moreover, much better models for redistribution
are urgently needed.  Lunine \& Lorenz (2002) have speculated that planetary atmospheres
with jet streams to redistribute dayside heat to the nightside adjust their wind dynamics 
to maximize the rate of entropy generation.  Their formalism suggests a value of P$_n$ of 
$\sim$0.2, which is not inconsistent with any of the currently known data on
EGP atmospheres.

The direct-detection data we have addressed in this paper are clearly only the 
first of many anticipated {\it Spitzer} contrast and light-curve measurements of 
strongly-irradiated EGPs. MOST will continue its campaign, future dedicated space 
missions will be proposed, and ground-based IR telescopes may have a role.  
Transiting EGPs are continuing to be discovered at an impressive rate that 
will not soon abate, providing an expanding catalog for follow-up and 
characterization.  JWST is in the wings to revolutionize the field, and will 
come on line in the middle of the next decade. This paper is meant to provide
a broad theoretical context for these initiatives and an interpretive vocabulary
with which to address the ongoing study of extrasolar planets in tight orbits
around their illuminating stars.


\acknowledgments

We thank Heather Knutson, Dave Charbonneau, Bill Hubbard, 
Mike Cushing, and Drew Milsom for helpful discussions and 
guidance and Drake Deming, Alex Sozzetti, and Jamie Matthews for 
the use of their data in advance of publication. This study was 
supported in part by NASA grants NNG04GL22G, NNX07AG80G, and NNG05GG05G 
and through the NASA Astrobiology Institute under Cooperative 
Agreement No. CAN-02-OSS-02 issued through the Office of Space
Science. In addition, the first author thanks the Image Processing and Analysis Center (IPAC)
and the Spitzer Science Center for hosting him during the preparation of this manuscript.
Model data will be available at http://zenith.as.arizona.edu/\~{}burrows/.

\begin{appendix}

\section{An Improved Treatment of the Redistribution of Stellar Irradiation from the Dayside to the Nightside}
\label{redist}

In this Appendix, we describe a slightly more physical
version of our previous treatment (Burrows, Sudarsky, \& Hubeny 2006)
of the day to night side redistribution.
The total incident stellar flux (expressed as the $H$-moment) at the planetary surface is
\begin{equation}
H_{\rm ext} = \frac{1}{2} \left(\frac{R_\ast}{a}\right)^2 \,
\frac{\sigma}{4\pi}\, T_{\rm eff}^4 = \frac{1}{2}
\left(\frac{R_\ast}{a}\right)^2 \, H_\ast\, ,
\end{equation}
where $R_\ast$ is the stellar radius, $T_{\rm eff}$ is the effective
temperature of the stellar surface, and $a$ is the planet-star distance.
The basic feature of our model is an assumption that out of this
total incident stellar flux, a fraction P$_n$ contributes an additional
source of energy on the nightside, and is removed from the dayside.
P$_n$ is bounded between 0.0 (no redistribution) and 0.5 (``complete" redistribution)
and is conceptually the same as the redistribution parameter employed in Burrows, Sudarsky,
\& Hubeny (2006), but is here implemented slightly differently. We introduce
\begin{equation}
H_{\rm irr} \equiv {\rm P}_n\, H_{\rm ext} =
\frac{{\rm P}_n}{2}\, \left(\frac{R_\ast}{a}\right)^2 \, H_\ast\, .
\end{equation}
Formally, we take P$_n > 0$ at the nightside (signifying a gain in
energy), and P$_n < 0$ at the dayside (signifying a sink of energy).

We define a local gain/sink of energy, $D(m)$, such that
\begin{equation}
\label{dint}
\int_0^\infty D(m) \, dm = H_{\rm irr}\, .
\end{equation}
We assume that $D(m)$ is non-zero only between column masses
$m_0$ and $m_1$ (specified through limiting pressures $P_0$ and
$P_1$, which for our default set of calculations are 0.05 and 0.5, bars, respectively).
We have studied the model dependence on the values of $P_0$ and $P_1$ 
around 0.1 to 1.0 bars. The results suggest only a modest effect on 
the overall spectra at secondary eclipse. However, as the top-left
panels of Figs. \ref{fig1} and \ref{fig4} suggest, one can optimally fit the IRAC 2 to
IRAC 1 flux ratio for HD 209458b by strategically placing the redistribution band between the 
corresponding photospheres, and thereby cool the IRAC 1 ``$\tau = 2/3$" surface relative
to the IRAC 2 ``$\tau = 2/3$" surface. 
While this did not motivate our default values of $P_0$ and $P_1$ (used, 
one notes, for all objects in this study, not just HD 209458b, and motivated by the desire
to redistribute heat near the $\tau_{\rm Rosseland} = 2/3$ level), the reader should 
be aware that detailed fits to the secondary eclipse data depend on the choice of $P_0$ and $P_1$.
Clearly, better models of heat redistribution than we have employed here are called for.  

With this caveat in mind, we consider two models for $D(m)$. The first
one assumes a constant $D(m)$ between $m_0$ and $m_1$:
\begin{equation}
\label{dmod1}
D(m) = \frac{H_{\rm irr}}{m_1-m_0}\, , \qquad {\rm (model\ 1)}\, .
\end{equation}
The second model, model 2, assumes a linearly decreasing $D(m)$ between $m_0$
and $m_1$, in such a way that $D(m)$ reaches $0$ at $m=m_1$. This
is our default model for the calculations of this paper and its
functional form is:
\begin{equation}
\label{dmod2}
D(m) = \frac{2 H_{\rm irr}}{m_1-m_0}\, \frac{m_1-m}{m_1-m_0}\, ,
\qquad {\rm (model\ 2)}\, .
\end{equation}
Therefore, $D(m)$ is non-negative on the nightside and
is non-positive on the dayside.

The first two moments of the transport equation read
\begin{equation}
\label{hmomf}
\frac{dH_\nu}{dm} = \kappa_\nu (J_\nu - B_\nu)\, ,
\end{equation}
and
\begin{equation}
\label{kmomf}
\frac{dK_\nu}{dm} = \chi_\nu H_\nu\, ,
\end{equation}
where $\kappa_\nu$ is the absorption coefficient per gram, $\chi_\nu$
is the total extinction coefficient (absorption + scattering), and $m$
is the column mass, related to the pressure by the relation $P=mg$, where $g$ is the
gravity.

Since we stipulate sinks or sources of energy at certain layers, the
usual radiative equilibrium (or radiative+convective) equilibrium
does not apply. Instead, it is replaced by the following energy equation,
which can be written in two different ways:

a) using the frequency-integrated first moment of the transfer equation,
\begin{equation}
\label{ener1}
\int_0^\infty\kappa_\nu (J_\nu - B_\nu)\, d\nu = -D(m)\, ,
\end{equation}
because the energy gained per unit mass, $D(m)$, is balanced
by the net radiation loss per unit mass, given by the integral on the
left-hand-side of eq. (\ref{ener1}).

b) using the equation for the frequency-integrated flux,
$H \equiv \int_0^\infty H_\nu d\nu$. From eqs. (\ref{hmomf}) and (\ref{ener1}), we have
\begin{equation}
\label{ener2}
\frac{dH}{dm} = - D(m)\, ,
\end{equation}
which we rewrite as an equation for the integrated $H$ as
\begin{equation}
\label{htot}
H(m) = H_{\rm eff} + \int_m^{m_d} D(m^\prime)\, dm^\prime\, ,
\end{equation}
where $m_d$ is a sufficiently large column mass at which one has
$D(m_d)=0$ (that is, deeper than the region of the sources/sinks), and
\begin{equation}
\label{heff}
H_{\rm eff} \equiv \frac{\sigma}{4\pi} T_{\rm int}^4\,
\end{equation}
is the nominal total flux deep in the atmosphere, expressed through an
effective temperature, $T_{\rm int}$.

In view of a simple linear form of $D(m)$, the integral in eq. (\ref{htot})
can be evaluated analytically. We obtain for model 1:
\begin{eqnarray}
\label{htotm1}
H(m) & = & H_{\rm eff}
\, , \qquad m \geq m_1\nonumber  \\
H(m) & = & H_{\rm eff} + H_{\rm irr}\frac{m_1-m}{m_1-m_0}
\, , \qquad  m_0 \leq m \leq m_1\nonumber \\
H(m) & = & H_{\rm eff} + H_{\rm irr} \qquad m\leq m_0\, ,
\end{eqnarray}
and for model 2:
\begin{eqnarray}
\label{htotm2}
H(m) & = & H_{\rm eff}
\, , \qquad m \geq m_1\nonumber  \\
H(m) & = & H_{\rm eff} + H_{\rm irr}\left(\frac{m_1-m}{m_1-m_0}\right)^2
\, , \qquad m_0 \leq m \leq m_1\nonumber  \\
H(m) & = & H_{\rm eff} + H_{\rm irr}  \qquad m\leq m_0\, .
\end{eqnarray}
As suggested by Hubeny \& Lanz (1995), it is numerically advantageous
to represent the energy balance equation as a linear combination of
eqs. (\ref{ener1}) and (\ref{htot}), where the $H$-moment is expressed as
$H_\nu = dK_\nu/dm = d(f_\nu J_\nu)/dm$, where $f_\nu$ is the Eddington
factor. Using the Eddington factor enables us to consider only one
radiation moment, $J_\nu$, as an unknown quantity. The Eddington factor
is not taken as an unknown; instead it is held fixed in linearization
and is recalculated in the formal solution step.

The above equations do not consider convection, so they apply in the
radiative zone. In the convection zone, eqs. (\ref{htot}), (\ref{htotm1}),
and (\ref{htotm2}) remain unchanged, provided we take
\begin{equation}
H_{\rm eff} = \frac{\sigma}{4\pi} T_{\rm int}^4 - \frac{F^{\rm conv}}{4\pi}\, ,
\end{equation}
where $F^{\rm conv}$ is the convective flux. Equation (\ref{ener1})
is modified in the presence of convection to read
\begin{equation}
\label{ener1c}
\int_0^\infty\kappa_\nu (J_\nu - B_\nu)\, d\nu = -D(m)
-\frac{1}{4\pi}\, \frac{d F^{\rm conv}}{dm}\, .
\end{equation}

{\sc CoolTLUSTY} solves the energy equation specified through
eqs. (\ref{heff}) - (\ref{ener1c}) numerically and self-consistently
with the set of equations of radiative transfer.

However, one can gain deeper physical insight by
developing a simplified gray model, in which we can actually derive analytic
expressions for the local temperature. More importantly, such a model
allows us to understand what values of the limiting pressures for
the sink region on the dayside and the corresponding optical depths
are physically acceptable.   We develop these expressions in 
Appendices \S\ref{app2} \& \S\ref{app3} below.

\section{A Semi-analytic, Gray model with Redistribution}
\label{app2}

In {\sc CoolTLUSTY}, the exact energy balance equation is solved self-consistently
with the radiation transport equation. However, it is very useful to
develop a simple gray model that allows us to study the conditions
under which the structural equations have a solution at all.
On the nightside, there is always a solution because we are adding energy
at certain layers. However, on the dayside, we remove energy
at certain layers. If we require that these layers are too deep in the
optically-thick part of the atmosphere, the only way energy can 
be removed is to create a negative temperature gradient. If,
moreover, we require the region with a negative gradient (the region of the
energy sink) to continue to depth, we would eventually reach a negative temperature,
which is clearly unphysical.

To demonstrate this and to construct an analytic atmosphere model with
redistribution, we closely follow the derivation of the analytic model given in Hubeny, Burrows, \&
Sudarsky (2003), generalizing it to account for departures from
radiative equilibrium due to the stipulated sources/sinks of energy.
We write the frequency-integrated moment equations 
(\ref{hmomf}) and (\ref{kmomf}) using the mean opacities:
\begin{equation}
\label{hmom}
\frac{dH}{dm} = \kappa_J J - \kappa_B B\, ,
\end{equation}
and
\begin{equation}
\label{kmom}
\frac{dK}{dm} = \chi_H H\, ,
\end{equation}
where $\kappa_J$, $\kappa_B$, and $\chi_H$ are the absorption-mean,
Planck-mean, and flux-mean opacities, respectively, and $J,H,K$ are
the frequency-integrated moments.

The energy balance equation reads (neglecting convection) 
\begin{equation}
\kappa_J J - \kappa_B B = -D(m)\, ,
\end{equation}
which is just another form of eq. (\ref{ener1}).

First, we obtain the solution for the second moment $K$. We write the second-moment
equation as
\begin{equation}
\frac{dK(\tau)}{d\tau} = H(\tau)\, ,
\end{equation}
where $d\tau \equiv \chi_H dm \approx \chi_{\rm ross} dm$ is the
flux-mean optical depth, which can be approximated as the Rosseland
optical depth. In order to integrate the second-moment
equation analytically, we introduce the limiting optical depths
$\tau_0 \equiv \tau(m_0)$ and  $\tau_1 \equiv \tau(m_1)$, and
adopt the following approximation for $H(\tau)$:
\begin{equation}
H(\tau) = H_{\rm eff} + H_{\rm irr}\left(\frac{\tau_1-\tau}{\tau_1-\tau_0}
\right)^n\, ,
\qquad {\rm for}\  \tau_0\leq \tau \leq \tau_1\, ,
\end{equation}
together with the exact expressions $H(\tau)=H_{\rm eff}$ for $\tau\geq \tau_1$
and $H(\tau) = H_{\rm eff} + H_{\rm irr}$ for $\tau\leq \tau_0$. Here,
$n=1$ for model 1, and $n=2$ for model 2. This equation is easily
solved and yields:
\begin{eqnarray}
K(\tau) & = & K(0) + (H_{\rm eff} + H_{\rm irr})\, \tau\, , 
\qquad \tau \leq \tau_0\, \nonumber \\
K(\tau) & = & K(0) + H_{\rm eff}\, \tau + H_{\rm irr}\left\{\tau_0 + 
\frac{\tau_1-\tau_0}{n+1}
\left[1-\left(\frac{\tau_1-\tau}{\tau_1-\tau_0}\right)^{n+1}
\right]\right\}\, , \qquad \tau_0 \leq \tau \leq \tau_1\, , \nonumber \\
K(\tau) & = & K(0) + H_{\rm eff}\,\tau + H_{\rm irr}
\left(\frac{1}{n+1}\, \tau_1  + \frac{n}{n+1}\, \tau_0 \right)\, , 
\qquad \tau \geq \tau_1\, .
\end{eqnarray}
With the LTE-gray model, this equation is in fact an equation for
the local temperature. We invoke the Eddington approximation,
$K = J/3$, and use the energy balance equation
\begin{equation}
\label{bloc0}
\kappa_B B = \kappa_J J + D(m)\, ,
\end{equation}
and the fact that $B=(\sigma/\pi) T^4$ to derive the 
local temperature. We introduce the following quantities:
\begin{equation}
J(0)=3K(0)=\alpha J_{\rm ext}\, ,\quad H_{\rm ext} = f_H J_{\rm ext}\, ,\quad
H_{\rm irr} = {\rm P}_n H_{\rm ext}, {\rm and} \quad w \equiv H_{\rm eff}/H_{\rm ext}\, ,
\end{equation} 
and we write $D(\tau)$ instead of $D(m)$ in eq. (\ref{bloc0}) as
an approximate expression
\begin{equation}
D(\tau) = \frac{H_{\rm irr}}{\tau_1-\tau_0}\, \frac{\bar\chi}{\bar\kappa_B}
\equiv \frac{H_{\rm irr}}{\tau_1-\tau_0}\, \epsilon\, ,
\end{equation}
where $\bar\chi$ and $\bar\kappa_B$ are the average values of the
flux-mean and the Planck-mean opacities in the interval $(\tau_0, \tau_1)$.
Note that in the strict gray model, $\epsilon=1$. Note also that the
ratio $w$ is given by
\begin{equation}
w = (T_{\rm int}/T_\ast)^4\, 2\,(R_\ast/a)^{-2}\, ,
\end{equation}
so that for the case of strong irradiation, $w \ll 1$.
Since here we are interested in deep layers, we use the approximation
$\kappa_J = \kappa_B$, (which is, however, not valid at the surface layers).
Using the above defined quantities, we can express the integrated Planck
function (i.e., temperature) through $H_{\rm ext}$ only and derive:
\begin{equation}
\label{btot}
B=\left[\beta + w \tau +3{\rm P}_n q(\tau) +\frac{\epsilon {\rm P}_n}{\tau_1 - \tau_0}
\right] H_{\rm ext}\, ,
\end{equation}
where we denoted $\beta=\alpha/(3f_H)$ (which is a constant of order
unity), and 
\begin{equation}
q(\tau) \equiv \tau_0 + \frac{1}{n+1}(\tau_1-\tau_0)\left[1 -
\left(\frac{\tau_1-\tau}{\tau_1-\tau_0}\right)^{n+1}\right]\, ,
\end{equation}
for $\tau_0 \leq \tau \leq \tau_1$; $q(\tau)=0$ for $\tau\leq\tau_0$,
and $q(\tau)=q(\tau_1)$ for $\tau\geq \tau_1$.
Notice also that since $B=(\sigma/\pi) T^4$, eq.(\ref{btot}) can be 
understood as an equation for the local temperature.

\section{Condition for the Existence of the Solution on the Dayside}
\label{app3}

We now investigate the existence conditions for the solution on the dayside.
We first make the following approximations: $\alpha=2$, $f_H=1/2$;
thus, $\beta=4/3$, $\epsilon=1$. 
We assume that $\tau_0 \ll \tau_1$, so we 
neglect it in the expression for $q(\tau)$. We also introduce
the notation $p_n = -{\rm P}_n$, noting that $p_n$ is a positive quantity.
Let us first take model 1, in which $n=1$. Using all the above
approximations, we obtain for the region between $\tau_0$ and $\tau_1$:
\begin{equation}
B(\tau)/H_{\rm ext} = \beta-\frac{p_n}{\tau_1} + w\tau - p_n \tau\left(
1-\frac{\tau}{2\tau_1}\right)\, .
\end{equation}
Since $\tau_1$ is typically larger than 1, we neglect 
the term $p_n/\tau_1$ 
compared to $\beta$. This is not necessary for the formal development,
but it simplifies the resulting expressions. The derivative 
$dB/d\tau = w-p_n(1-\tau/\tau_1)$; therefore, the local minimum of $B(\tau)$
is at $\tau \equiv \tau_{\rm min} = \tau_1(1-w/p_n)$. 
Since the most interesting case is
for strong irradiation where $w \ll 1$, we see that the minimum
of $B$ is close to $\tau_1$. The value of $B(\tau)$ at the local minimum is
\begin{equation}
B_{\rm min} = B(\tau_{\rm min}) = \beta - \frac{\tau_1 p_n}{2}
\left(1 - \frac{w}{p_n}\right)^2\, .
\end{equation}
The condition for the existence of the solution is that $B_{\rm min} > 0$,
and, thus,
\begin{equation}
\tau_1 < \frac{2\beta}{p_n}\, \left(1-\frac{w}{p_n}\right)^2\, .
\end{equation}
Taking the most interesting case of strong irradiation, we can
neglect the second factor, and write simply
\begin{equation}
\tau_1 < \frac{2\beta}{p_n} \approx \frac{8}{3 p_n}\, .
\end{equation}
An analogous analysis for model 2 gives a similar condition,
\begin{equation}
\label{result}
\tau_1 < \frac{4}{p_n}\, .
\end{equation}
For instance, this demonstrates that for $p_n=0.5$, the case with the maximum energy sink on 
the dayside and the maximum degree of redistribution, the deeper (high-pressure) limit of the sink region must
be at optical depths less than 8.  Eq. \ref{result} indicates that this limit is larger
for smaller $p_n$ (e.g., it is 40 for $p_n=0.1$).  Hence, we have derived consistency conditions
for our redistribution algorithm that have physical content and in our choices 
for $P_0$ and $P_1$, we are careful not to exceed this condition.  Our default 
values of $P_0$ and $P_1$, 0.05 and 0.5 bars, respectively,
translate into an optical depth range of a few$\times$0.1 to $\sim$a few, within the
consistency constraints for all the P$_n$s.  Indeed, if we were to exceed the consistency constraints
in our atmosphere calculations, they would not converge and the simulations would crash numerically.

\section{The Origin of the $\lowercase{f} = 2/3$ Term}
\label{app5}

To derive the proper $f$ factor, we now expand upon the formalism of Appendix \S\ref{redist}.
The total energy flux received by a unit area on the planetary
surface at angle $\theta_0$ from the substellar
point is given by
\begin{equation}
F(\mu_0) = 4\pi\left(\frac{R_\ast}{a}\right)^2\,H_\ast \mu_0\, ,
\end{equation}
where $\mu_0 = \cos\theta_0$ and $\theta_0$ is the angle between
the normal to the planetary surface and the direction toward the 
star. For simplicity, we assume that the angular diameter of
the star is small. Consequently, all rays coming from the star
are parallel. $H_\ast = F_\ast/4\pi$ and $F_\ast$ is 
the radiation flux at the surface of the star.
The average flux received by the planet is then
\begin{equation}
\label{fav}
F_{\rm av} = \int_0^1 F(\mu_0)\, d\mu_0 = \frac{1}{2}\,
4\pi H_\ast \left(\frac{R_\ast}{a}\right)^2\, ,
\end{equation}
which explains the origin of the $f=1/2$ ansatz.

To improve upon this, we assume the Eddington approximation, $J=3K$, and
ignore convection, which allows us to write down an analytic solution 
of eq. (B4) for $J$:
\begin{equation}
\label{jeq}
J(\tau) = J_0 + 3 H \tau\, ,
\end{equation}
where $4\pi H$ is the interior planetary flux.
To obtain the constant $J_0 = J(0)$, we employ the formal solution
of the transfer equation for the specific intensity:
\begin{equation}
\label{ifor}
I(0, \mu) = \int_0^\infty (\kappa_J/\kappa_B)\, J(t)\,  e^{-t/\mu}\, dt/\mu \, ,
\end{equation}
since the source function, $S$, is given by $S=B=(\kappa_J/\kappa_B)J$.
As shown by Hubeny et el. (2003), $\kappa_j/\kappa_B$ can differ 
significantly from unity, but only for low values of the flux-mean
optical depth, $\tau \ll 1$. Therefore, we set  $\kappa_J/\kappa_B=1$, and,
using eq. (\ref{jeq}), we integrate eq. (\ref{ifor}) to obtain
\begin{equation}
I(0,\mu) = J_0 + 3 H \mu\, ,
\end{equation}
which is the well-known Eddington-Barbier relation.
The mean intensity on the planetary surface is given by
\begin{equation}
J(0) = (1/2)\int_0^1 I(0, \mu)\, d\mu + (1/2)\int_{-1}^0 
I^{\rm ext}(\mu) d\mu\, ,
\end{equation}
where
\begin{equation}
I^{\rm ext}(\mu) = \delta(-\mu-\mu_0) F(\mu_0)/\pi\, ,
\end{equation}
because we have assumed all incident rays from the star are parallel.
$\delta()$ is the Dirac $\delta$-function.
Therefore,
\begin{equation}
J_0 = J(0) = \frac{1}{2} J_0 + \frac{3}{2} H + \frac{F(\mu_0)}{2\pi}\, ,
\end{equation}
and, consequently
\begin{equation}
J_0 =  3H + \frac{F(\mu_0)}{\pi}\, .
\end{equation}
The specific intensity is then given by
\begin{equation}
I(\mu, \mu_0) = \frac{F(\mu_0)}{\pi} + 3 H (\mu + 1)\, ,
\end{equation}
where we have given the explicit dependence of the emergent specific 
intensity on $\mu_0$.

For close-in planets, the irradiation flux is much larger than the
intrinsic flux, $4\pi H$, so we neglect the second term.
The local atmosphere characterized by angle $\mu_0$ exhibits,
within the present approximations, an essentially isotropic emergent radiation
pattern, independent of the local polar angle $\mu$ and dependent only on the
angular distance from the substellar point, $\mu_0$.
For the total planetary flux close to secondary eclipse received 
by a observer at a distance $D$, we have the expression
\begin{equation}
\label{fobs1}
(D/{\rm R}_p)^{2} F_{\rm obs} = \int_0^1 I(\mu_0, \mu_0) \mu_0\, d\mu_0 =
\frac{1}{\pi} \int_0^1 F(\mu_0) \mu_0 d\mu_0 =
\frac{4}{3} \left( \frac{R_\ast}{a}\right)^2 H_\ast \, .
\end{equation}
For an average atmosphere characterized by the parameter, $f$,
the external flux is given by eq. (\ref{fav}), where we replace $1/2$ by
$f$. We obtain
\begin{equation}
I(\mu, \mu_0) = \frac{F(\mu_0)}{\pi} = 4\left(\frac{R_\ast}{a}\right)^2
H_{\ast}\, f\, ,
\end{equation}
and, therefore, 
\begin{equation}
\label{fobs2}
(D/{\rm R}_p)^{2} F_{\rm obs} = \int_0^1 I(\mu_0, \mu_0) \mu_0\, d\mu_0 =
4\left(\frac{R_\ast}{a}\right)^2 H_{\ast}\, f \int_0^1\mu_0\, d\mu_0 =
\frac{4f}{2}\left(\frac{R_\ast}{a}\right)^2 H_{\ast}\,  .
\end{equation}
In order to get agreement between eqs. (\ref{fobs1}) and 
(\ref{fobs2}), we have to set $f=2/3$. This is the origin of our use of this value.  

\end{appendix}

{}

\clearpage

\begin{table*}
\small
\begin{center}
\caption{Transiting Planet Data\tablenotemark{1}}
\tablewidth{17.0cm}
\begin{tabular}{ccccccc}
\hline\hline
    Planet&      a   & Period    &  M$_{p}$                & R$_{p}$   &F$_{p}$& Ref.   \\
               &   (AU)     &  (day)      &  ($M_{J}$)  & ($R_{J}$)   &(${\rm 10^9\ erg\ cm^{-2}\ s^{-1}}$)&    \\
\hline
   OGLE-TR-56b &  0.0225  & 1.2119 &  $1.29\pm0.12         $ & $1.30\pm0.05            $& 5.912 & 1,2,3,4,5 \\
       TrES-3  &  0.0226  & 1.3062 &  $1.92\pm0.23         $ & $1.295\pm0.081          $& 1.567 & 31 \\
  OGLE-TR-113b &  0.0229  & 1.4325 &  $1.32\pm0.19         $ & $1.09\pm0.03            $& 0.739 & 1,2,4,6,7,8 \\
  CoRoT-Exo-1b &  0.026   & 1.5    &  $1.3$                  & $1.65-1.78$                   & $\cdots$ & $\cdots$    \\
GJ 436b  &  0.0285  & 2.6438 &  $0.071\pm0.006   $ & $0.386\pm{0.016}   $& 0.044 & 32,35,36(8 $\mu$m) \\
  OGLE-TR-132b &  0.0306  & 1.6899 &  $1.14\pm0.12         $ & $1.18\pm0.07            $& 3.500 & 26 \\
       WASP-2b &  0.0307  & 2.1522 &  $0.88\pm0.11         $ & $1.04\pm0.06            $& 0.579 & 10,11  \\
    HD 189733b &  0.0313  & 2.2186 &  $1.15\pm0.04         $ & $1.15\pm0.03            $& 0.468 & 4,12,13 \\
       WASP-3b &  0.0317  & 1.8463 &  $1.76^{+0.08}_{-0.14}$ & $1.31^{+0.07}_{-0.14}$   & 3.520 & 44  \\
       TrES-2  &  0.0367  & 2.4706 &  $1.28^{+0.09}_{-0.04}$ & $1.24^{+0.09}_{-0.06}   $& 1.150 & 14,27    \\
         XO-2b &  0.0369  & 2.6158 &  $0.57\pm0.06         $ & $0.973^{+0.03}_{-0.008} $& 0.759 & 29    \\
       WASP-1b &  0.0379  & 2.5199 &  $0.79^{+0.13}_{-0.06}$ & $1.40\pm0.08            $& 2.488 & 10,15,28  \\
     HAT-P-3b &  0.0389  & 2.8997 &  $0.599\pm0.026       $ & $0.89\pm{0.046}          $& 0.395 & 37      \\
       TrES-1  &  0.0393  & 3.0301 &  $0.75\pm0.07         $ & $1.08\pm0.3             $& 0.428 & 1,2,4,16,17 \\
    HAT-P-5b   & 0.0408  & 2.7885 &  $1.06\pm{0.11}$     & $1.26\pm{0.05}$            & 1.259 & 41      \\
   OGLE-TR-10b & 0.0416   & 3.1013 &  $0.63\pm0.14         $ & $1.26\pm0.07            $& 1.344 & 1,2,4,5,18 \\
    HD 149026b &   0.042  & 2.8766 &  $0.36\pm0.03$       & $0.73\pm0.03$              & 2.089 & 4,19    \\
    HAT-P-4b   & 0.0446   & 3.0565 &  $0.68\pm{0.04}$     & $1.27\pm{0.05}$            & 1.833 & 39      \\
    HD 209458b &   0.045  & 3.5247 &  $0.64\pm0.06         $ & $1.32\pm0.03            $& 1.074 & 4,20,21 \\
  OGLE-TR-111b &   0.047  & 4.0144 &  $0.52\pm0.13         $ & $1.07\pm0.05            $& 0.248 & 1,2,4,22 \\
         XO-3b &  0.0477  & 3.1915 &  $13.24\pm0.64        $ & $1.92\pm{0.16}          $& 4.156 & 33\\ 
         TrES-4 &  0.0488  & 3.5539 &  $0.84\pm0.10        $ & $1.674\pm{0.094}        $& 2.306 & 38\\ 
         XO-1b &  0.0488  & 3.9415 &  $0.90\pm0.07         $ & $1.18^{+0.03}_{-0.02}   $& 0.485 & 23,24    \\

    OGLE-TR-211b &  0.051  & 3.6772 &  $1.03\pm0.20         $ & $1.36^{+0.18}_{-0.09}   $& 2.034 & 45 \\

    OGLE-TR-182b &  0.051  & 3.9791 &  $1.01\pm{0.15}$      & $1.13^{+0.24}_{-0.08}$        & 0.755 & 43 \\

    HAT-P-6b   & 0.0526  & 3.8530 &  $1.06\pm{0.12}$     & $1.33\pm{0.06}$            & 1.755 & 42      \\
     HAT-P-1b &  0.0551  & 4.4653 &  $0.53\pm0.04         $ & $1.203\pm{0.051}   $& 0.681 & 25,34      \\
     HAT-P-2b &  0.0685  & 5.6334 &  $8.17\pm0.72         $ & $1.18\pm{0.16}          $& 1.326 & 30      \\

     HD 17156b &  0.1594  & 21.2173 &  $3.11^{+0.035}_{-0.013}   $ & $0.964^{+0.016}_{-0.027}  $& 0.161 & 40,46,47      \\ 


\hline
\end{tabular}
\tablenotetext{1}{Data, plus representative references, for 29 of the known transiting EGPs
with measured M$_{\rm p}$ and R$_{\rm p}$.  The list is in order of increasing semi-major axis.  F$_p$ is the
stellar flux at the planet's substellar point, given the stellar luminosities provided in Table \ref{t2}.
}
\bigskip
\tablerefs{
(1) Santos et al. (2006a),
(2) Santos et al. (2006b),
(3) Vaccaro \& Van Hamme (2005),
(4) Melo et al. (2006),
(5) Pont et al. (2007a),
(6) Gillon et al. (2006),
(7) Bouchy et al. (2004),
(8) Konacki et al. (2004),
(9) Moutou et al. (2004),
(10) Cameron et al. (2007),
(11) Charbonneau et al. (2007),
(12) Bouchy et al. (2005),
(13) Bakos et al. (2006),
(14) O'Donovan et al. (2006),
(15) Shporer et al. (2007),
(16) Alonso et al. (2004),
(17) Winn, Holman, \& Roussanova  (2007),
(18) Holman et al. (2007),
(19) Sato et al. (2005),
(20) Santos, Israelian, \& Mayor (2004),
(21) Knutson et al. (2007a),
(22) Winn, Holman, \& Fuentes (2007),
(23) Holman et al. (2006),
(24) McCullough et al. (2006),
(25) Bakos et al. (2007a),
(26) Gillon et al. (2007a), 
(27) Sozzetti et al. (2007), 
(28) Stempels et al. (2007),
(29) Burke et al. (2007),
(30) Bakos et al. (2007b),
(31) O'Donovan et al. (2007),
(32) Gillon et al. (2007b),
(33) Johns-Krull et al. (2007),
(34) Winn et al. (2007), 
(35) Gillon et al. (2007c), 
(36) Deming et al. (2007), 
(37) Torres et al. (2007), 
(38) Mandushev et al. (2007), 
(39) Kov\'{a}cs et al. (2007), 
(40) Barbieri et al. (2007), 
(41) Bakos et al. (2007c), 
(42) Noyes et al. (2007), 
(43) Pont et al. (2007b),  
(44) Pollacco et al. (2007), 
(45) Udalski et al. (2007), 
(46) Gillon et al. (2007d), 
(47) Fischer et al. (2007) 
}
\label{t1}
\end{center}
\end{table*}

\clearpage

\begin{table*}
\small
\begin{center}
\caption{Data on Parent Stars\tablenotemark{1}}
\tablewidth{17.0cm}
\begin{tabular}{llllllllll}
\hline\hline
   Star & Sp.T.  & $R_*$              &$T_{\rm eff}$&$\log g$&[Fe/H]$_*$ & M$_*$ & L$_*$    & Age             & Dist  \\
              &        &  ($R_{\odot}$)             &  (K)   & (cgs)  &       &($M_{\odot}$)&($L_{\odot}$) &(Gyr)&(pc) \\
\hline
   OGLE-TR-56 &      G & $1.32\pm0.06            $& 6119 & 4.21 &  0.25 & 1.04 & 2.20   &$2.5^{+1.5}_{-1.0} $& 1600 \\
   TrES-3     &    G3V & $0.80\pm0.05            $& 5650 & 4.60 & $-0.19$ & 0.90 & 0.59 &$\cdots$& $\cdots$ \\
  OGLE-TR-113 &      K & $0.77\pm0.02            $& 4804 & 4.52 &  0.15 & 0.78 & 0.29  &$5.35\pm{4.65}    $&  550 \\
  CoRoT-Exo-1 &      G & $1.2\pm0.2$             & $\cdots$ & $\cdots$ &  $\cdots$ & $\cdots$ & $\cdots$ & $\cdots$ &$>460$ \\
  GJ 436      &  M2.5V & $0.44\pm0.04            $& 3500 & 4.5 &  0.0 & 0.44 & 0.026  &$> 3$              &  10.2 \\

  OGLE-TR-132 &      F & $1.34\pm0.08            $& 6210 & 4.51 &  0.37 & 1.26 & 2.41   &$1.25\pm{0.75}    $& 2200 \\ 
       WASP-2 &    K1V & $0.81\pm0.03          $& 5200 & 4.50  &   $\cdots$ & 0.79 & 0.44 & $\cdots$ & $\cdots$  \\
    HD 189733 &  K1.5 & $0.76\pm0.02          $& 5050 & 4.53 & $-0.03$ & 0.82 & 0.34  &$5.25\pm{4.75}  $& 19.3 \\
       WASP-3 &    F7V & $1.31^{+0.05}_{-0.12}    $& 6400 & 4.30  & $\cdots$   & 1.24 & 2.60   & $\cdots$ &  223  \\
       TrES-2 &    G0V & $1.00^{+0.06}_{-0.04}   $& 5960 & 4.40 & $-0.15$ &1.08& 1.14 &$4.9^{+2.9}_{-2.0}  $& $\cdots$ \\
         XO-2 &    K0V & $0.964^{+0.02}_{-0.009}$& 5400 & 4.62 & 0.45 & 0.98 & 0.76  &$5.0^{+1.0}_{-0.5}$ &  149 \\
       WASP-1 &    F7V & $1.45\pm0.03          $& 6110 & 4.28  &  0.26   & 1.3 & 2.67   & $2.0\pm{1.0}  $ &  $\cdots$   \\
       HAT-P-3 &    KV & $0.824^{+0.036}_{-0.062}  $& 5185 & 4.61  &  0.27   & 0.94 & 0.44  & $0.4^{+6.5}_{-0.3}$ &  140   \\
       TrES-1 &    K0V & $0.81\pm0.02           $& 5226 & 4.40 &  0.06 & 0.88 & 0.49  &$4.0\pm{2.0}      $&  143 \\
      HAT-P-5 &    G0   & $1.17\pm{0.05}$     & 5960 & 4.37  &  0.24   & 1.16 & 1.54  & $2.6\pm{1.8}$ &  340   \\
   OGLE-TR-10 & G & $1.16\pm0.06            $& 6075 & 4.54 &  0.28 & 1.02 & 1.65   &$2.0\pm1.0        $& 1300 \\
    HD 149026 &  G0 IV & $1.45\pm0.10            $& 6147 & 4.26 &  0.36 & 1.3  & 2.71   &$2.0\pm{0.8}      $& 78.9 \\
      HAT-P-4 &    G0   & $1.59\pm{0.07}$     & 5860 & 4.14  &  0.24   & 1.26 & 2.68  & $4.2^{+2.6}_{-0.6}$ &  310   \\
    HD 209458 &   G0 V & $1.13\pm0.02            $& 6117 & 4.48 &  0.02 & 1.10& 1.60   &$5.5\pm{1.5}      $&   47 \\
  OGLE-TR-111 & G/K & $0.83\pm0.03          $& 5044 & 4.51 &  0.19 & 0.81 & 0.40  &$5.55\pm{4.45}    $& 1000 \\
         XO-3 &   F5V & $2.13\pm{0.21}$        & 6429 & 3.95 & $-0.18$ & 1.41 & 6.96  &$2.8\pm{0.15}$ &  260 \\ 
         TrES-4 &   F & $1.74\pm{0.09}$        & 6200 & 4.05 & 0.14 & 1.22 & 4.04 &$4.7\pm{2.0}$ & 440 \\ 
         XO-1 &    G2V & $0.93^{+0.02}_{-0.01}$& 5750 & 4.53 & 0.015 & 1.00 & 0.85  &$4.6\pm{2.3}      $&  200 \\

         OGLE-TR-211 &  F7V & $1.64^{+0.21}_{-0.07}$& 6325 & 4.22 & 0.11 & 1.33 & 3.88  &$\cdots$   &  $\cdots$ \\

         OGLE-TR-182 &  G0V & $1.14^{+0.23}_{-0.06}$& 5924 & 4.47 & 0.37 & 1.14 & 1.44  &$\cdots$   &  $\cdots$ \\

         HAT-P-6 &    F8 & $1.46\pm{0.06}$ & 6570 & 4.22 & $-0.13$ & 1.29 & 3.57  &$2.3^{+0.5}_{-0.7}$    &  260 \\
     HAT-P-1 &    G0V & $1.15^{+0.10}_{-0.07}   $& 5975 & 4.45 &  0.13 & 1.12 & 1.52    &$3.6\pm{1.0}      $&  139 \\
     HAT-P-2 &    F8V & $1.80\pm0.25            $& 6290 & 4.22 &  0.12 & 1.35 & 4.58   &$2.7^{+1.4}_{-0.6}     $&  135 \\
     HD 17156 &    G0V & $1.354^{+0.012}_{-0.037}  $& 6079 & 4.29 &  0.24 & 1.20 & 2.66 &$4.7^{1.3}_{1.9}$&  78.24 \\


\hline
\end{tabular}
\tablenotetext{1}{A compilation of the physical parameters derived for the parents of 
29 of the known transiting EGPs.  The error bars have been rounded from those found 
in the literature.  The ages, the least well-known quantities, should be taken 
with caution. The stellar metallicities are given without
error bars, which are assumed to be large. Due to their great distances
(rightmost column), the stellar types of the OGLE objects are not well constrained.
Refer to Table \ref{t1} for the corresponding references.}
\label{t2}
\end{center}
\end{table*}

\clearpage
\begin{deluxetable}{lllllll}
\tabletypesize{\tiny}
\tablewidth{0pt}
\tablecaption{Spectral, Photometric, and Composition Measurements of Extrasolar Giant Planets\label{t3}}
\tablehead{
\colhead{Planet} &
\colhead{$\lambda$} &
\colhead{Telescope/} &
\colhead{${\rm F}_{p}/{\rm F}_{\star}$} &
\colhead{$\Delta {\rm F}_{p}/{\rm F}_{\star}$} &
\colhead{Comments} &
\colhead{Reference} \\
\colhead{ } &
\colhead{$\mu$m} &
\colhead{Instrument} &
\colhead{Second.   Eclipse} &
\colhead{ } &
\colhead{ } &
\colhead{ }
}
\startdata
HD 189733 b  & 7.5-14.7 & Spitzer/IRS   & Spectrum               &                        & No $H_{2}O?,\,CH_{4}?$ & Grillmair et al. (2007)\tablenotemark{1}  \\
             & 8.0      & Spitzer/IRAC4 & \bf 0.003392(55)       & \bf 0.0012(2)          & Light curve            & Knutson et al. (2007b)\tablenotemark{2}   \\
             & 16.0     & Spitzer/IRS   & \bf 0.00551(30)        &                        & Peak-up mode           & Deming et al. (2006)        \\ 
             & 3.6      & Spitzer/IRAC1 & 0.02356(020)       &                        &  Primary Transit Depth $-$ & Beaulieu et al. (2007)   \\
             & 5.8      & Spitzer/IRAC3 & 0.02436(020)       &                        &  (Tinetti et al. 2007) & Beaulieu et al. (2007)   \\
             & 8.0      & Spitzer/IRAC4 & 0.0239(02)       &                          &  "                     & Knutson et al. (2007b)\\
TrES-1       & 4.5      & Spitzer/IRAC2 & \bf 0.00066(13)        &                        &  $H_2O$ identified $-$ & Charbonneau et al. (2005)   \\
             & 8.0      & Spitzer/IRAC4 & \bf 0.00225(36)        &                        &  (Burrows et al. 2005)  & Charbonneau et al. (2005)   \\
HD 209458 b &Ly $\alpha$& HST/STIS      &                        &                        & HI, R$_{p}=4.3\,R_{J}$     & Vidal-Madjar et al. (2003)\tablenotemark{4}\\
             & 0.12-0.17& HST/STIS      &                        &                        & C, O                   & Vidal-Madjar et al. (2004)\tablenotemark{5}\\
             & 0.3-0.5  & HST/STIS      &                        &                        & R$_P=1.3300(6)\,R_{J}$   & Ballester et al. (2007)\tablenotemark{6} \\
             &          &               &                        &                        & at Balmer cont.?        &                             \\
             & 0.58-0.64& HST/STIS      &                        &                        & Na D (transit)                  & Charbonneau et al. (2002)\tablenotemark{7} \\
             & $\sim$0.94 & HST/STIS      &                      &                        & $H_2O$ identified (transit)               & Barman 2007  \\
             & 0.4-0.7  & MOST          & $< 1.34\times 10^{-4}~(3\sigma)$ &              &$A_{g}<0.68~(3\sigma)$  & Rowe et al. (2006)\tablenotemark{8} \\
             &          &               & $< 4.88\times 10^{-5}~(1\sigma)$ &              &$A_{g}<0.25~(1\sigma)$  & Rowe et al. (2006)          \\
             & 0.4-0.7  & MOST          & $< 3.9\times 10^{-5}~(1\sigma)$  &              &$A_{g}\sim 4.0\pm 4.0$\%& Rowe et al. (2007)          \\
             & 2.2      & IRTF/SpeX     & $< 0.0003~(1\sigma) $       &                        &                        & Richardson et al. (2003)    \\  
             & 3.6      & Spitzer/IRAC1 &                        & $< 0.015~(2\sigma)$    & Light curve            & Cowan et al. (2007)         \\
             & 8.0      & Spitzer/IRAC4 &                        & $< 0.0015~(2\sigma)$   & Light curve            & Cowan et al. (2007)         \\
             &          &               &                        &                        &$P_{n}\ \sgreat\ 0.32\,(1 \sigma)$&                             \\
             & 3.6      & Spitzer/IRAC1 & \bf 0.00094(9) &                        & Temperature Inversion $-$ & Knutson et al. (2007c)\\
             & 4.5      & Spitzer/IRAC2 & \bf 0.00213(15) &                        &  $H_{2}O$ in emission $-$  & Knutson et al. (2007c)\\
             & 5.8      & Spitzer/IRAC3 & \bf 0.00301(43) &                        &  (Burrows et al. (2007b)   & Knutson et al. (2007c)\\
             & 8.0      & Spitzer/IRAC4 & \bf 0.00240(26) &                        &   "                    & Knutson et al. (2007c)\\
             & 7.5-13.2 & Spitzer/IRS   & Spectrum               &                        & Features at            & Richardson et al. (2007)\tablenotemark{9} \\
             &          &               &                        &                        & 7.78, 9.65 $\mu$m ?      &                                                  \\
             &          &               &                        &                        & No $H_{2}O$, $CH_4$?   &                                                  \\
             & 8.2-13.2 & Spitzer/IRS   & Spectrum               &                        &$F_{p}=0.40\pm 0.19$    & Swain et al. (2007)\tablenotemark{10}  \\
             &          &               &                        &                        & mJy at 12 $\mu$m       &                             \\
            & 24.0     & Spitzer/MIPS  & \bf 0.00260(46)        &                        &                        & Deming et al. (2005)        \\
             & 24.0     & Spitzer/MIPS  & \bf 0.0033(3) (?)     &                        &                        & Deming (private communication)        \\ 
HD 149026 b   & 8.0      & Spitzer/IRAC4 & \bf 0.00084(11)        &                        &                        & Harrington et al. (2007)    \\
$\upsilon$ And b    & 24.0     & Spitzer/MIPS  & No transit             & \bf 0.0029(7)          & Light curve            & Harrington et al. (2006)    \\
HD 179949 b  & 3.6      & Spitzer/IRAC1 & No transit             & $< 0.019~(2\sigma)$    & Light curve            & Cowan et al. (2007)         \\  
             & 8.0      & Spitzer/IRAC4 & No transit             & \bf 0.00141(33)        & Light curve            & Cowan et al. (2007)\tablenotemark{3} \\  
             &          &               &                        &                        &$P_{n}\ \sles\ 0.30\,(1 \sigma)$&                             \\
GJ 436b   & 8.0      & Spitzer/IRAC4 & \bf 0.00057(8)        &                        &                        & Deming et al. (2007)    \\
   &     &  &         &                        &                        & Demory et al. (2007)    \\
51 Peg b     & 4.5      & Spitzer/IRAC2 & No transit             & $< 0.017~(2\sigma)$    & Light curve            & Cowan et al. (2007)         \\
             & 8.0      & Spitzer/IRAC4 & No transit             & $< 0.0007~(2\sigma)$   & Light curve            & Cowan et al. (2007)         \\
OGLE-113 b   & 2.2      & NTT/SOFI      & 0.0017(5)~$(3\sigma)$? &                        &                        & Snellen \& Covino (2007)    \\         
\enddata
\tablecomments{Notes $-$ This table contains the following columns:
the name of the planet, spectral region of the observations, 
the telescope with the instrument used for the observations,
the planet to star flux ratio during the secondary eclipse (${\rm F}_{p}/{\rm F}_{\star}$), 
the amplitude or peak to trough variations of the light curve ($\Delta {\rm F}_{p}/{\rm F}_{\star}$),
brief comments, and the reference. Numbers in {\bf bold} are measurements (not upper limits)
at secondary eclipse or actual light curve ($\Delta {\rm F}_{p}/{\rm F}_{\star}$) measurements.
Models for all these bolded data (except the 8-$\mu$m point for GJ 436b at secondary 
eclipse) are presented in this paper.  Numbers in parentheses are errors in the 
last digits. \tablenotetext{1} Grillmair et al. (2007) argue that a flat spectrum is caused by 
the lack of significant water or methane absorption. However, this is not 
consistent with the conclusions of Tinetti et al. (2007), using the primary 
eclipse data of Beaulieu et al. (2007) and Knutson et al. (2007b);
\tablenotetext{2} Knutson et al. (2007b) also derived the longitudinal dependence of the surface brightness
and found a hot spot shifted by $16\pm 6$ degrees east of the substellar point while
the coolest region was shifted about 30 degrees west of the anti-stellar point.
They also found an indication of nonzero eccentricity with $e{\cos {\omega}}=0.0010\pm 0.0002$,
a transit radius at 8 $\mu$m of $1.137\pm 0.006\, R_{J}$, a stellar radius of
$0.757\pm 0.003\, R_{\odot}$ and an inclination of $85.61\pm 0.04$ degrees, 
where $\omega$ is the longitude of periastron;
\tablenotetext{3} Cowan et al. (2007) constrain P$_{n}$ for HD 209458b and HD 179949b to be 
P$_{n}>0.32~(1\sigma)$ and P$_{n}<0.30~(1\sigma)$, respectively;
P$_{n}$ is the fraction of the total energy incident on the dayside of the planet
which is transfered to and radiated out on the night side of the planet;
\tablenotetext{4} Vidal-Madjar et al. (2003) detected atomic hydrogen in the planet's atmosphere
with a transit absorption depth of $15\pm 4 (1\sigma)$\%,
and evaporation at the rate of $\ge 10^{10}g\,s^{-1}$;
\tablenotetext{5} Vidal-Madjar et al. (2004) detected oxygen and carbon in the planet's atmosphere;
\tablenotetext{6} Ballester et al. (2007) claim to have identified HI absorption in the Balmer continuum,
but Barman (2007) challenges this interpretation; \tablenotetext{7} Charbonneau et al. 
(2002) found that the transit depth at the Na D feature is deeper by 
$2.32\pm 0.57\times 10^{-4}$ than in the continuum,
which is interpreted as a detection of Na in the planet's atmosphere;
\tablenotetext{8} Rowe et al. (2006) constrained $A_{g}<0.25~(1\sigma)$, or $A_{g}<0.68~(3\sigma)$,
but Rowe et al. (2007) constrained $A_{g}$ to be below 8\% to 1$-\sigma$.  $A_{g}$ is
the geometric albedo in the optical;
\tablenotetext{9} Richardson et al. (2007) claim to have detected a broad emission peak centered near 9.65 $\mu$m 
which they attribute to the emission by silicate clouds, and a narrow unidentified 
emission feature at 7.78 $\mu$m. They say that models with water absorption fit the data poorly.  
However, Burrows et al. (2007b) conclude that water is in fact seen in emission, not absorption;
\tablenotetext{10} Swain et al. (2007) determined the planet flux at 12 $\mu$m to be $0.40\pm 0.19$mJy
and the normalized secondary eclipse depth to be $0.0046\pm 0.0006$.  They are updating 
their absolute calibrations.}
\end{deluxetable}

\clearpage
\thispagestyle{empty}
\setlength{\voffset}{-10mm}

\clearpage
\setlength{\voffset}{0mm}

\clearpage

\clearpage

\clearpage

\clearpage
\thispagestyle{empty}
\setlength{\voffset}{-25mm}

\clearpage
\setlength{\voffset}{0mm}

\clearpage

\end{document}